\documentclass[preprint,12pt]{elsarticle}

\usepackage{amssymb,bm,amsmath}
\usepackage{svg}
\usepackage{booktabs}
\usepackage{multirow}
\usepackage{array} % Added for p{} column specifier
\usepackage{mwe}
\usepackage{graphbox} %loads graphicx package\usepackage{caption}
\usepackage{geometry}
\newcolumntype{M}[1]{>{\centering\arraybackslash}m{#1}}
\usepackage{subcaption}
\usepackage{comment}
\usepackage[linesnumbered,ruled,vlined]{algorithm2e}
\usepackage{array}
\usepackage{ragged2e}
\usepackage{float}

\DeclareMathOperator*{\argmax}{argmax}

\newcommand{\revminor}[1]{#1}

\journal{Journal of Computational Physics}

\begin{document}

\begin{frontmatter}



%\title{Eulerian Interface Tracking with Triangle Edge Cuts}
\title{An Interface Tracking Method with Triangle Edge Cuts}


\author[inst1]{Mengdi Wang\corref{cor1}}

\cortext[cor1]{Corresponding author.}

\fntext[emails]{E-mail addresses: mengdi.wang@gatech.edu (Mengdi Wang), matthew.d.cong@gmail.com (Matthew Cong), bo.zhu@gatech.edu (Bo Zhu).}

\affiliation[inst1]{organization={School of Interactive Computing},%Department and Organization
            addressline={Georgia Institute of Technology}, 
            city={Atlanta},
            postcode={30332}, 
            state={Georgia},
            country={USA}}

\author[inst2]{Matthew Cong}

\affiliation[inst2]{organization={NVIDIA Corporation},%Department and Organization
            addressline={2788 San Tomas Expy}, 
            city={Santa Clara},
            postcode={95051}, 
            state={California},
            country={USA}}

\author[inst1]{Bo Zhu}

\begin{abstract}
This paper introduces a volume-conserving interface tracking algorithm on unstructured triangle meshes. We propose to discretize the interface via \textit{triangle edge cuts} which represent the intersections between the interface and the triangle mesh edges using a compact $6$ numbers per triangle. This enables an efficient implicit representation of the sub-triangle polygonal material regions without explicitly storing connectivity information. Moreover, we propose an efficient advection algorithm for this interface representation that is based on geometric queries and does not require an optimization process. This advection algorithm is extended via an area correction step that enforces volume-conservation of the materials. We demonstrate the efficacy of our method on a variety of advection problems on a triangle mesh and compare its performance to existing interface tracking methods including VOF and MOF.
\end{abstract}




\end{frontmatter}


\section{Introduction}

The interface between immiscible fluids plays a fundamental role in understanding many real-world phenomena such as bubble dynamics, fluid fragmentation, and ocean surface swirls. \textit{Interface tracking}, i.e., numerically modeling and evolving the interface, therefore has been a long-standing challenge in computational fluid dynamics.\cite{tryggvason2011direct,luo2004computation,pan2023edge}
The physical simulator requires precise information about the interface's location and various related physical quantities, such as total mass within a cell, and/or boundary conditions at the liquid-air interface. Essentially, interface tracking encompasses two primary components: A data structure for representing the interface and an advection algorithm for evolving the interface over time. In a typical simulation setup, the interface is discretized using primitives such as points, segments, and polygons and evolved intending to accurately preserve both the interface geometry and the total mass of materials.

Traditionally, interface tracking methods are classified as either Lagrangian or Eulerian. In a typical Lagrangian method, the interface geometry is represented by a low-dimensional Lagrangian data structure independent of the computational grid, \textit{e.g.}, marker points \cite{shin2002modeling,chen1997surface}, connected meshes and/or curves \cite{popinet1999front,tryggvason2001front,unverdi1992front}, and various other structures \cite{chen2022characterizing,chirco2022manifold}. While these methods can represent complex interfaces, accurately time-stepping the Lagrangian data structures often requires frequent remeshing to represent topological transitions and adequately sample large geometric distortions. These geometric operations may hinder the efficiency of implementation, especially on parallel systems in the presence of topology changes which often require (global) remeshing. On the other hand, Eulerian methods involve representing the interface on an Eulerian grid or mesh fixed in the space. For example, the level set (LS) method \cite{osher2004level,osher1988fronts,osher2001level,gibou2018review} adopts an implicit function $\phi$ that defines the interface as the iso-surface $\phi(\bm{x})=0$. 
The level set method has been widely used in computational physics applications to track and solve dynamic interface problems and has proven its efficacy in handling complex topological and geometrical evolution without tackling local mesh repairments. However, it also suffers from volume loss, especially where/when thin features occur, due to the inherent limitations of representing these features on a grid with fixed resolution. 

\begin{figure}
    \centering
    \begin{subfigure}{0.24\textwidth}
        \includegraphics[width=\linewidth]{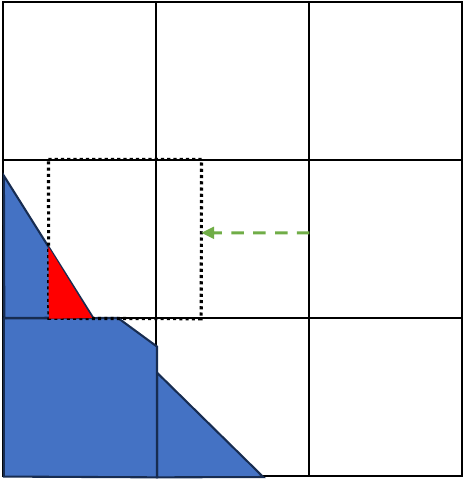}
        \caption{split PLIC-VOF}
    \end{subfigure}
    \begin{subfigure}{0.24\textwidth}
        \includegraphics[width=\linewidth]{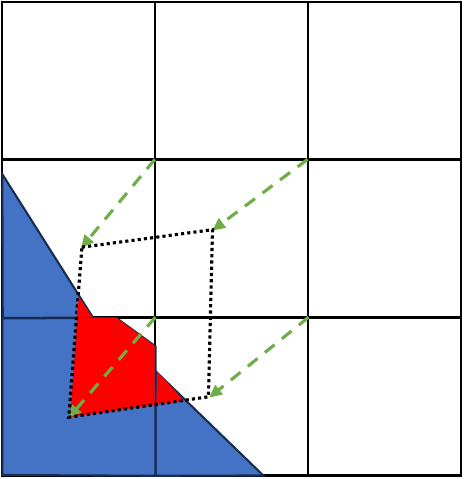}
        \caption{MOF}
    \end{subfigure}
    \begin{subfigure}{0.24\textwidth}
        \includegraphics[width=\linewidth]{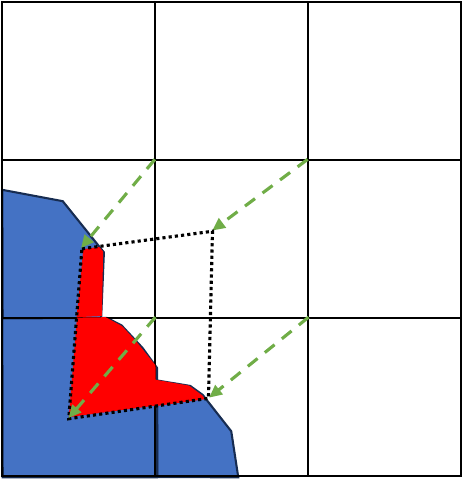}
        \caption{PAM}
    \end{subfigure}
    \begin{subfigure}{0.24\textwidth}
        \includegraphics[width=\linewidth]{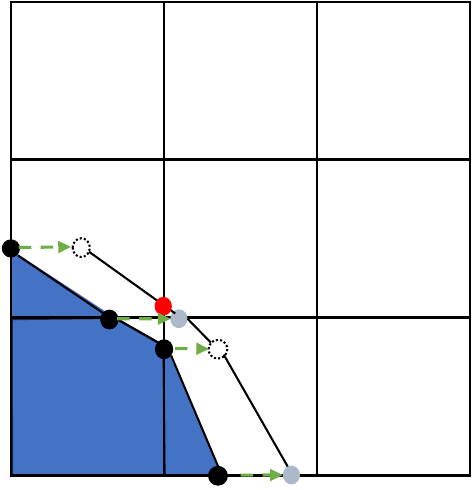}
        \caption{EBIT}
    \end{subfigure}
    \caption{Different advection methods of direction-splitting PLIC-VOF, MOF, PAM, and EBIT.  (a) The cell is translated along the $x$-axis, and its intersection with the interface (red area) is calculated as the flux. (b) \revminor{In} MOF advection, \revminor{the pre-image $\overleftarrow{C_{i,j}}$ of cell $(i,j)$ is first calculated}, and then the zeroth and first moments \revminor{are calculated} by intersecting it with the liquid region at \revminor{the} last time step. (c) PAM advection is similar to MOF, but it \revminor{accommodates} more interface segments inside a cell. (d) EBIT advection along \revminor{the} $x-$axis. Red and grey points \revminor{represent} new marker points \revminor{located} on grid lines.}
    \label{fig:advection-methods}
\end{figure}

\revminor{The} volume of fluid (VOF) \cite{hirt1981volume,brackbill1992continuum,chiodi2020advancement,rider1998reconstructing} is an Eulerian method that has gained \revminor{the} most significant attention for its ability to conserve fluid volume. The volume of fluid $F_{i,j}\in [0,1]$ is defined as the fraction of liquid inside a cell $(i,j)$, and its value is tracked throughout the algorithm. The commonly used piece-wise linear interface calculation (PLIC-VOF) method \cite{scardovelli1999direct,popinet2009accurate} assumes that the interface in a cell $(i,j)$ is a line segment, which is reconstructed from $F_{i,j}$ using certain numerical schemes \cite{aulisa2007interface,gueyffier1999volume}.  In the traditional direction-splitting \cite{debar1974fundamentals} advection method for PLIC-VOF, the advection problem of a cell $(i,j)$ is decomposed into two simple one-dimensional advection problems along the $x$ and $y$ axes respectively, and their fluxes are combined to obtain the change of $F_{i,j}$. Figure~\ref{fig:advection-methods}(a) illustrates an advection step of the central cell along the $x$-axis in split PLIC-VOF advection.

The PLIC method often suffers from the artifacts of the interface reconstruction, especially for thin geometric features. Typically, the PLIC method reconstructs a thin fluid sheet as a group of broken droplets. L\'opez et al.\ have proposed a double-PLIC method~\cite{lopez2005improved} to tackle this problem by allowing up to two parallel interface segments to coexist in one cell and adding a marker point at the midpoint of each interface segment.  These marker points are connected as segment meshes, whose topology will be used for spline interpolation. While double-PLIC is a hybrid Lagrangian-Eulerian approach and challenging for parallelization, it is important to recognize that representing the interface within a cell using only a single line segment is inherently inadequate for capturing thin interfaces. By placing two interface segments within a cell, double-PLIC effectively achieves sub-grid accuracy, improving the representation of these thin interfaces.

\begin{figure}[htbp]
  \centering
  \includegraphics[width=0.8\textwidth]{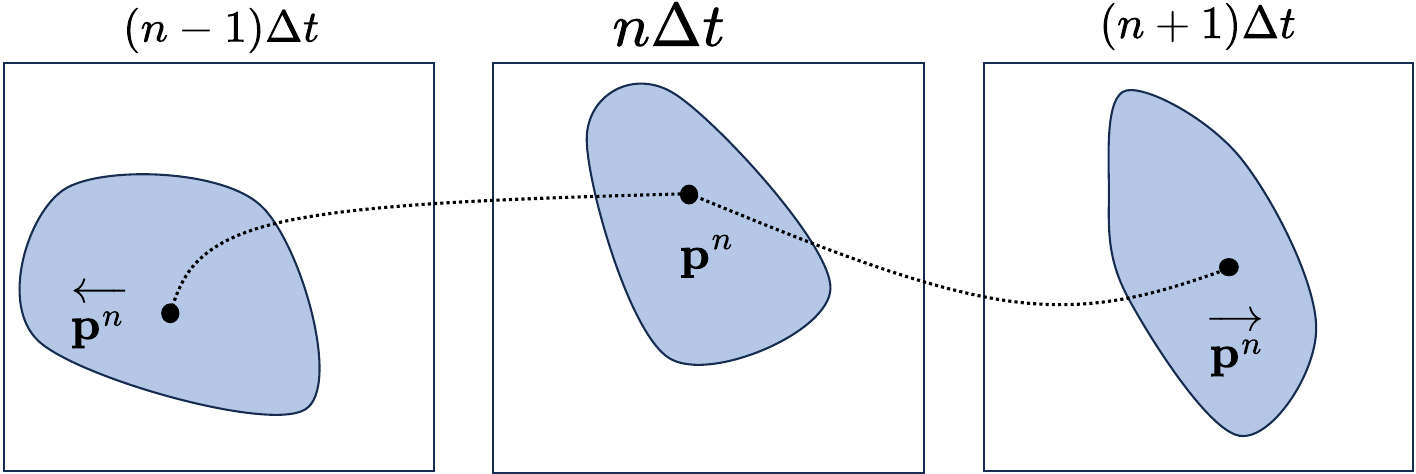}
  \caption{\revminor{The} pre-image of a point $\bm{p}^n$ at the previous time step $n-1$, and its image at \revminor{the subsequent} time step $n+1$.}
  \label{fig:image-preimage}
\end{figure}

In recent years, many improvements to VOF method have emerged, aiming to enhance representation and advection accuracy~\cite{mulbah2022review,maric2020unstructured}.
Notably, many of these methods address the advection problem by back-tracking a cell through the flow map to acquire its shape in the last frame. We formally describe this process using the concepts of \textit{image} and \textit{pre-image}. \revminor{As} illustrated in Figure~\ref{fig:image-preimage}, with superscript $n$ denoting the discretized time step $t=n\Delta t$, we define the \textit{image} $\overrightarrow{\bm{p}^n}$ and \textit{pre-image} $\overleftarrow{\bm{p}^n}$ of a point $\bm{p}^n$ as a mapping of point positions between different time steps~\cite{zhang2008new},
\begin{equation}
\begin{aligned}
\overrightarrow{\bm{p}^n} &= \bm{p}^{n+1} &= \bm{p}^n+\int_{t^n}^{t^{n}+\Delta t}\bm{u}(\bm{p}(t),t)\mathrm{d}t,\\
\overleftarrow{\bm{p}^n} &= \bm{p}^{n-1} &= \bm{p}^n+\int_{t^n}^{t^{n}-\Delta t}\bm{u}(\bm{p}(t),t)\mathrm{d}t.
\end{aligned}
\end{equation}
The current state-of-the-art VOF methods adopt unsplit advection schemes~\cite{maric2020unstructured}, which significantly reduce the geometric errors introduced by operator splitting. This error arises because the split \revminor{advection} steps can distort the interface, leading to artifacts such as the staircase phenomenon~\cite{pilliod2004second} and errors from clamping the VOF values to $[0,1]$.
Denoting the rectangle shape defined by cell $(i,j)$ as $C_{i,j}$, a typical unsplit advection scheme like EMFPA~\cite{lopez2004volume} and CCU~\cite{comminal2015cellwise} calculates the new value of $F_{i,j}$ by \revminor{determining} the fluid area inside its pre-image $\overleftarrow{C_{i,j}}$ at the last time step, ensuring volume conservation. Other methods~\cite{ivey2017conservative,maric2020unstructured} may use different types of control regions, but \revminor{ultimately} also involve intersecting their pre-images with the liquid region.

The MOF method~\cite{dyadechko2005moment,mukundan20203d,anbarlooei2009moment,shashkov2023moments,cutforth2021efficient} is also an improvement to \revminor{the} VOF method that operates explicitly on the pre-images of cells. In addition to the volume of fluid, \revminor{also known as} the zeroth moment of the fluid, the MOF method also tracks the first moment of fluid in a cell $C_{i,j}$, which is defined as $M_1=\int_{C_{i,j}}\bm{x}\mathrm{d}\bm{x}$, for higher accuracy of reconstruction. During the MOF advection, the moments of fluid are calculated from the intersection between the pre-image $\overleftarrow{C_{i,j}}$ and the liquid region at the last time step, which results in a set of polygons. The zeroth moment of fluid is given by summing up the areas of these polygons similar to unsplit VOF advection. However, the first moment is calculated by forward-tracking these polygons to get their images and then performing a weighted sum of the images' centroids~\cite{ahn2009adaptive}. This distinction \revminor{arises because} advection does not conserve the first moment. We illustrate the MOF advection process in Figure~\ref{fig:advection-methods}(b), where dotted lines draw the pre-image of the central cell, and its intersection with the liquid region is shown in red. The reconstruction step of the MOF method preserves the zeroth moment and tries to make the first moment of the reconstructed liquid polygon as close to the predicted value as possible, through a \revminor{nonlinear} optimization process~\cite{nocedal1999numerical}.

Another interface tracking method, known as the \textbf{polygon area mapping} (PAM) method~\cite{zhang2008new,zhang2009hypam,zhang2014fourth}, goes a step further by not explicitly relying on VOF values. 
Instead, it adopts a combined Lagrangian-Eulerian perspective by storing the liquid region inside each cell as a set of polygons. The advection \revminor{process} of PAM begins with a similar intersection calculation of \revminor{the} cell pre-image and \revminor{the} liquid polygons and uses the images of the resulting polygons as an initial approximation of the liquid region inside the cell. \revminor{Subsequently}, PAM applies a correction step involving \revminor{removing} small polygons and vertices~\cite{zhang2008new} to limit the number of polygons and vertices inside a cell. The advection step of PAM is illustrated in Figure~\ref{fig:advection-methods}(c). \revminor{PAM} is very similar to MOF, but \revminor{it} allows \revminor{for} a more complex interface in a cell.

Motivated by the \revminor{previously mentioned} methods, including double PLIC, unsplit VOF advection, MOF, and PAM, we aspire to formulate a sub-grid\revminor{-accurate} interface tracking algorithm by combining sub-grid interface and pre-image geometric calculation methods, which is particularly \revminor{suited} for handling thin fluid features. At the same time, we aim to design an algorithm \revminor{that has} lower complexity, a characteristic \revminor{that is} less emphasized in many existing methods. \revminor{Although} the PLIC interface can be reconstructed solely from VOF values, in practice, \revminor{simulators often require an} explicit interface, especially when calculating geometric quantities like interface curvature in surface tension modeling~\cite{popinet2009accurate, karnakov2020hybrid}. Therefore, PLIC-VOF requires $3$ DoFs for each cell. In order to represent two parallel interfaces, double-PLIC takes $4$ DoFs for two segments and $4$ for two marker points, \revminor{resulting in} a total of $8$ \revminor{DoFs per cell}.

MOF shows higher accuracy than the original version of split PLIC-VOF advection, at the cost of having $5$ DoFs for each cell ($2$ additional DoFs for the first moment), while still \revminor{representing only} one interface segment inside a cell. Based on MOF, \revminor{Shashkov and Kikinzon}~\cite{shashkov2023moments} propose MOF2, a second-order MOF that models sub-grid geometry by representing the liquid polygon as the union or intersection of two MOF-style liquid polygons. MOF2 tracks $3$ second moments of fluid, increasing the DoFs \revminor{per} cell to $8$. The complexity of the MOF method is also reflected in its interface reconstruction process which requires a costly nonlinear optimization. The complexity of moment tracking and interface reconstruction will further increase when \revminor{extending these} methods to 3D space. Similarly, although the PAM method can model complex sub-grid interfaces within a cell, its explicit definition of Lagrangian-style liquid polygons \revminor{significantly} increases \revminor{the} complexity of \revminor{its} data structure. For example, the maximum number of polygon vertices inside a cell can be up to $10$~\cite{zhang2008new}, making a total of $20$ DoFs per cell.

Recently, an efficient interface tracking method, \textbf{edge-based interface tracking} (EBIT)~\cite{chirco2023edge, pan2023edge}, has \revminor{gained community attention}. EBIT delineates the interface by a combination of vertex materials and a set of marker points along grid lines \revminor{that indicate} the intersection between the interface and the grid. In each cell, the intersection points implicitly define the interface \revminor{within} it as a segment, therefore, the interface in the whole computational field can be represented as a segment mesh. EBIT adopts a direction-splitting advection scheme, \revminor{where} the interface is moved along $x,y$ axes sequentially, and the new set of marker points is produced \revminor{through} calculating the intersection between the interface and the grid. Figure~\ref{fig:advection-methods}(d) \revminor{illustrates} the advection in \revminor{the} $x$ direction. EBIT has \revminor{only} $2$ DoFs for each cell, \revminor{as} each marker point is shared by two cells, highlighting its ability to represent the interface with a small amount of memory usage. However, EBIT struggles to conserve volume, and since it only allows one intersection on one edge, it often \revminor{fails} to represent features \revminor{like} thin fluid sheets. These shortcomings will be addressed in our proposed method.

The EBIT method inspired us to represent the interface elegantly using only a few degrees of freedom based on its intersections with the grid edges.
As we will discuss in Section~\ref{sec:conclusion}, we found that designing such an algorithm is more straightforward on a triangle mesh compared to a lattice grid.

Therefore, we propose an Eulerian interface-tracking algorithm that accurately tracks sub-triangle geometric features and \revminor{ensures mass conservation}. Our algorithm circumvents the issues related to maintaining connectivity information, \revminor{requires} less memory, and is \revminor{easy} to implement.
In our proposed algorithm, the interface representation on an unstructured triangle mesh is divided into the interface representation inside each triangle element \revminor{using} \textit{triangle edge cuts}, a novel data structure for representing two material regions inside the triangle. This data structure defines the intersections between the interface and the triangle \revminor{edges}. Assuming at most $2$ intersections on each edge, we can store the interface \revminor{using} no more than $6$ \revminor{values per} triangle, and the material regions are automatically reconstructed as polygons without ambiguity (see Section~\ref{sec:triangle-edge-cut}). The triangle edge cuts representation enables us to model sub-triangle interface geometries, including arbitrarily thin fluid features, with a low memory footprint. Moreover, we describe an efficient interface advection method for this representation in Section~\ref{sec:interface-advection}, \revminor{which queries only} segment intersections between the pre-images of triangle edges and the interface. To \revminor{ensure} higher accuracy and mass conservation, we propose an area correction method in Section~\ref{sec:area-correction} based on pre-image polygon intersections similar in spirit to MOF and PAM. In Section~\ref{sec:implementation}, we discuss the implementation details of our proposed algorithm. \revminor{We then} validate our algorithm using several static reconstruction and dynamic advection tests, \revminor{comparing} its accuracy and efficiency with state-of-the-art algorithms in Section~\ref{sec:experiments}. Finally, Section~\ref{sec:discussion-3d} discusses how to extend our proposed algorithm to 3D, while Section~\ref{sec:conclusion} provides a summary of this paper \revminor{and outlines} our future plans.

\section{Triangle Edge Cut Interface Representation}
\label{sec:triangle-edge-cut}

\begin{table}[h]
    \centering
    \begin{tabular}{|l|p{0.6\textwidth}|}
        \hline
        Variable & Definition \\
        \hline
        $\chi$ & Indicator function \\
        $\Omega$ & Material region\\
        $\mathcal{P}$ & Polygon\\
        $\mathcal{T}$ & Triangle\\
        $\mathcal{E}$ & Triangle edge cut\\
        $A$ & Liquid area\\
        $F$ & Relative area fraction\\
        $E_g$ & Absolute error\\
        $E_r$ & Relative error\\
        \hline
    \end{tabular}
    \caption{Terminology Table}
    \label{tab:terminology}
\end{table}

In this paper, we focus on the interface tracking problem in a physical system consisting of two materials in $\mathbb{R}^2$. Without loss of generality, we denote materials as $0$ (air) and $1$ (liquid) using the indicator function $\chi$ defined for any position $\bm{p}$:
\begin{equation}
    \chi(\bm{p})=
    \begin{cases}
        0,\ \text{if }\bm{p}\text{ is air},\\
        1,\ \text{if }\bm{p}\text{ is liquid},
    \end{cases}
\end{equation}
We denote the air and liquid regions respectively as:
\begin{equation}
    \begin{aligned}
        \Omega_0&=\{\bm{p}: \chi(\bm{p})=0\},\\
        \Omega_1&=\{\bm{p}: \chi(\bm{p})=1\},
    \end{aligned}
\end{equation}
and the liquid-air interface is the boundary of \revminor{the} material regions, which is a codimension-1 \revminor{geometric} structure, \textit{i.e.}, a curve in $\mathbb{R}^2$, or a surface in $\mathbb{R}^3$.

In our algorithm, we use the \textit{triangle edge cut} to represent the material region as polygons inside a triangle. A triangle edge cut $\mathcal{E}$ is defined as:
\begin{equation}
\label{eqn:triangle-edge-cut-representation}
    \mathcal{E}=(\mathcal{T},c,R).
\end{equation}
Here $\mathcal{T}=\Delta\bm{v}_1\bm{v}_2\bm{v}_3$ is a triangle defined by its three vertices $(\bm{v}_1,\bm{v}_2,\bm{v}_3)$. In the following \revminor{sections} of the paper, we will define its three edges as
\begin{equation}
    \begin{aligned}
        e_1&=\overline{\bm{v}_1\bm{v}_2},\\
        e_2&=\overline{\bm{v}_2\bm{v}_3},\\
        e_3&=\overline{\bm{v}_3\bm{v}_1}.
    \end{aligned}
\end{equation}
The boolean variable $c$ indicates the material of $\bm{v}_1$:
\begin{equation}
    c=\chi(\bm{v}_1).
\end{equation}
$R$ is a $3\times 2$ matrix that satisfies the following conditions:
\begin{equation}
    \begin{aligned}
        R_{i,j}&\in [0,1],\\
        R_{i,1}&\leq R_{i,2}.
    \end{aligned}
\end{equation}
Each row $R_i$ represents the vertices of material polygons on $e_i$ as interpolation \revminor{factors}. Specifically, $R_{i,j}$ corresponds to a vertex \revminor{defined as}
\begin{equation}
    \bm{r}_{i,j}=(1-R_{i,j})\bm{v}_i+R_{i,j}\bm{v}_{i+1}.
\end{equation}
We also call it a \textit{cut} because $\bm{r}_{i,j}$ is essentially an intersection between the triangle edge and the interface. If $R_{i,j}\in\{0,1\}$, we \revminor{refer to it as} an \textit{invalid cut} because, in that case, $\bm{r}_{i,j}$ coincides with a triangle vertex and does not serve as a vertex of a material polygon; \revminor{thus,} we simply ignore it. Otherwise, it's a \textit{valid cut}. For convenience, if there is only one valid cut on $e_i$, we will assume it to be $R_{i,1}$. For example, $R_i=(0,0.4)$ and $R_i=(0.4,1)$ represent the same polygon vertices, and we will always use the second one. Note that by the definition of $R$, we restrict each edge to have at most $2$ valid cuts.

\begin{figure}[htbp]
  \centering
  \includegraphics[width=0.4\textwidth]{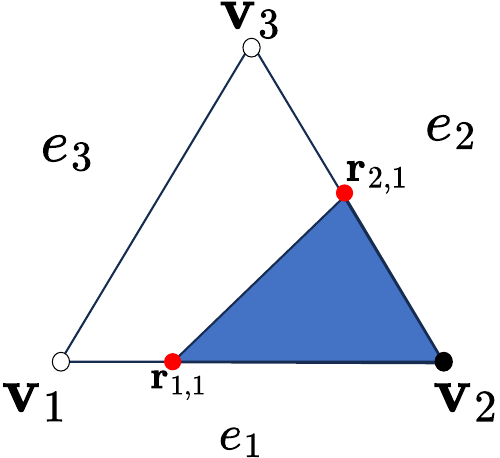}
  \caption{A simple example of a triangle edge cut. The small blue triangle \revminor{denotes} the liquid region.}
  \label{fig:simple-recon}
\end{figure}

Figure \ref{fig:simple-recon} depicts an example of a triangle edge cut with
\begin{equation}
    c = 0, 
    R = \begin{bmatrix}
      0.3 & 1 \\
      0.5 & 1 \\
      0 & 1
    \end{bmatrix}.
\end{equation}
\revminor{Here,} $R_{1,1},R_{1,2}=(0.3,1)$ indicates that there is one valid cut $\bm{r}_{1,1}$ on $e_1$, represented as a red dot.
\revminor{This} implies that $\chi(\bm{v}_2)=1$, because there must be a liquid-air interface intersecting $e_1$. Similarly, $\bm{r}_{2,1}$ is the only valid cut on $e_2$, and we can deduce that $\chi(\bm{v}_3)=0$. Therefore, we can reconstruct the liquid region as a triangle $\Delta \bm{r}_{1,1}\bm{v}_2\bm{r}_{2,1}$, as shown in the figure, where the small blue triangle \revminor{denotes the} liquid \revminor{region, while} the remaining white part \revminor{represents} air.

We can further perform similar interface reconstructions for all possible combinations of $(c,R)$. Fortunately, we don't need to specify all the reconstructions explicitly, because it's easy to see that exchanging the roles of liquid and air, or cyclically permuting the vertices does not really change the results of interface reconstruction. Now consider the three vertices of the triangle: \revminor{either} all \revminor{of them} have the same material, or two vertices have the same material while the \revminor{third} one is different. Therefore, without loss of generality, we can assume that either all three vertices are air, \revminor{\textit{i.e.}}, $c=\chi(\bm{v}_1)=\chi(\bm{v}_2)=\chi(\bm{v}_3)=0$, or $\bm{v}_1$ is the only liquid vertex, \revminor{\textit{i.e.}}, $c=\chi(\bm{v}_1)=1$ and $\chi(\bm{v}_2)=\chi(\bm{v}_3)=0$.

For the first case \revminor{where} $c=0$, \revminor{any} two vertices \revminor{on the same edge} are of the same material, implied by the Jordan curve theorem~\cite{hales2007jordan}, \revminor{which states} that the line segment connecting them must intersect the interface an even number of times. \revminor{Under our restriction that each edge has at most two valid cuts, this number} must be $0$ or $2$. Therefore, there can be $0,1,2$, or $3$ edges \revminor{with} $2$ valid cuts, \revminor{while the others} have no valid \revminor{cuts}. Due to cyclic symmetry, we can again assume that if there is \revminor{one} edge with $2$ valid cuts, it must be $e_1$, and if there are \revminor{two} edges with $2$ valid cuts, they must be $e_1$ and $e_2$. \revminor{In} the second case \revminor{where} $c=1$, there must be \revminor{one} valid cut on \revminor{both} $e_1$ and $e_3$, since \revminor{these} edges connect two vertices \revminor{of} different materials. \revminor{Edge} $e_2$ \revminor{can have either} $0$ or $2$ valid cuts.

\begin{figure}[htbp]
  \centering
  \includegraphics[width=0.8\textwidth]{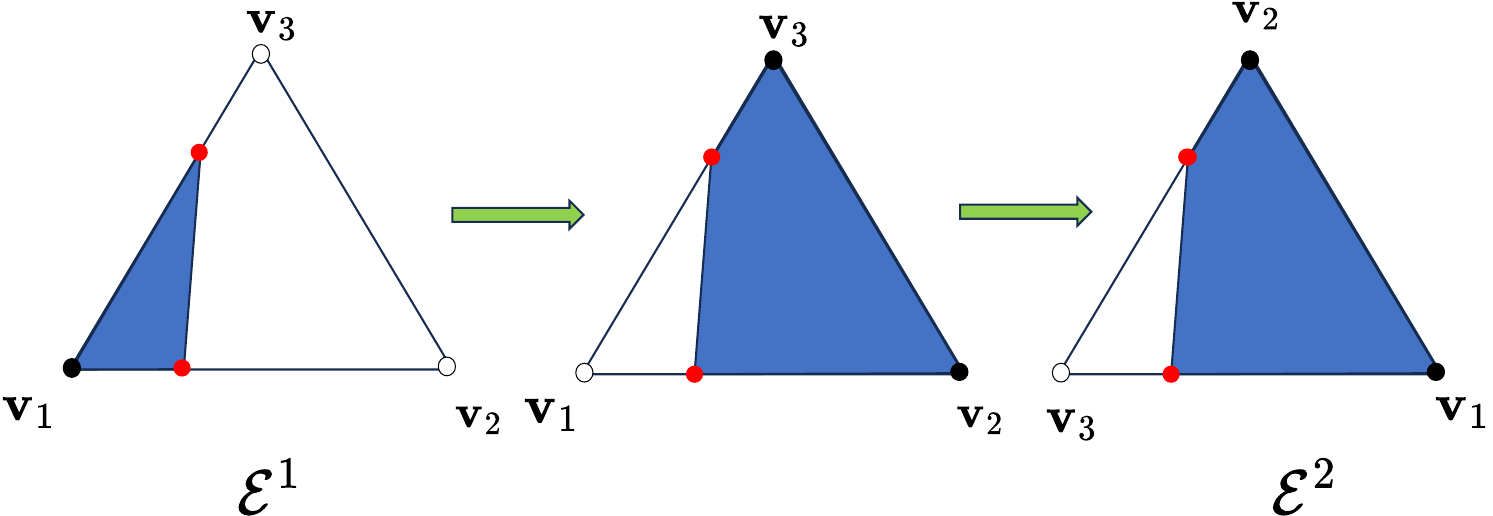}
  \caption{An example of deriving the interface reconstruction of a triangle edge cut $\mathcal{E}^2$ from a basic case $\mathcal{E}^1$. First, we exchange the roles of \revminor{the} liquid and \revminor{the} air, \revminor{changing the value of} $c$ from $0$ to $1$; \revminor{accordingly,} the air polygon is swapped with the liquid polygon. Second, we apply a cyclic permutation of vertex indices $1\ 2\ 3\to 3\ 1\ 2$ to \revminor{obtain} $\mathcal{E}^2$.}
  \label{fig:recon-equivalence}
\end{figure}

\begin{figure}[htbp]
  \centering
  \includegraphics[width=\textwidth]{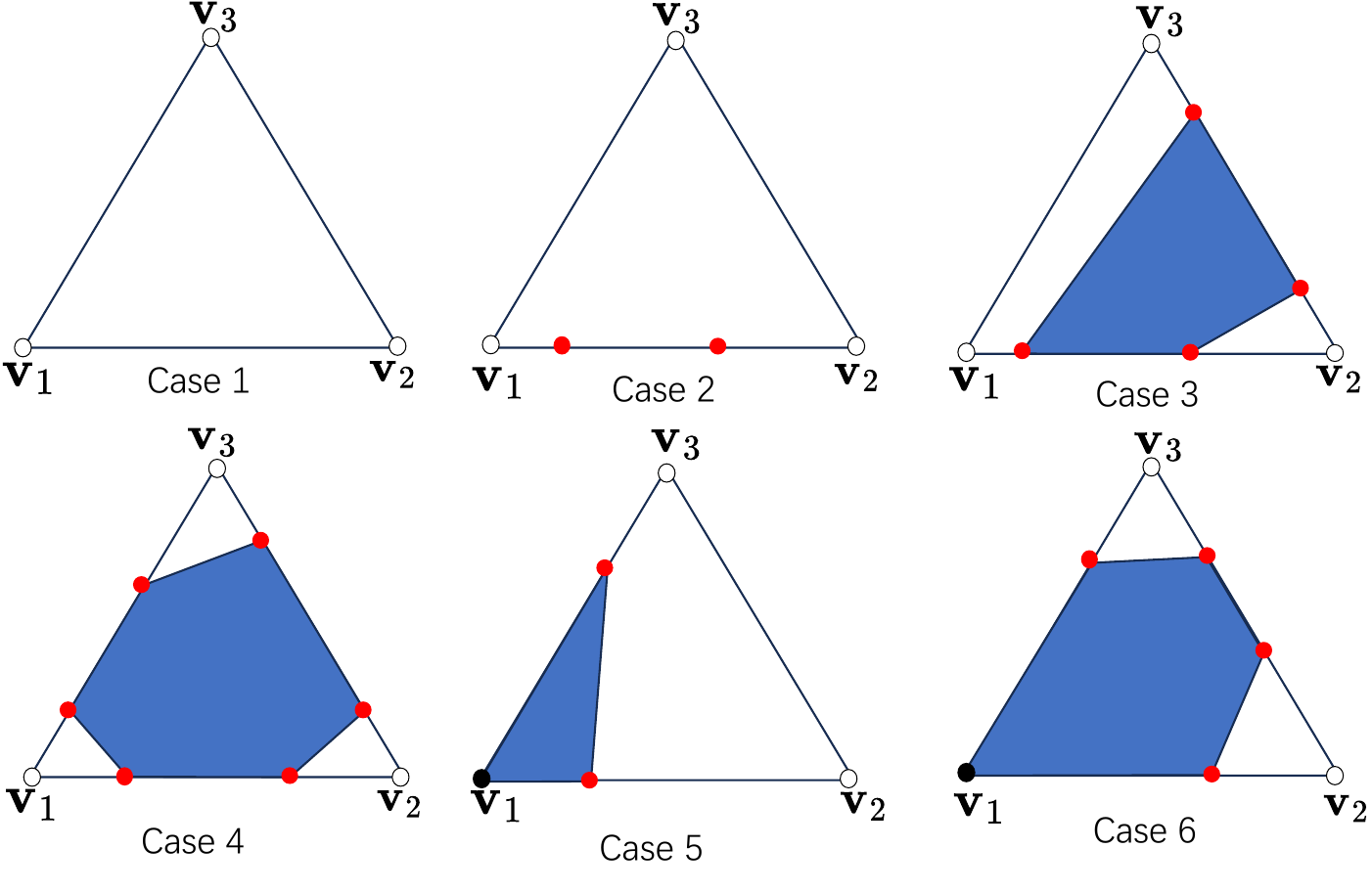}
  \caption{Six basic cases of triangle edge cuts. Black dots represent liquid, white dots represent air, and red dots represent valid cuts. Blue polygons \revminor{represent the} liquid regions inside triangles. In \revminor{cases} $1,2,3,4$, all vertices are air, \revminor{while} in \revminor{cases} $5,6$, there is one liquid vertex and two air vertices.}
  \label{fig:cases-nofix}
\end{figure}

\revminor{Figure~\ref{fig:cases-nofix} lists the} four cases with $c=0$ and \revminor{the} two cases with $c=1$, which we refer to as the basic cases. The interface reconstruction of any possible triangle edge cut $\mathcal{E}^2$ can be derived from some $\mathcal{E}^1$ in one of \revminor{the} six basic cases, \revminor{and} an example of such derivation is shown in Figure~\ref{fig:recon-equivalence}. \revminor{In the remainder} of \revminor{this} paper, we \revminor{will focus on} the interface tracking algorithms \revminor{on} these basic cases. On an unstructured triangle mesh, the interface in the whole computational domain can be represented by defining the edge cut $\mathcal{E}_i$ \revminor{for} each triangle $\mathcal{T}_i$. We denote the liquid polygon in $\mathcal{E}_i$ as $\mathcal{P}$; therefore, the liquid region in the whole computational domain \revminor{consists of} a set of polygons $\{\mathcal{P}_i\}$. 

\section{Interface Advection}
\label{sec:interface-advection}

Now consider the interface advection inside a triangle $\mathcal{T}$. We want to calculate $\mathcal{E}^{n+1}=(\mathcal{T}, c^{n+1}, R^{n+1})$ \revminor{using} the interface $\{\mathcal{E}_i^{n}\}$ defined at time step $n$. \revminor{Note that} the Eulerian triangle $\mathcal{T}$ doesn't have a \revminor{time step} superscript. Using the interface reconstruction described in Section~\ref{sec:triangle-edge-cut}, we can explicitly represent the liquid region at time step $n$ as a set of polygons $\{\mathcal{P}_i^n\}$.
Algorithm~\ref{alg:edgecut-simple-advection} describes a simple advection algorithm \revminor{that} is carried out in three steps:

\paragraph{Step 1} Calculate pre-image $\mathcal{T}'$ of the triangle by advecting all vertices $\bm{v}_i$ back by $-\Delta t$ \revminor{using} the 4th order Runge-Kutta method (RK4), and connecting these three advected vertices $\overleftarrow{\bm{v}_1},\overleftarrow{\bm{v}_2},\overleftarrow{\bm{v}_3}$:
\begin{equation}
\label{eqn:triangle-preimage}
    \mathcal{T}'=\Delta \overleftarrow{\bm{v}_1}\overleftarrow{\bm{v}_2}\overleftarrow{\bm{v}_3}.
\end{equation}
Throughout the evolution of the interface, each particle \revminor{retains} its material, \textit{i.e.}, the Lagrangian transportation equation
\begin{equation}
    \frac{\mathrm{D}\chi(\bm{p})}{\mathrm{D}t}=0
\end{equation}
is satisfied. \revminor{Therefore,} we can directly calculate the material $c^{n+1}$ from the interface at the $n$-th time step with
\begin{equation}
\label{eqn:material-calculation}
    c^{n+1}=\chi^n\left(\overleftarrow{\bm{v}_1}\right),
\end{equation}
which is \revminor{done} by checking if $\overleftarrow{\bm{v}_1}$ \revminor{lies} inside any polygon in $\{\mathcal{P}_i^n\}$ (including edges). If so, it will be \revminor{classified} as a liquid vertex; otherwise, it's air. Then we calculate the intersections of edges of $\mathcal{T}'$ with the edges of $\{\mathcal{P}_i^n\}$: 
\begin{equation}
\label{eqn:preimgae-edge-intersection}
     \{\bm{r}_{i,j}'\}=e_i'\cap \{\mathcal{P}_i^n\}.
\end{equation}
We will provide more implementation details about \eqref{eqn:material-calculation} and \eqref{eqn:preimgae-edge-intersection} in Section~\ref{sec:implementation}.

\paragraph{Step 2} Followed by the first step, we use \revminor{the} RK4 \revminor{method} again to calculate the images of the intersections $\{\bm{r}_{i,j}'\}$ as the predicted edge cuts for $\mathcal{E}^{n+1}$, that
\begin{equation}
\label{eqn:rk4-post-cuts}
    \{\bm{r}_{i,j}^{n+1}\}=\{\overrightarrow{\bm{r}_{i,j}'}\}.
\end{equation}
However, due to the non-linear nature of RK4 time integration, $\bm{r}_{i,j}^{n+1}$ may not \revminor{lie} on edge $e_i$, necessitating the next step.

\paragraph{Step 3} In the last step, \revminor{the} predicted edge cuts $\{\bm{r}_{i,j}^{n+1}\}$ are projected \revminor{onto} corresponding edges to calculate the final result $R^{n+1}$:
\begin{equation}
    R_{i,j}^{n+1}=\frac{(\bm{r}_{i,j}^{n+1}-\bm{v}_{i})\cdot(\bm{v}_{i+1}-\bm{v}_i)}{\left|\bm{v}_{i+1}-\bm{v}_i\right|^2}.
\end{equation}
\revminor{Thus, we acquire} the new triangle edge cut $\mathcal{E}^{n+1}=(\mathcal{T}, c^{n+1}, R^{n+1})$.

\begin{algorithm}
    \caption{InterfaceAdvection}
    \label{alg:edgecut-simple-advection}
    \KwData{$\mathcal{T}$, $\{\mathcal{E}_i^{n}\}$}
    \KwResult{$\mathcal{E}^{n+1}=(\mathcal{T},c^{n+1},R^{n+1})$}
    
    \tcp{Step 1 (pre-image query)}
    $\mathcal{T}'\leftarrow \Delta \overleftarrow{\bm{v}_1}\overleftarrow{\bm{v}_2}\overleftarrow{\bm{v}_3}$\; 
    $c^{n+1}\leftarrow\chi^n\left(\overleftarrow{\bm{v}_1}\right)$\;
    $\{\bm{r}_{i,j}'\}=e_i'\cap \{\mathcal{P}_i^n\}$\;
    \tcp{Step 2 (advection)}
    $\{\bm{r}_{i,j}^{n+1}\}\leftarrow\{\overrightarrow{\bm{r}_{i,j}'}\}$\;
    \tcp{Step 3 (reconstruction)}
    $R_{i,j}^{n+1}\leftarrow (\bm{r}_{i,j}^{n+1}-\bm{v}_{i})\cdot(\bm{v}_{i+1}-\bm{v}_i)/\left|\bm{v}_{i+1}-\bm{v}_i\right|^2$\;

    \textbf{return} $(\mathcal{T},c^{n+1},R^{n+1})$\;
\end{algorithm}

This interface advection algorithm only performs segment-segment intersections and cross products (to check the material of a point) in the pre-image; however, it exhibits appreciable errors and does not preserve the liquid area. In Section~\ref{sec:area-correction}, we will \revminor{discuss} the area-correction step we used to \revminor{address these issues}. \revminor{Furthermore, in} Section~\ref{sec:implementation}, we will provide more details about the implementation of this method.

\section{Area Correction}
\label{sec:area-correction}

\begin{figure}[htbp]
  \centering
  \includegraphics[width=0.7\textwidth]{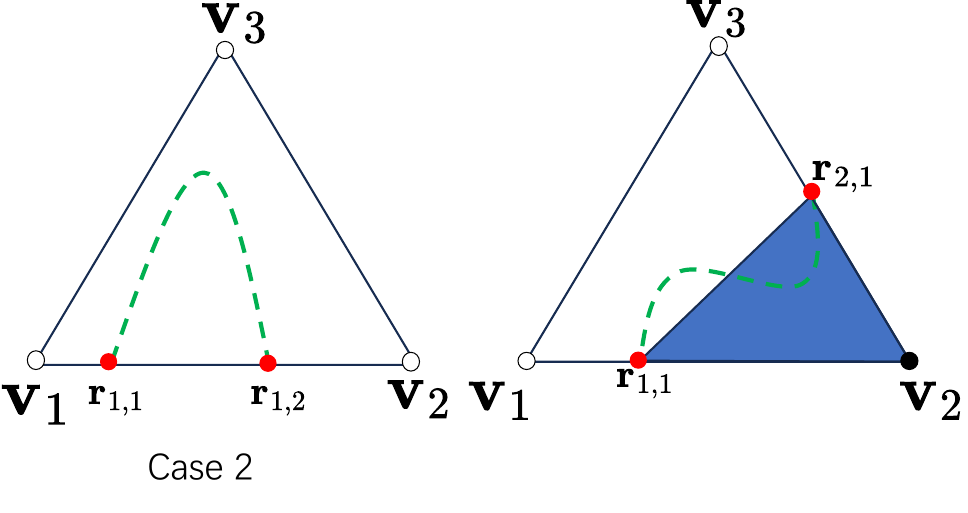}
  \caption{Two main reasons for shape errors. Left: incorrect interface reconstruction in case $2$. Right: errors from segment simplification of \revminor{the} interface inside $\mathcal{T}$. \revminor{The} green dashed curve \revminor{represents} the actual \revminor{interface}, and the blue region \revminor{illustrates} the reconstructed liquid region.}
  \label{fig:reconstruction-errors}
\end{figure}

There are two major reasons for the area error. First, the interface reconstruction of case $2$ in Figure~\ref{fig:cases-nofix} \revminor{underestimates} the liquid area. As shown in \revminor{the left part} of Figure \ref{fig:reconstruction-errors}, two cuts $\bm{r}_{1,1}, \bm{r}_{1,2}$ \revminor{suggest the presence of a liquid region}, depicted as the green dashed curve, however, it's reconstructed as pure air. Second, the triangle edge cut representation simplifies the interfaces inside $\mathcal{T}$ \revminor{to linear} segments. As \revminor{depicted} in the right \revminor{part} of Figure~\ref{fig:reconstruction-errors}, the actual interface is \revminor{represented by} the green dashed curve, but it is reconstructed as a straight segment, resulting in area errors.

\subsection{Additional Vertex for Case $2$}
\label{subsec:case2-additional-vertex}

\begin{figure}[htbp]
  \centering
  \includegraphics[width=0.35\textwidth]{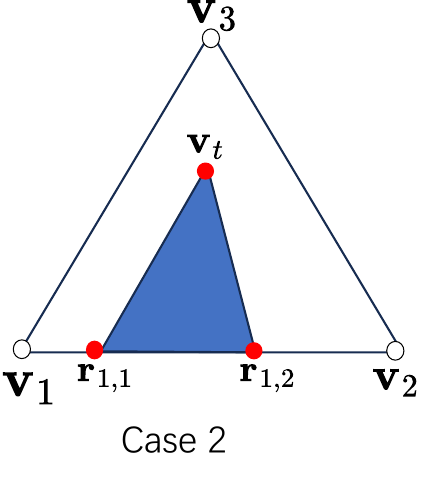}
  \caption{Case $2$ after adding the additional vertex inside the triangle. The liquid area is \revminor{now represented by} the blue triangle $\Delta \bm{r}_{1,1}\bm{r}_{1,2}\bm{v}_t$.}
  \label{fig:case2-fix}
\end{figure}

To preserve the liquid area in case $2$, we \revminor{introduce} a new liquid polygon vertex, \revminor{denoted as} $\bm{v}_t$, inside $\mathcal{T}$. The liquid region is \revminor{then} reconstructed as a triangle $\Delta \bm{r}_{1,1}\bm{r}_{1,2}\bm{v}_t$, as shown in Figure~\ref{fig:case2-fix}.
\begin{table}[h]
    \centering
    \begin{tabular}{|c|c|c|c|}
        \hline
        Case \# & $c$ & $t$ & Liquid Polygon $\mathcal{P}$\\
        \hline
        $1$ & $0$ & $(0,0,0)$ & $\emptyset$ \\
        \hline
        $2$ & $0$ & $(2,0,0)$ & $(\bm{r}_{1,1},\bm{r}_{1,2},\bm{v}_t)$\\
        \hline
        $3$ & $0$ & $(2,2,0)$ & $(\bm{r}_{1,1},\bm{r}_{1,2},\bm{r}_{2,1},\bm{r}_{2,2})$ \\
        \hline
        $4$ & $0$ & $(2,2,2)$ & $(\bm{r}_{1,1},\bm{r}_{1,2},\bm{r}_{2,1},\bm{r}_{2,2},\bm{r}_{3,1},\bm{r}_{3,2})$\\
        \hline
        $5$ & $1$ & $(1,0,1)$ & $(\bm{v}_1,\bm{r}_{1,1},\bm{r}_{3,1})$\\
        \hline
        $6$ & $1$ & $(1,2,1)$ & $(\bm{v}_1,\bm{r}_{1,1},\bm{r}_{2,1},\bm{r}_{2,2},\bm{r}_{3,1})$\\
        % Add more rows as needed
        \hline
    \end{tabular}
    \caption{Corresponding liquid polygon of each basic case, including the additional vertex $\bm{v}_t$ in case $2$.}
    \label{tab:basic-cases}
\end{table}
Taking $\bm{v}_t$ into account, the liquid polygons in all basic cases are summarized in Table~\ref{tab:basic-cases}. Here, $t$ is defined as a $3\times 1$ vector indicating the number of valid cuts on each edge:
\begin{equation}
\label{eqn:def-of-t}
    t_i=\left|(0,1)\cap\{R_{i,1},R_{i,2}\}\right|.
\end{equation}

When advecting the interface, we \revminor{determine} the position of $\bm{v}_t^{n+1}$ in the occurrence of case $2$ using the method \revminor{outlined} in Algorithm~\ref{alg:case2-calc-vt}. First, we calculate the intersection of $\mathcal{T}'$ defined in \eqref{eqn:triangle-preimage} with the liquid polygons at time step $n$:
\begin{equation}
\label{eqn:preimage-intersection-polys}
    \{\Pi_k\}=\mathcal{T}'\cap\{\mathcal{P}_i^n\}.
\end{equation}
Different from \eqref{eqn:preimgae-edge-intersection} that \revminor{yields} a set of points $\{\bm{r}_{i,j}'\}$, the polygon-polygon intersections here result in a set of polygons $\{\Pi_k\}$.
Similar to MOF~\cite{dyadechko2005moment} and PAM~\cite{zhang2008new}, the image of a polygon $\Pi_k$ is \revminor{obtained} by connecting its RK4-advected vertices forward in time $\Delta t$. The images of $\{\Pi_k\}$ are referred to as $\{\overrightarrow{\Pi_k}\}$, which serves as an approximation of the liquid region inside $\mathcal{T}$ at time step $n+1$. The centroid $\bm{x}_c$ of this liquid region is then \revminor{computed as}
\begin{equation}
\label{eqn:post-centroid}    \bm{x}_c=\frac{\sum_k\int_{\overrightarrow{\Pi_k}}\bm{x}\mathrm{d}\bm{x}}{\sum_k |\overrightarrow{\Pi_k}|}.
\end{equation}
Unlike the liquid area calculation \eqref{eqn:liquid-area} \revminor{which will be discussed} later \revminor{and} is performed in the pre-image, we advect the liquid polygons $\{\Pi_k\}$ forward in time in \eqref{eqn:post-centroid} only in the presence of case $2$. The reason is the same as in the MOF method's first moment calculation, namely, the first moment conservation is not guaranteed.

We define the first tentative value $\bm{v}_t^*$ of the additional vertex $\bm{v}_t^{n+1}$ such that the reconstructed liquid polygon has a centroid $\bm{x}_c$. \revminor{This} can be achieved by taking
\begin{equation}
    \bm{v}_t^*=3\bm{x}_c-\bm{r}_{1,1}^{n+1}-\bm{r}_{1,2}^{n+1},
\end{equation}
where $\bm{r}_{1,1}^{n+1},\bm{r}_{1,2}^{n+1}$ are given by \eqref{eqn:rk4-post-cuts}. However, $\bm{v}_t^*$ may fall outside $\mathcal{T}$, prohibiting us from taking it as the additional vertex. If that happens, we then try to find two lines $L_1,L_2$ in $\{\mathcal{P}_i^n\}$ that generate \revminor{the} edge cuts $\bm{r}_{1,1},\bm{r}_{1,2}$. \revminor{The} image of their intersection \revminor{is then calculated} as a second tentative value $\bm{v}_t^{**}$:
\begin{equation}
    \bm{v}_t^{**}=\overrightarrow{L_1\cap L_2}.
\end{equation}
If the line intersection also fails, we will try to find a vertex of $\{\overrightarrow{\Pi_k}\}$ \revminor{that lies} inside \revminor{$\mathcal{T}$} and \revminor{is} farthest from $e_1$. \revminor{This vertex is considered} as the third tentative value $\bm{v}_t^{***}$:
\begin{equation}
\label{eqn:case2-top-vertex}
    \bm{v}_t^{***}=\argmax_{\bm{p}} \mathrm{D}_{\perp}(\bm{p}, e_1), \bm{p} \in\mathcal{T} \land \bm{p} \text{ is a vertex of}\{\overrightarrow{\Pi_k}\}.
\end{equation}
If that fails again, \revminor{meaning} there \revminor{are no vertices} of $\{\overrightarrow{\Pi_k}\}$ falling in $\mathcal{T}$, we will consider it as a degenerate case and fall back to case $1$.

\begin{algorithm}
    \setlength{\baselineskip}{1.3\baselineskip}
    \caption{FindAdditionalVertex}
    \label{alg:case2-calc-vt}
    \KwData{$\mathcal{T}$, $\bm{r}_{1,1}^{n+1}$, $\bm{r}_{1,2}^{n+1}$, $\{\mathcal{P}_i^n\}$ }
    \KwResult{$\bm{v}_t^{n+1}$}
    
    \tcp{Centroid reconstruction}
    $\{\Pi_k\}\leftarrow\mathcal{T}'\cap\{\mathcal{P}_i^n\}$\;
    $\bm{x}_c\leftarrow\sum_k\int_{\overrightarrow{\Pi_k}}\bm{x}\mathrm{d}\bm{x}/\sum_k |\overrightarrow{\Pi_k}|$\;
    $\bm{v}_t^*\leftarrow 3\bm{x}_c-\bm{r}_{1,1}^{n+1}-\bm{r}_{1,2}^{n+1}$\;
    \If{$\bm{v}_t^*\in \mathcal{T}$}{
        \textbf{return} $\bm{v}_t^*$\;
    }
    \Else{
        \tcp{Line intersection}
        $L_1,L_2\leftarrow $ two lines in $\{\mathcal{P}_i^n\}$ corresponding to $\bm{r}_{1,1},\bm{r}_{1,2}$\;
        $\bm{v}_t^{**}\leftarrow \overrightarrow{\revminor{L_1}\cap L_2}$\;
        \If{$\bm{v}_t^{**}\in \mathcal{T}$}{
            \textbf{return} $\bm{v}_t^{**}$\;
        }
        \Else{
            $\bm{v}_t^{***}\leftarrow \argmax_{\bm{p}} \mathrm{D}_{\perp}(\bm{p}, e_1)$, where $\bm{p} \in\mathcal{T}$ and $\bm{p}$ is a vertex of $\{\overrightarrow{\Pi_k}\}$\;
            \If{$\nexists\bm{v}_t^{***}$}{
                \tcp{Algorithm fails, fall back to case $1$}
                \textbf{return} $\emptyset$\;
            }
            \Else{
                \textbf{return} $\revminor{\bm{v}_t^{***}}$\;
            }
        }
    }
\end{algorithm}

\subsection{Edge Cut Correction}
\label{subsec:edgecut-correction}

In this section, we will \revminor{discuss} the quadratic edge cut correction step we used to preserve the material areas, which is applied at the end of the advection algorithm. Before that, we will first briefly \revminor{discuss} the material areas in a triangle edge cut. We can easily observe that the relative material areas, \textit{i.e.}, the proportion of air and liquid in $\mathcal{T}$ only \revminor{depend} on $c$ and $R$. We denote the relative air area by $F_0$, and the relative liquid area by $F_1$. Thus
\begin{equation}
\begin{aligned}
    F_0&=F_0(c,R),\\
    F_1&=F_1(c,R).
\end{aligned}
\end{equation}
In all basic cases, $F_0, F_1$ are quadratic functions of elements in $R$, as listed in Table~\ref{tab:cases-rel-areas}, where $(u,v,w)$ satisfying 
\begin{equation}
\begin{aligned}
\label{eqn:barycentric-coordinates}
    &\bm{v}_t=w\bm{v}_1+u\bm{v}_2+v\bm{v}_3,\\
    &u+v+w=1,
\end{aligned}    
\end{equation}
are the barycentric coordinates of the additional vertex $\bm{v}_t$ in case $2$.

\begin{table}[h]
    \centering
    \begin{tabular}{|c|>{\RaggedRight}p{4.2cm}|>{\RaggedRight}p{4.2cm}|}
        \hline
        Case \# & $F_0$ & $F_1$ \\ \hline
        $1$ & $1$ & $0$ \\ \hline
        $2$ & $1-F_1$ & $v(R_{1,2}-R_{1,1})$\\ \hline
        $3$ & $1-F_1$ & $(1-R_{1,1})R_{2,2}-(1-R_{1,2})R_{2,1}$\\ \hline
        $4$ & $R_{1,1}(1-R_{3,2})+(1-R_{1,2})R_{2,1}+(1-R_{2,2})R_{3,1}$ & $1-F_0$ \\ \hline
        $5$ & $1-F_1$ & $R_{1,1}\revminor{(1-R_{3,1})}$ \\ \hline
        $6$ & $(1-R_{1,1})R_{2,1}+(1-R_{2,2})R_{3,1}$ & $1-F_0$ \\ \hline
    \end{tabular}
    \caption{Relative material areas of different cases}
    \label{tab:cases-rel-areas}
\end{table}

Suppose that we have established a triangle edge cut 
\[\mathcal{E}^{n+1}=(\mathcal{T},c^{n+1},R^{n+1})\]
at the end of Algorithm~\ref{alg:edgecut-simple-advection}. Then, we want to further modify $\mathcal{E}^{n+1}$ to preserve the liquid area:
\begin{equation}
\label{eqn:liquid-area}
    A = \sum_{k} |\Pi_k|,
\end{equation}
with $\Pi_k$ defined in \eqref{eqn:preimage-intersection-polys}. Note that we calculate the liquid area in the pre-image similarly to MOF~\cite{dyadechko2005moment}, \revminor{for} the same reason \revminor{of} mass conservation.

Our edge cut correction only involves modifying the elements in $R^{n+1}$. We will not change $c^{n+1}$ or the number of valid cuts along each edge. In other words, we want to find a $3\times 2$ matrix $R'$ satisfying:
\begin{equation}
\label{eqn:correction-requirements}
\begin{aligned}
    t(R')&=t(R^{n+1}),\\
    F_1(c^{n+1},R')&=\frac{A}{|\mathcal{T}|},
\end{aligned}
\end{equation}
\revminor{where} $t$ is defined in \eqref{eqn:def-of-t}.

\begin{figure}[htbp]
  \centering
  \includegraphics[width=0.35\textwidth]{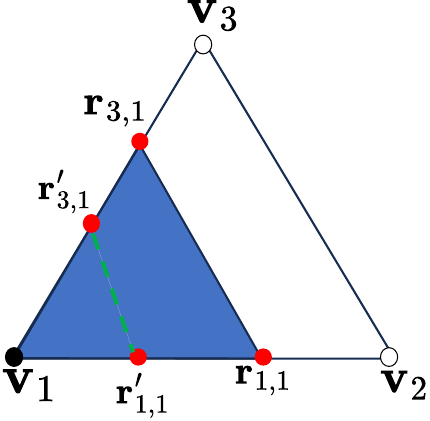}
  \caption{An example of edge cut correction. The initial liquid region (indicated by blue) is too large, and we want to make it smaller by moving the interface to a new position (green dashed line).}
  \label{fig:case5-area-correction}
\end{figure}

The way we decide $R'$ is modeled by a parameter $\tau\in [0,1)$ such that 
\begin{equation}
\label{eqn:correction-interpolation}
    R'=(1-\tau)R^{n+1}+\tau R^*.
\end{equation}
Clearly, $\tau=0$ corresponds to the unchanged situation $R'=R^{n+1}$, \revminor{that occurs only} when
\begin{equation}
    F_1(c^{n+1},R^{n+1})=\frac{A}{|\mathcal{T}|}
\end{equation}
is already satisfied. $R^*$ in \eqref{eqn:correction-interpolation} is the limit of $R'$ when $\tau\to 1$, a degenerate case that we shall never actually reach. For example, in Figure~\ref{fig:case5-area-correction} we have
\begin{equation}
    c = 0, 
    R = \begin{bmatrix}
      0.7 & 1 \\
      0 & 1 \\
      0.4 & 1
    \end{bmatrix},
\end{equation}
and we want to make the liquid area smaller by moving $\bm{r}_{1,1}$ to $\bm{r}_{1,1}'$, and $\bm{r}_{3,1}$ to $\bm{r}_{3,1}'$, forming the new interface indicated by the green dashed line. In this case, we shall have
\begin{equation}
    R^*= \begin{bmatrix}
      0 & 1 \\
      0 & 1 \\
      1 & 1
    \end{bmatrix}.
\end{equation}
Intuitively, we're trying to move $\bm{r}_{1,1},\bm{r}_{3,1}$ toward $\bm{v}_1$, shrinking the liquid region. However, if they do reach $\bm{v}_1$ at $\tau=1$, the interface will degenerate to pure air.
For all general cases, Table~\ref{tab:cases-rel-areas} and \eqref{eqn:correction-interpolation} \revminor{imply} that $F_1(c^{n+1},R')$ is a quadratic function of $\tau$. The geometric meaning of $\tau$ indicates that $F_1(\tau)$ is monotonic on $[0,1)$, therefore we either have zero or one solution $\tau_0\in [0,1)$ satisfying \eqref{eqn:correction-requirements}. 
If the solution $\tau_0$ is present, we will update $R^{n+1}$ accordingly with \eqref{eqn:correction-interpolation}. Otherwise, we consider the area correction step to be failed and leave $R^{n+1}$ untouched.

Next, we will \revminor{discuss} how to find $R^*$. For an edge $e_i$, if there is only one valid cut $\bm{r}_{i,1}$ on it, then we will want to move it either toward $\bm{v}_i$ or $\bm{r}_{i+1}$, thus $R_{i,1}^*\in\{0,1\}$. If there are two valid cuts $\bm{r}_{i,1},\bm{r}_{i,2}$ and we want to separate them further apart, we have
\begin{equation}
\begin{aligned}
    R_{i,1}^*&=0,\\
    R_{i,2}^*&=1.
\end{aligned}
\end{equation}
\revminor{This means that we are} moving them toward the two ends of the edge. Otherwise, if we want to move them closer, we will \revminor{set}
\begin{equation}
    R_{i,1}^*=R_{i,2}^*=s_i,
\end{equation}
with
\begin{equation}
\label{eqn:correction-split-point}
    s_i=\frac{R_{1,1}}{R_{1,1}+1-R_{1,2}}.
\end{equation}
Equation \eqref{eqn:correction-split-point} ensures that we always have
\begin{equation}
    \frac{R_{1,1}^{n+1}}{1-R_{1,2}^{n+1}}=\frac{R'_{1,1}}{1-R'_{1,2}},
\end{equation}
for any value of $\tau$, keeping the ratio of \revminor{the} lengths of \revminor{the} two outer segments constant.

Special care must be taken for case $2$ with the additional vertex $\bm{v}_t$. If we want to expand the liquid region, we will move $\bm{v}_t$ toward $\bm{v}_3$ in the same manner parameterized by $\tau$ as \revminor{in} \eqref{eqn:correction-interpolation}. \revminor{Conversely}, if we want to shrink the liquid region, we will calculate a target point
\begin{equation}
\label{eqn:case2-st-def}
    \bm{s}_t=\frac{w}{u+w}\bm{v}_1+\frac{u}{u+w}\bm{v}_2,
\end{equation}
and move $\bm{v}_t$ toward it. The barycentric coordinates of $\bm{s}_t$ are $\left(\frac{u}{u+w},0,\frac{w}{u+w}\right)$, \revminor{which means that} we will keep the ratio $u/w$ of the barycentric coordinates constant during the movement of $\bm{v}_t$.

\begin{figure}[htbp]
  \centering
  \includegraphics[width=0.8\textwidth]{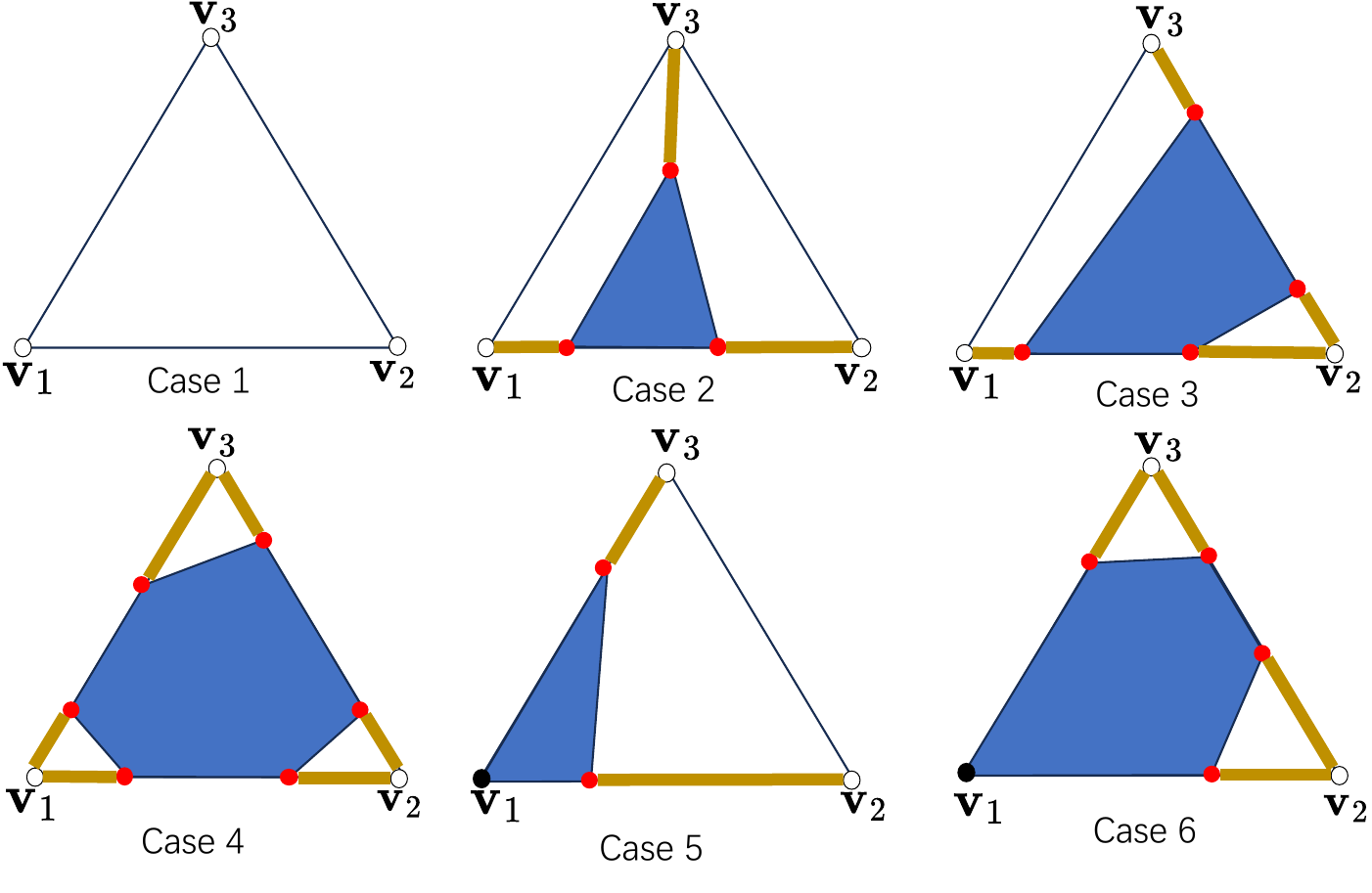}
  \caption{Area corrections when trying to expand the liquid region. Brown lines indicate the trajectories of cut points with $\tau\in [0,1)$.}
  \label{fig:correction-expand}
\end{figure}

\begin{figure}[htbp]
  \centering
  \includegraphics[width=0.8\textwidth]{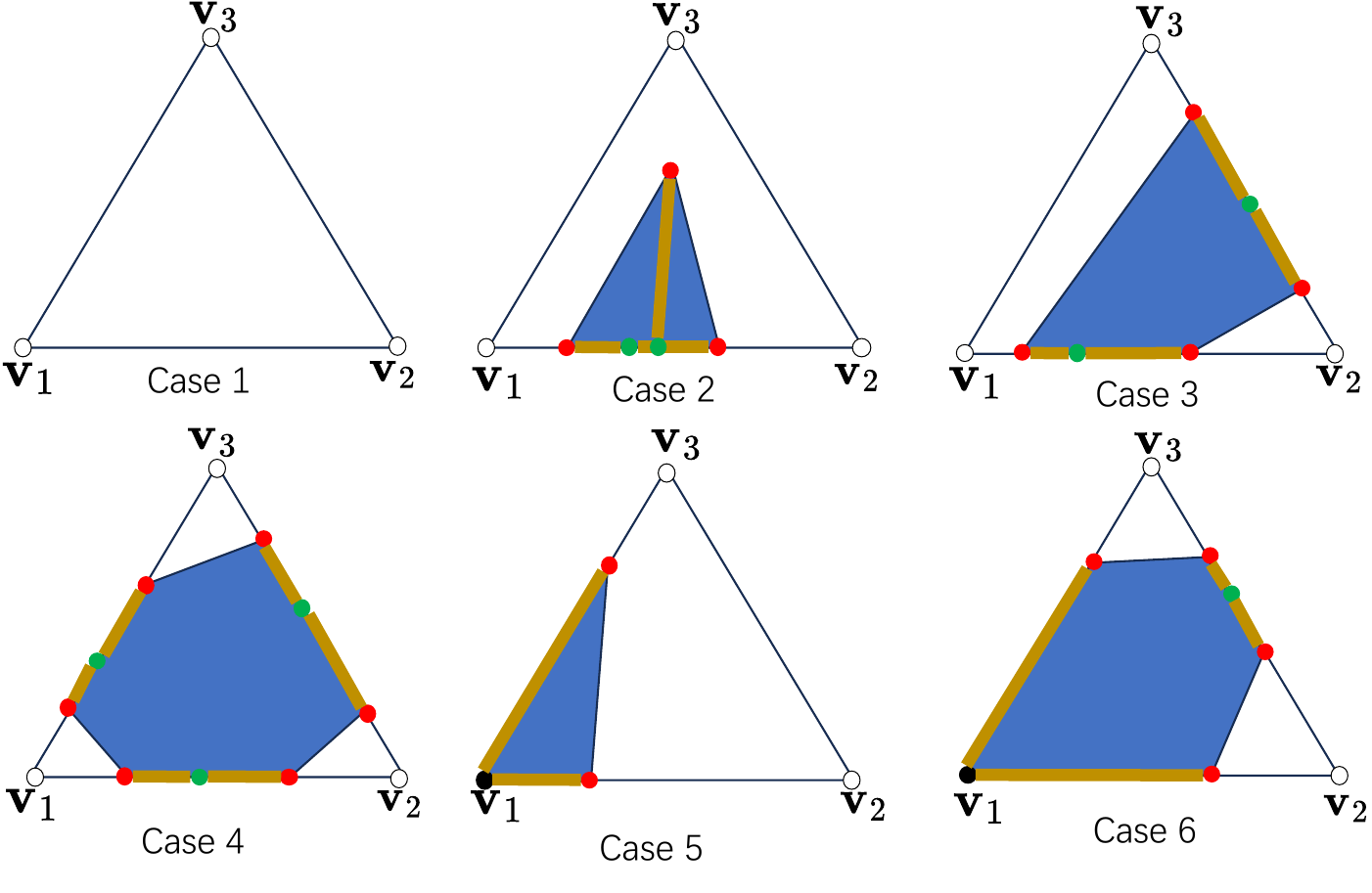}
  \caption{Area corrections when trying to shrink the liquid region. Brown lines indicate the trajectories of cut points with $\tau\in [0,1)$. Green dots correspond to $s_i$ defined in \eqref{eqn:correction-split-point}, or $\bm{s}_t$ defined in \eqref{eqn:case2-st-def}.}
  \label{fig:correction-shrink}
\end{figure}

Table~\ref{tab:edge-cut-movements} shows the area correction formulas. In case $2$, the target points of $\bm{v}_t$ are indicated by $\bm{v}_t^*$. The illustrations of the quadratic area correction for expanding and shrinking the liquid region are presented in Figure \ref{fig:correction-expand} and Figure \ref{fig:correction-shrink}.
\begin{table}[h]
    \centering
    \begin{tabular}{c|c|c|c|c|c|c|c|c|}
        \toprule
        & Case & $R_{1,1}^*$ & $R_{1,2}^*$ & $R_{2,1}^*$ & $R_{2,2}^*$ & $R_{3,1}^*$ & $R_{3,2}^*$ & $\bm{v}_t^*$ \\
        \midrule
        \multirow{4}{*}{Expand} & $2$ & $0$ & $1$ & --- & --- & --- & --- & $\bm{v}_3$\\
        & $3$ & $0$ & $1$ & $0$ & $1$ & --- & --- & ---\\ 
        & $4$ & $0$ & $1$ & $0$ & $1$ & $0$ & $1$ & ---\\ 
        & $5$ & $1$ & --- & --- & --- & $0$ & --- & ---\\
        & $6$ & $1$ & --- & $0$ & $1$ & $0$ & --- & ---\\
        \hline
        \multirow{4}{*}{Shrink} & $2$ & $s_1$ & $s_1$ & --- & --- & --- & --- & $\bm{s}_t$\\
        & $3$ & $s_1$ & $s_1$ & $s_2$ & $s_2$ & --- & --- & ---\\ 
        & $4$ & $s_1$ & $s_1$ & $s_2$ & $s_2$ & $s_3$ & $s_3$ & ---\\ 
        & $5$ & $0$ & --- & --- & --- & $1$ & --- & ---\\
        & $6$ & $0$ & --- & $s_2$ & $s_2$ & $1$ & --- & ---\\
        \bottomrule
    \end{tabular}
    \caption{Movements of edge cuts}
    \label{tab:edge-cut-movements}
\end{table}
\begin{algorithm}
    \caption{EdgeCutCorrection}
    \label{alg:quadratic-area-correction}
    \tcp{$F_1^*$ is the target value of liquid proportion}
    \KwData{$c^{n+1}$, $R^{n+1}, F_1^*$}
    \KwResult{$R'$}
    
    $R^*\leftarrow$ Table \ref{tab:edge-cut-movements}\;
    $f(\tau)\leftarrow F_1\left(c^{n+1}, (1-\tau)R^{n+1}+\tau R^*\right)$\;
    \tcp{Solve for the quadratic equation}
    $\tau_0\leftarrow f(\tau_0)=F_1^*$\;
    
    \textbf{return} $(1-\tau_0)R^{n+1}+\tau_0 R^*$\;
\end{algorithm}
The quadratic edge cut correction is summarized in Algorithm~\ref{alg:quadratic-area-correction}. For \revminor{cases} $2,3,5,6$, we can exactly recover any relative liquid area in $(0,1)$, because
\begin{equation}
    \begin{aligned}
        \lim_{t\to 0}F_1&=0 \text{(pure air)}, \\
        \lim_{t\to 1}F_1&=1 \text{(pure liquid)},
    \end{aligned}
\end{equation}
and the monotonicity of $F_1(\tau)$ guarantees that there must be one solution. However, for case $4$ we have
\begin{equation}
    \begin{aligned}
        \lim_{t\to 0}F_1&=0 \text{(pure air)}, \\
        \lim_{t\to 1}F_1&<1,
    \end{aligned}
\end{equation}
because even in the limit case $\tau=1$, there will \revminor{still} be a liquid triangle $\bm{s}_1\bm{s}_2\bm{s}_3$ inside the triangle. \revminor{This} is the only case where our algorithm fails to preserve the liquid area. \revminor{Fortunately}, case $4$ is very uncommon because it means there are $2$ cuts on all three edges of $\mathcal{T}^{n+1}$, indicating thin intricate geometries of the interface that rarely \revminor{occur}.

\revminor{By} adding the additional vertex in case $2$ and the \revminor{incorporating the} quadratic edge cut correction step \revminor{into} Algorithm~\ref{alg:edgecut-simple-advection}, we have \revminor{completed} our interface advection algorithm with area correction in Algorithm~\ref{alg:edgecut-advection-full}. Compared to Algorithm~\ref{alg:edgecut-simple-advection}, we \revminor{have introduced a} fix for case $2$ (lines $6-7$) and \revminor{included} an extra edge cut correction step (lines $8-11$).

\begin{algorithm}
    \caption{InterfaceAdvectionWithAreaCorrection}
    \label{alg:edgecut-advection-full}
    \KwData{$\mathcal{T}$, $\{\mathcal{E}_i^{n}\}$}
    \KwResult{$\mathcal{E}^{n+1}=(\mathcal{T},c^{n+1},R^{n+1})$}

    \tcp{Step 1 (pre-image query)}
    $\mathcal{T}'\leftarrow \Delta \overleftarrow{\bm{v}_1}\overleftarrow{\bm{v}_2}\overleftarrow{\bm{v}_3}$\; 
    $c^{n+1}\leftarrow\chi^n\left(\overleftarrow{\bm{v}_1}\right)$\;
    $\{\bm{r}_{i,j}'\}\leftarrow e_i'\cap \{\mathcal{P}_i^n\}$\;
    \tcp{Step 2 (advection)}
    $\{\bm{r}_{i,j}^{n+1}\}\leftarrow\{\overrightarrow{\bm{r}_{i,j}'}\}$\;
    \tcp{Step 3 (reconstruction)}
    $R_{i,j}^{n+1}\leftarrow (\bm{r}_{i,j}^{n+1}-\bm{v}_{i})\cdot(\bm{v}_{i+1}-\bm{v}_i)/\left|\bm{v}_{i+1}-\bm{v}_i\right|^2$\;
    \If{$(\mathcal{T},c^{n+1},R^{n+1})$ falls into case $2$}{
        \tcp{Algorithm \ref{alg:case2-calc-vt}}
        $\bm{v}_t^{n+1}\leftarrow\text{FindAdditionalVertex($\mathcal{T},\bm{r}_{1,1}^{n+1},\bm{r}_{1,2}^{n+1},\{\mathcal{P}_i^n\}$)}$\;
    }
    \tcp{Step 4 (edge cut correction)}
    $\{\Pi_k\}\leftarrow\mathcal{T}'\cap\{\mathcal{P}_i^n\}$\;
    $F_1^*\leftarrow \sum_k |\Pi_k|/|\mathcal{T}|$\;
    \If{$F_1(c^{n+1},R^{n+1})\neq F_1^*$}{
        \tcp{Algorithm \ref{alg:quadratic-area-correction}}
        $R^{n+1}\leftarrow\text{EdgeCutCorrection}(c^{n+1},R^{n+1},F_1^*)$\;
    }
    \textbf{return} $(\mathcal{T},c^{n+1},R^{n+1})$\;
\end{algorithm}
 
\section{Implementation}
\label{sec:implementation}

In this section, we talk about the program implementation of our interface tracking algorithm. For each triangle $\mathcal{T}$ and its triangle edge cut $\mathcal{E}=(\mathcal{T},c,R)$, we store the $R$ matrix with \revminor{six floating-point} variables. Additionally, $c$ can be put into the sign bit of $R_{1,1}$, \revminor{which avoids any extra memory usage}. For case $2$ in Figure~\ref{fig:case2-fix}, \revminor{noting} that we have all $R_{i,j}\in [0,1]$, we can store the barycentric coordinates $(u,v)$ defined in \eqref{eqn:barycentric-coordinates} for the extra vertex by setting $R_{2,1}=u+2, R_{2,2}=v+2$ to distinguish \revminor{them from} valid cuts. Table~\ref{tab:advection-algorithms} \revminor{compares} the memory \revminor{costs} of different interface-tracking methods, including our proposed \revminor{approach}. With the triangle edge cut representation, our method enables sub-triangle modeling \revminor{with} a low memory cost of $6$ DoFs for each element.

\begin{table}[htbp]
    \centering
    \begin{tabular}{|c|c|c|}
        \hline
        Algorithm & Element DoFs& Sub-grid \\
        \hline
        Level Set & 1 & No \\
        PLIC-VOF & 3 & No \\
        MOF & 5 & No \\
        Double-PLIC & 8 & Yes\\
        MOF2 & 8 & Yes\\
        PAM & 20 & Yes\\
        Proposed & 6 & Yes\\
        \hline
    \end{tabular}
    \caption{Comparison of memory \revminor{costs} of different interface tracking algorithms.}
    \label{tab:advection-algorithms}
\end{table}

\begin{figure}[htbp]
  \centering
  \includegraphics[width=0.4\textwidth]{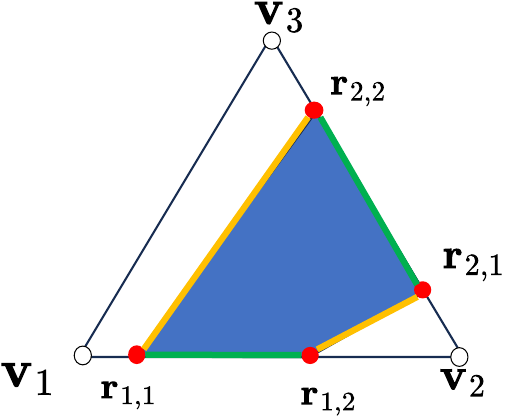}
  \caption{Difference between interior segments and boundary segments of the liquid polygon. Yellow lines $\overline{\bm{r}_{1,2}\bm{r}_{2,1}}$ and $\overline{\bm{r}_{2,2}\bm{r}_{1,1}}$ are interior segments. Green lines $\overline{\bm{r}_{1,1}\bm{r}_{1,2}}$ and $\overline{\bm{r}_{2,1}\bm{r}_{2,2}}$ are boundary segments.}
  \label{fig:interior-boundary-edges}
\end{figure}

For clarity, before introducing the implementation of the interface advection algorithm, we \revminor{first} distinguish between two types of liquid polygon edges \revminor{shown} in Figure~\ref{fig:interior-boundary-edges}: \textit{interior segments} and \textit{boundary segments}. An interior segment, as the name suggests, is fully inside a triangle, while a boundary segment \revminor{lies along} the boundary of a triangle. In Figure~\ref{fig:interior-boundary-edges}, interior segments are drawn with yellow lines, and boundary segments with green lines.

\begin{figure}[h!]
    \centering
    \includegraphics[width=0.4\textwidth]{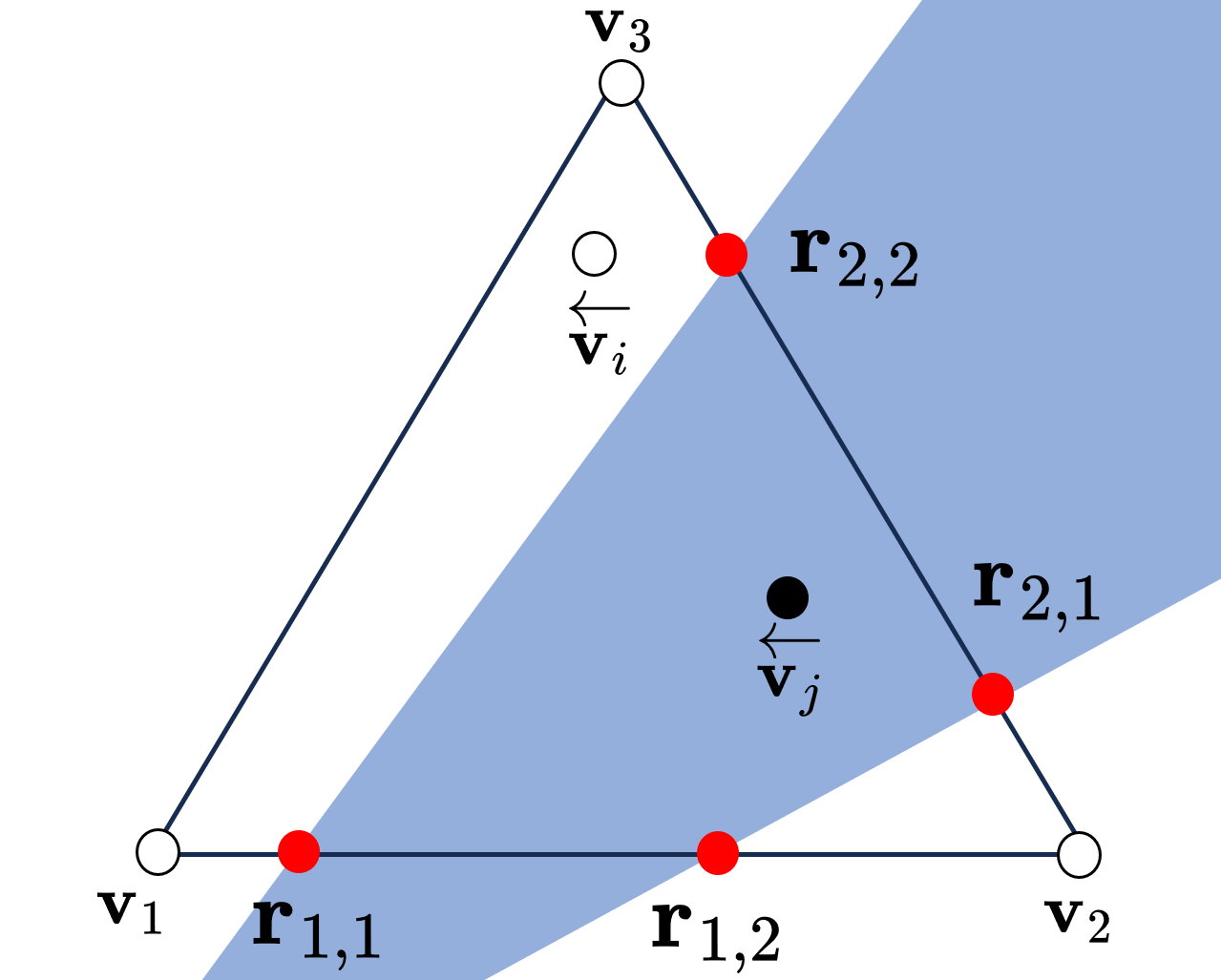}
    \caption{\revminor{Determining} the materials of $\overleftarrow{\bm{v}_i}$ and $\overleftarrow{\bm{v}_j}$ by checking if they're on the right side of interior segment $\overline{\bm{r}_{1,2}\bm{r}_{2,1}}$ or $\overline{\bm{r}_{2,2}\bm{r}_{1,1}}$. }
    \label{fig:quick-material-check}
\end{figure}

The most important steps in the interface advection Algorithm~\ref{alg:edgecut-advection-full} are the first two: calculating the materials of \revminor{vertex} images \eqref{eqn:material-calculation} and \revminor{finding} the intersections between edges and the interface \eqref{eqn:preimgae-edge-intersection}. To \revminor{determine} the material $\chi^n(\overleftarrow{\bm{v}})$ of \revminor{a} vertex image $\overleftarrow{\bm{v}}$, we first \revminor{locate its} belonging triangle $\mathcal{T}$ of $\overleftarrow{\bm{v}}$ in the mesh, and check if $\overleftarrow{\bm{v}}$ \revminor{lies} inside any liquid polygon defined by its corresponding edge cut $\mathcal{E}^{n}=(\mathcal{T}, c^{n}, R^{n})$. In practice, we don't need to perform the inclusion test for all the liquid polygons. Instead, for each interior segment, it is sufficient to check whether $\overleftarrow{\bm{v}}$ is on its left or right side. By using this method, we can reduce the computational effort. For example, in Figure~\ref{fig:quick-material-check}, we only need to check if $\overleftarrow{\bm{v}}$ is on the right side of $\overline{\bm{r}_{1,2}\bm{r}_{2,1}}$ or $\overline{\bm{r}_{2,2}\bm{r}_{1,1}}$. If so,  we \revminor{conclude} that $\overleftarrow{\bm{v}}$ is air, like the case of $\overleftarrow{\bm{v}_i}$. Otherwise, $\overleftarrow{\bm{v}}$ is in the shadowed area, indicating it's liquid like $\overleftarrow{\bm{v}_j}$. To avoid \revminor{degeneracies}, if $\overleftarrow{\bm{v}}$ \revminor{lies} exactly on an interior segment, we also consider it to be liquid.

\begin{figure}[htbp]
  \centering
  \includegraphics[width=0.4\textwidth]{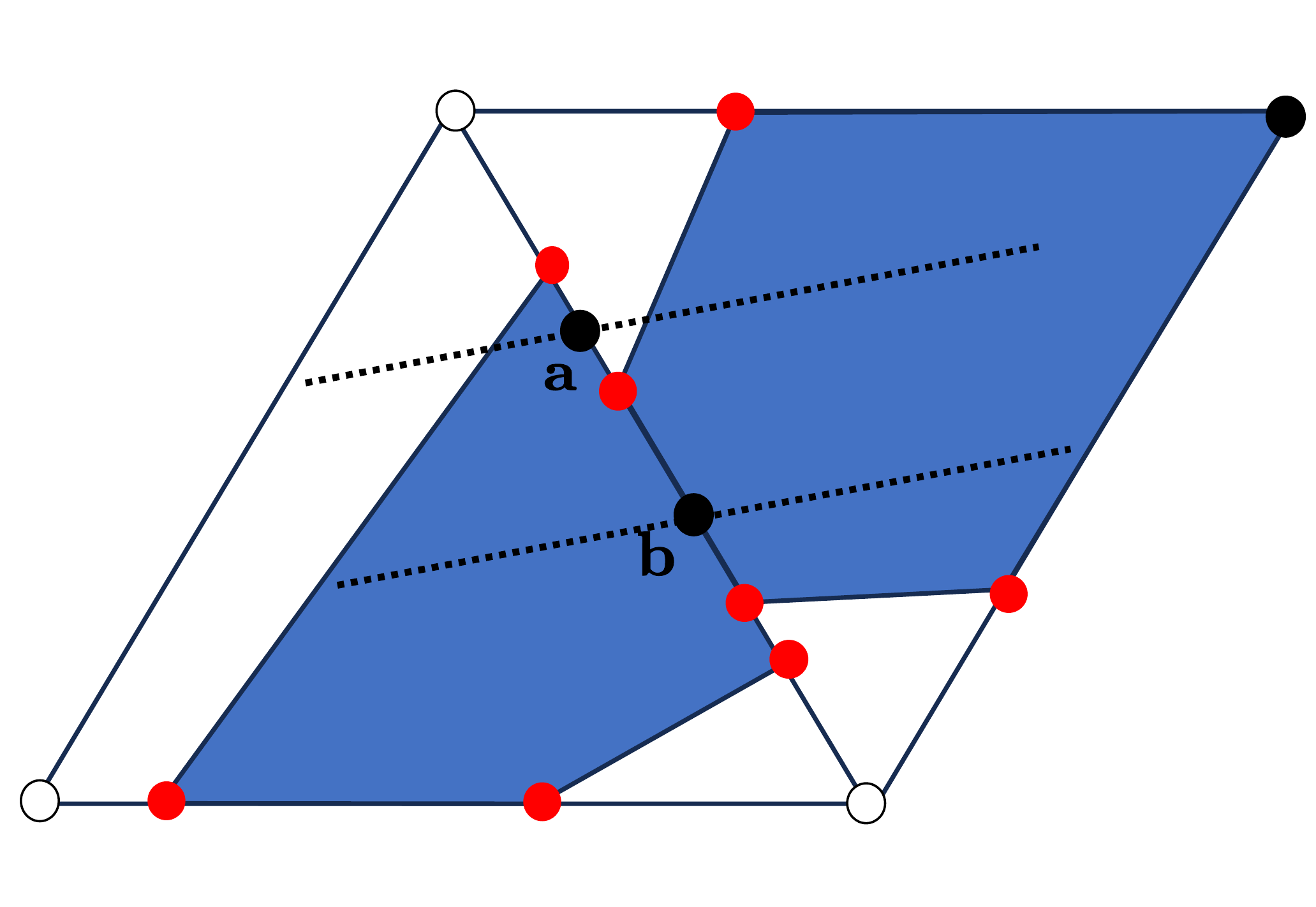}
  \caption{Calculating possible edge cuts between an edge $e_i'$ and boundary segments. \revminor{Point} $\bm{a}$ is considered an edge cut since it has different materials in two adjacent triangles. However, $\bm{b}$ is not considered an edge cut because it \revminor{lies} inside a larger liquid polygon.}
  \label{fig:edge-boundary-intersection}
\end{figure}

There are two types of edge-interface intersections \eqref{eqn:preimgae-edge-intersection}: \revminor{intersections} with interior segments, and intersections with boundary segments. For an edge $e_i'$ in \eqref{eqn:preimgae-edge-intersection}, we simply iterate over all interior segments in its intersecting triangles in the mesh to perform segment-segment intersections. However, \revminor{extra attention must be paid} to the boundary segments to avoid generating duplicate edge cuts, \revminor{as} an edge is shared by two triangles in the mesh. First, we calculate the intersections between edge $e_i'$ and the edges of its intersecting triangles. If an intersection $\bm{p}$ is found, we will find its two neighboring triangle edge cuts $\mathcal{E}^1$, $\mathcal{E}^2$, and calculate the materials using the method described in Figure~\ref{fig:quick-material-check}. If the results in $\mathcal{E}^1$, $\mathcal{E}^2$ differ, $\bm{p}$ is considered as an edge cut, like point $\bm{a}$ in Figure~\ref{fig:edge-boundary-intersection}. However, if the results \revminor{are the same}, $\bm{p}$ shall be inside of a larger liquid or air polygon, and thus not considered as an edge cut, like point $\bm{b}$.

In our implementation, we also take \revminor{efforts} to avoid degenerate cases. For example, if an edge of $\mathcal{T}'$ in \eqref{eqn:preimgae-edge-intersection} coincides with an edge in \revminor{the} liquid polygons $\{\mathcal{P}_i^n\}$, the intersection will be a segment instead of a single cut point. Or, if an edge of $\mathcal{T}'$ passes through a vertex of a liquid polygon $\mathcal{P}_i^n$, we may find two duplicate cuts on it, however, none of them should be taken into account because, in this situation, the edge does not \revminor{actually} cut through the liquid region. We use two efforts to deal with degeneracy. First, before the backward RK4 to calculate $\mathcal{T}'$, we add a random perturbation in the magnitude of $10^{-6}\Delta x$ to each vertex $\bm{v}_i$. This random perturbation largely eliminated the edge coincidence cases, especially when the velocity field is $\bm{0}$. Then, we also use the \revminor{materials of the vertices} as an additional safeguard for degenerate cases when performing edge-interface intersections \eqref{eqn:preimgae-edge-intersection}. We first query the materials
\begin{equation}
\begin{aligned}
    c_1&=\chi^n\left(\overleftarrow{\bm{v}_i}\right),\\
    c_2&=\chi^n\left(\overleftarrow{\bm{v}_{i+1}}\right).
\end{aligned}
\end{equation}
\revminor{According to} the Jordan curve theorem, if $c_1=c_2$, there are either $0$ or $2$ cuts on $e_i'$, and if $c_1\neq c_2$, there is only one cut. In the case of $c_1=c_2$, if only one cut \revminor{is} found, we discard it. If we find more than two cuts, we \revminor{keep} only the first one and the last one, sorted by their distances \revminor{from} $\overleftarrow{\bm{v}_i}$. Similarly, in the case of $c_1\neq c_2$, we will only keep the first cut.

\section{Experiments}
\label{sec:experiments}

In this section, we evaluate our interface advection algorithm with area correction on a static reconstruction test and four dynamic advection tests.

\subsection{Static Reconstruction}

We begin the numerical experiments with the convergence analysis of static reconstruction tests with three shapes: a circle, a snake shape, and a heart shape. In our numerical tests, we first divide the computational domain into a lattice grid and then divide each grid cell into a lower-right triangle and an upper-left triangle. The triangle edge cut is defined in \revminor{each} triangle, representing the interface.

In the circle test, we place a liquid circle with a radius \revminor{of} $r=0.15$ at the center of the computational domain $[0,1]^2$.
The snake shape test is introduced by \cite{shashkov2023moments}. It also takes a $[0,1]^2$ computational domain. In this test, the liquid region is the area between
\begin{equation}
    y=0.5+0.3\sin(2\pi x),
\end{equation}
and
\begin{equation}
    y=0.5+0.3\sin(2\pi 6).
\end{equation}
\revminor{The heart shape}, also adopted by \cite{shashkov2023moments}, is defined by a curve
\begin{equation}
    \begin{aligned}
        x&=\frac{16\sin^3(\theta)}{40}+0.52,\\
        y&=\frac{13\cos(\theta)-5\cos(2\theta)-2\cos(3\theta)-cos(4\theta)}{40}+0.55,
    \end{aligned}
\end{equation}
for $\theta\in [-\pi,\pi]$.

\begin{figure}
    \centering
    \begin{subfigure}{0.3\textwidth}
        \includegraphics[width=\linewidth]{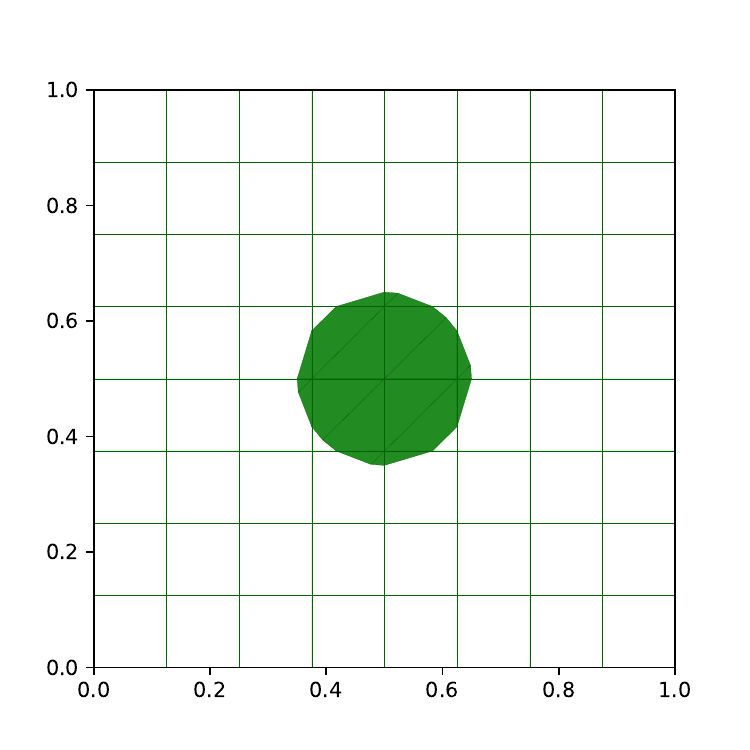}
        \caption{Circle, $8\times 8$}
    \end{subfigure}
    \begin{subfigure}{0.3\textwidth}
        \includegraphics[width=\linewidth]{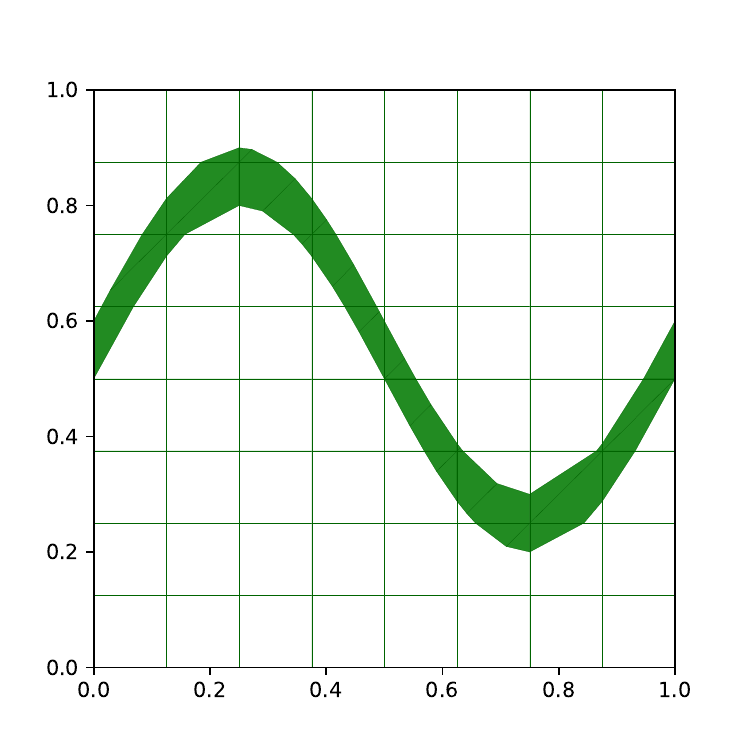}
        \caption{Snake, $8\times 8$}
    \end{subfigure}
    \begin{subfigure}{0.3\textwidth}
        \includegraphics[width=\linewidth]{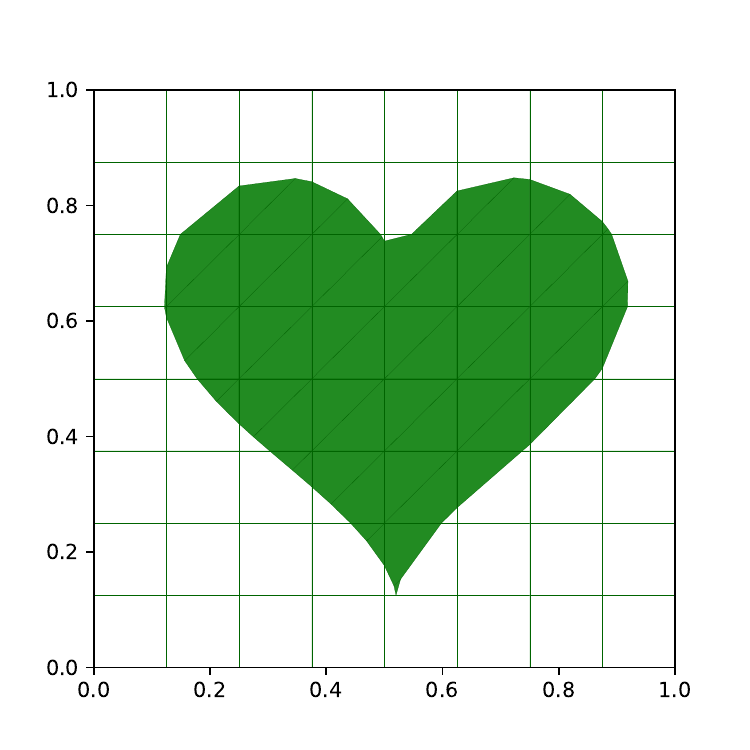}
        \caption{Heart, $8\times 8$}
    \end{subfigure}
    \begin{subfigure}{0.3\textwidth}
        \includegraphics[width=\linewidth]{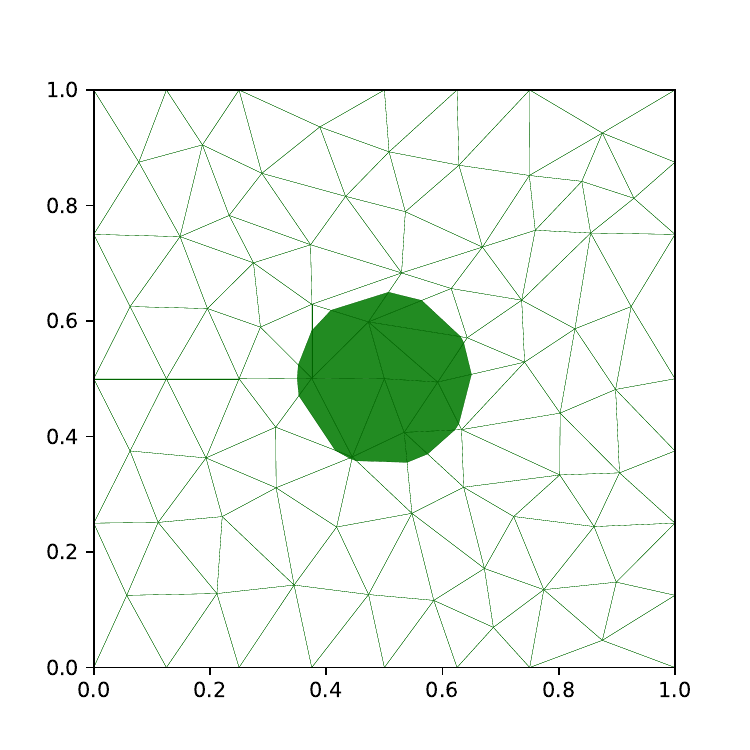}
        \caption{Circle, $l=0$}
    \end{subfigure}
    \begin{subfigure}{0.3\textwidth}
        \includegraphics[width=\linewidth]{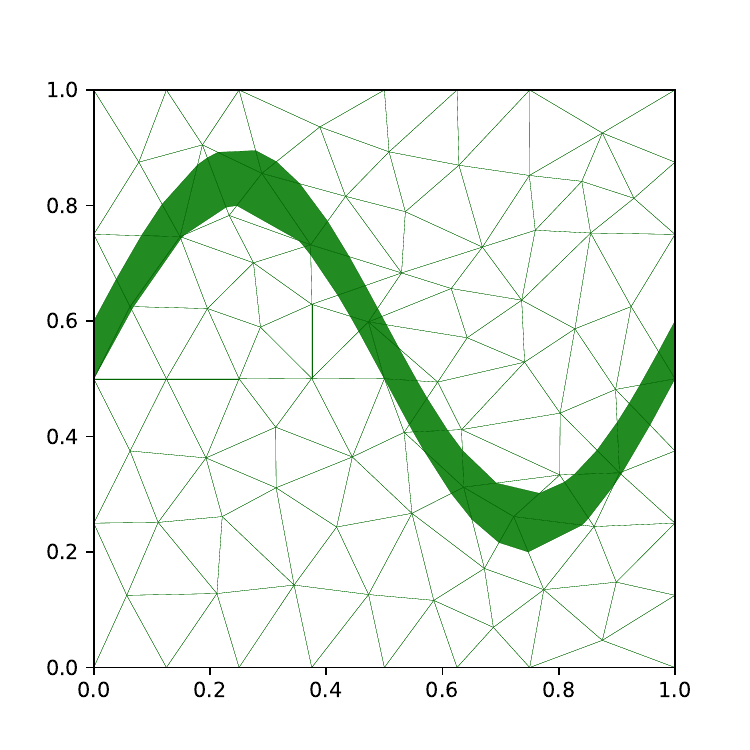}
        \caption{Snake, $l=0$}
    \end{subfigure}
    \begin{subfigure}{0.3\textwidth}
        \includegraphics[width=\linewidth]{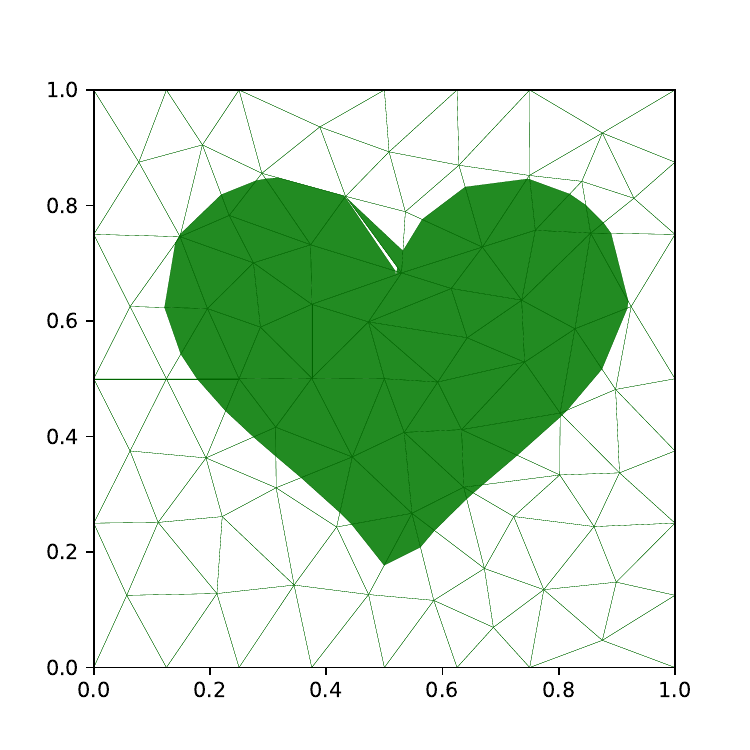}
        \caption{Heart, $l=0$}
    \end{subfigure}
    \caption{Static reconstruction results for circle, snake, and heart shapes using \revminor{an} $8\times 8$ lattice mesh (top row) and \revminor{an} unstructured triangle mesh (bottom row) at refinement level $0$. Our interface representation method can effectively recover the shape even at such low resolutions.}
    \label{fig:static-recon-results}
\end{figure}

The reconstruction results of three shapes at $8\times 8$ resolution are shown in the top row of Figure~\ref{fig:static-recon-results}.
To quantify the accuracy of our method, we calculate the absolute shape error $E_g$ using the expression \cite{lopez2005improved}
\begin{equation}
E_g = \sum_{i,j} |A_{i,j} - \tilde{A}_{i,j}|,
\label{eqn:abs-error}
\end{equation}
where $\tilde{A}_{i,j} = A(\mathcal{E}_1) + A(\mathcal{E}2)$ represents the sum of liquid areas in two triangles of cell $(i,j)$, and $A_{i,j}$ is the liquid area in cell $(i,j)$, calculated \revminor{using} a high-resolution \revminor{representation} of the liquid shape consisting of $1000$ vertices.

\begin{figure}
  \centering
  \includegraphics[width=0.75\textwidth]{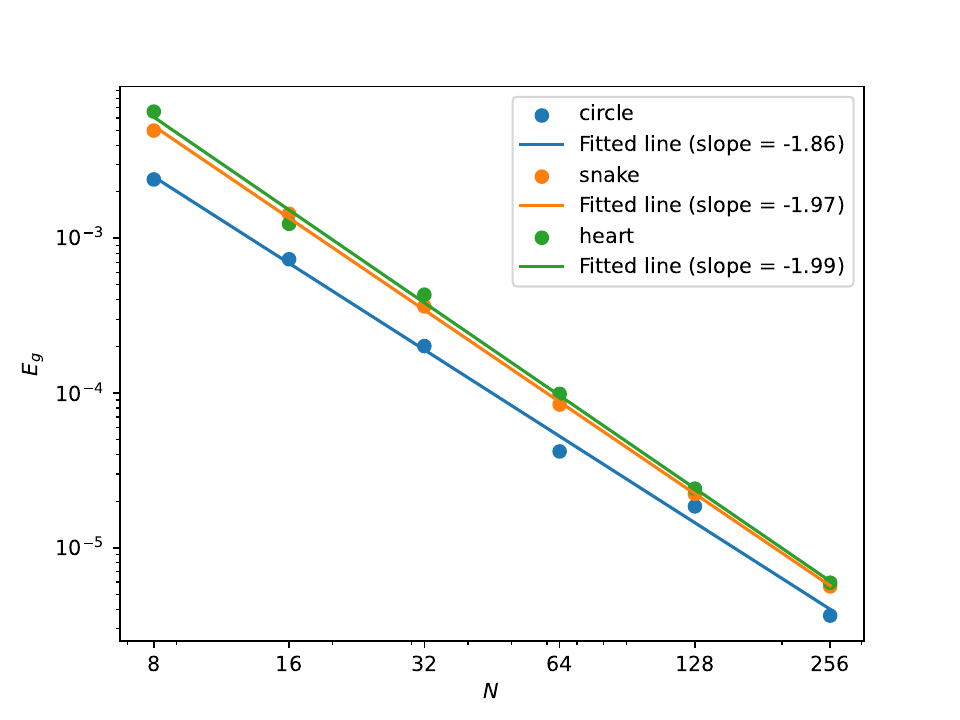}
  \caption{Absolute shape errors \revminor{for} static reconstruction tests \revminor{of} three different shapes. Our proposed method \revminor{demonstrates the} expected second-order convergence.}
    \label{fig:static-recon-fits}
\end{figure}

To assess the convergence rate across different resolutions, we employ the following formula \cite{zhang2008new}:
\begin{equation}
\label{eqn:convergence-formula}
\mathcal{O}_n = \log_2\left(\frac{E_g(n/2)}{E_g(n)}\right).
\end{equation}
We tested these three shapes at $N\times N$ resolutions for $N=\{8,16,32,64,128,256\}$. Equation~\eqref{eqn:convergence-formula} implies that the convergence order corresponds to the slope in the $\log_2(N)-\log_2(E_g)$ diagram, which is shown in Figure~\ref{fig:static-recon-fits}. We use least squares fitting to estimate the slope, and the results indicate that our proposed algorithm achieves second-order accuracy, which is expected because we use straight lines to represent the interface. Higher convergence order requires using higher-order discretizations of the interface, such as curves.

We also evaluated our algorithm's static reconstruction results on unstructured meshes, which \revminor{were} generated by the Triangle mesh generator \cite{shewchuk1996triangle,shewchuk2002delaunay}. For a refinement level $l=\{0,1,2,3\}$, we restrict the maximum triangle area in the mesh to be $0.01/(2^l)^2$, and \revminor{thus,} the convergence order can be calculated as
\begin{equation}
\label{eqn:convergence-formula-triangles}
\mathcal{O}_n = \log_2\left(\frac{E_g(l-1)}{E_g(l)}\right).
\end{equation}
The numbers of triangles inside the $[0,1]^2$ domain are $\{150,620,2472,9899\}$ respectively for four refinement levels. The \revminor{reconstruction} results for $l=0$ are shown in the bottom row of Figure \ref{fig:static-recon-results}.
Based on the surface reconstruction approach proposed in this paper, we can use a relatively simple method to compute curvature: For each interior segment $\overline{\bm{p}\bm{q}}$, we find the nearest interior segment within the two adjacent triangles \revminor{to $\overline{\bm{p}\bm{q}}$}. Then, using all the vertices of these segments, including $\bm{p}$ \revminor{and} $\bm{q}$, we fit a parabola $y=ax^2+bx+c$ and compute the curvature at the midpoint $(x_p+x_q)/2$ with the formula \[\kappa=\frac{2a}{(1+(2ax+b)^2)^{3/2}}.\]
If the range of $y$ coordinates for given points is larger than the range of $x$ \revminor{coordinates}, we fit a parabola $x=ay^2+by+c$ instead and compute $\kappa$ similarly.
We measure the $L^{\infty}$ error \revminor{for the} curvature computation as \[E_{\kappa}=\max_{e_i} \frac{|\tilde{\kappa_i}-\kappa_i|}{|\kappa_i|}\] for all interior segments $e_i$, where $\tilde{\kappa_i}$ is the computed curvature for $e_i$, and $\kappa_i$ is the ground truth. Table \ref{tab:triangle-mesh-static-recon} summarizes the shape error $E_g$ and the curvature error $E_\kappa$, \revminor{along with} their convergence orders for the same three shapes on unstructured meshes. The shape error $E_g$ shows a similar second-order convergence as lattice grids. The simple curvature scheme achieves first-order convergence, and its performance decreases at \revminor{higher resolutions}, because the points used for parabolic fitting are close to each other, leading to reduced accuracy.

Higher-order curvature computation schemes are beyond the scope of this paper. However, we adopted an interface discretization scheme similar to PLIC-VOF, and in our method, each edge cut corresponds to an edge in the triangle mesh. This correspondence is not altered during area correction, allowing us to establish a local segment mesh for each interface segment. As a result, the curvature computation schemes based on PLIC-VOF and the segment mesh are compatible with our algorithm. Readers may refer to \cite{mulbah2022review,crane2018discrete} for more details.

\begin{table}[ht]
\centering
\begin{tabular}{|cc|c|c|c|c|}
\hline
\multirow{2}{*}{Shape} & \multirow{2}{*}{} & \multicolumn{4}{c|}{Refinement level($l$)} \\
\cline{3-6}
& & 0 & 1 & 2 & 3 \\
\hline
\multirow{4}{*}{circle} & $E_g$ & $3.34 \times 10^{-3}$ & $9.30 \times 10^{-4}$ & $2.09 \times 10^{-4}$ & $4.56 \times 10^{-5}$ \\
& $O_n$ & --- & $1.84$ & $2.15$ & $2.20$ \\
& $E_{\kappa}$ & $6.43 \times 10^{-1}$ & $3.25 \times 10^{-1}$ & $1.25 \times 10^{-1}$ & $8.62 \times 10^{-2}$ \\
& $O_\kappa$ & --- & $0.99$ & $1.38$ & $0.53$ \\
\hline
\multirow{2}{*}{snake} & $E_g$ & $4.55 \times 10^{-3}$ & $1.20 \times 10^{-3}$ & $3.14 \times 10^{-4}$ & $8.01 \times 10^{-5}$ \\
& $O_n$ & --- & $1.92$ & $1.93$ & $1.97$ \\
\hline
\multirow{2}{*}{heart} & $E_g$ & $7.30 \times 10^{-3}$ & $1.22 \times 10^{-3}$ & $2.81 \times 10^{-4}$ & $8.70 \times 10^{-5}$ \\
& $O_n$ & -- & $2.58$ & $2.12$ & $1.69$ \\
\hline
\end{tabular}
\caption{Static reconstruction results \revminor{and convergence orders} for three shapes on unstructured triangle meshes. $O_\kappa$ is the convergence order of $E_\kappa$.}
\label{tab:triangle-mesh-static-recon}
\end{table}

\subsection{Rider–Kothe Reversed Single Vortex}

\label{subsec:single-vortex}

\begin{table}[ht]
  \centering
    \begin{tabular}{c|@{}c@{}c@{}c@{}}
        & $64\times 64$ & $128\times 128$ & $256\times 256$ \\
        \hline
        
        $\frac{1}{2}T$ & \includegraphics[width=4cm, align=c]{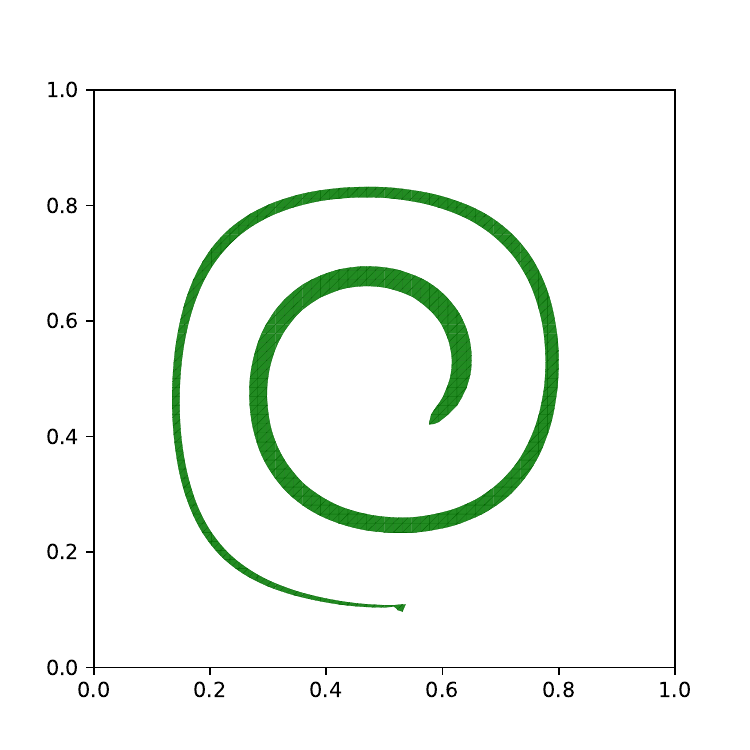} & \includegraphics[width=4cm, align=c]{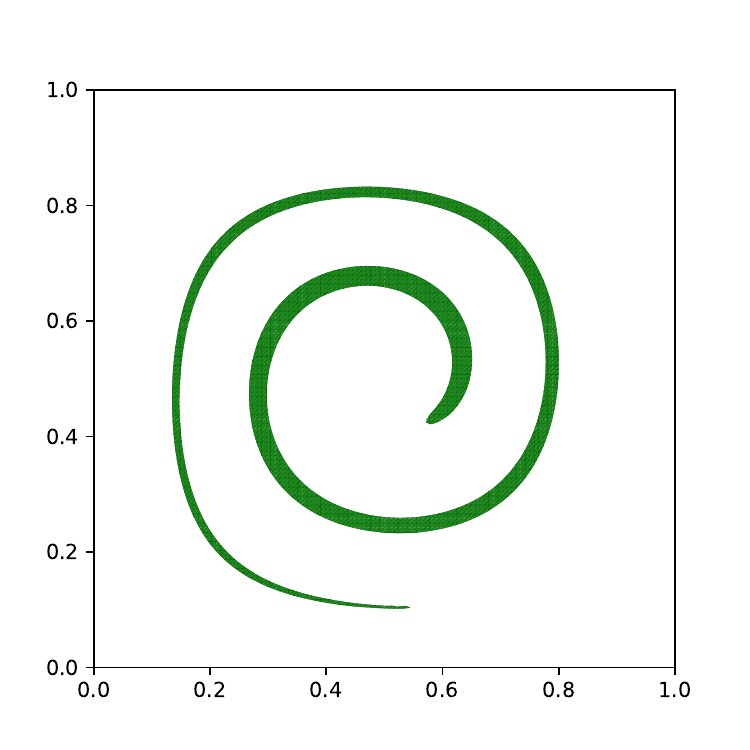} & \includegraphics[width=4cm, align=c]{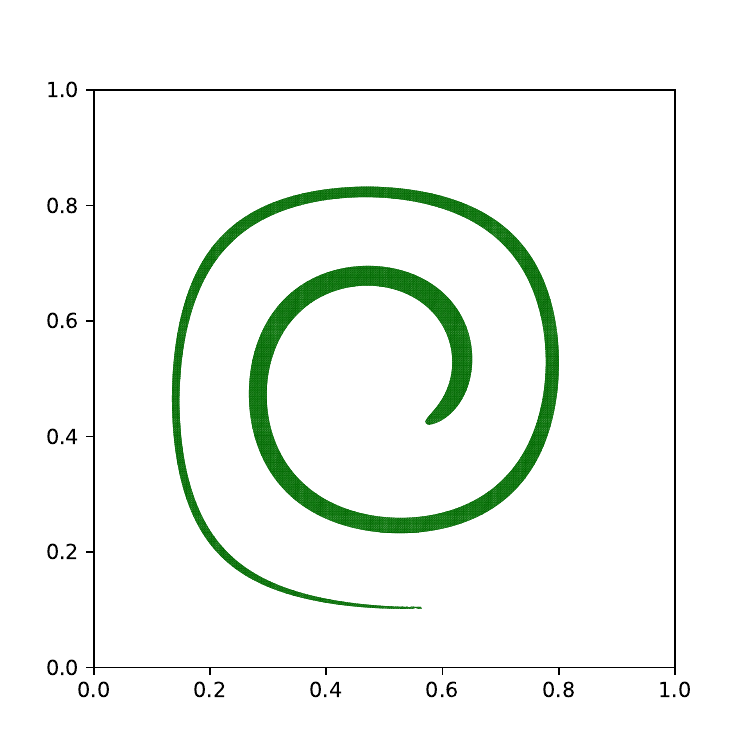} \\
        
        $\frac{3}{4}T$  & \includegraphics[width=4cm, align=c]{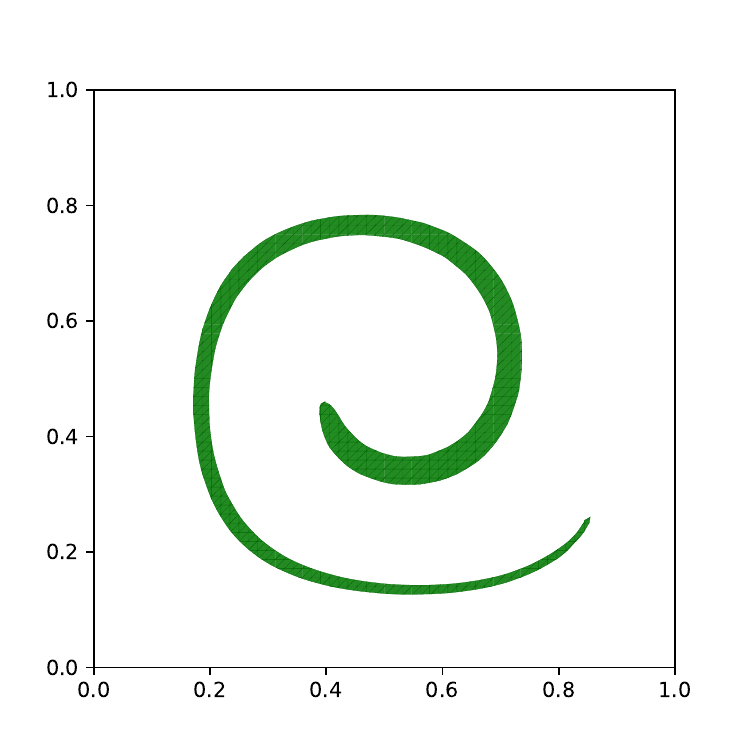} & \includegraphics[width=4cm, align=c]{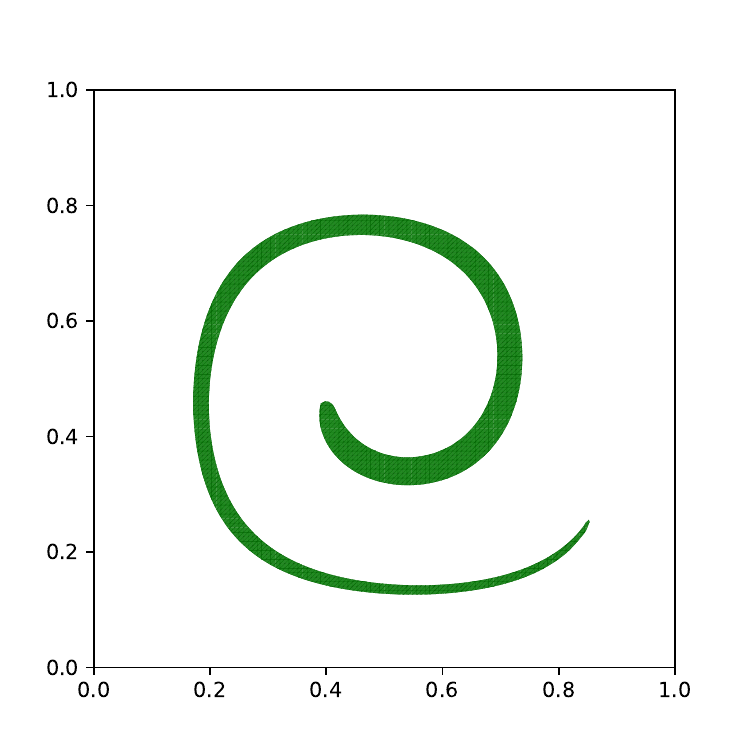} & \includegraphics[width=4cm, align=c]{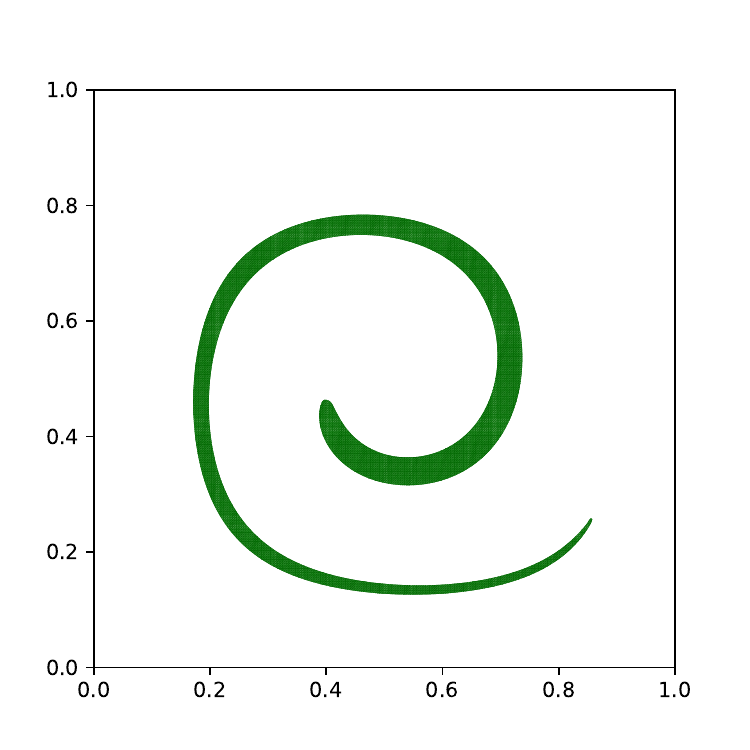} \\
        
        $T$  & \includegraphics[width=4cm, align=c]{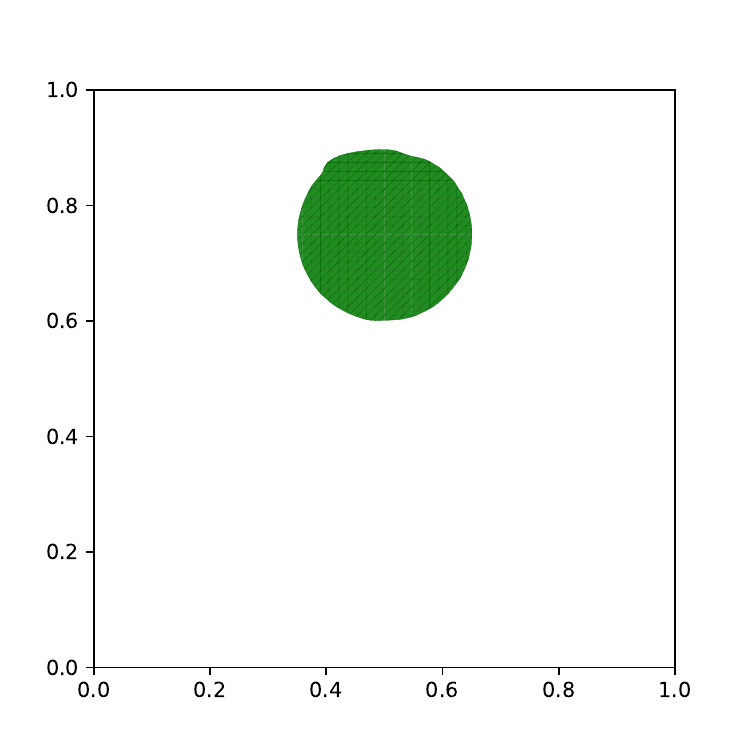} & \includegraphics[width=4cm, align=c]{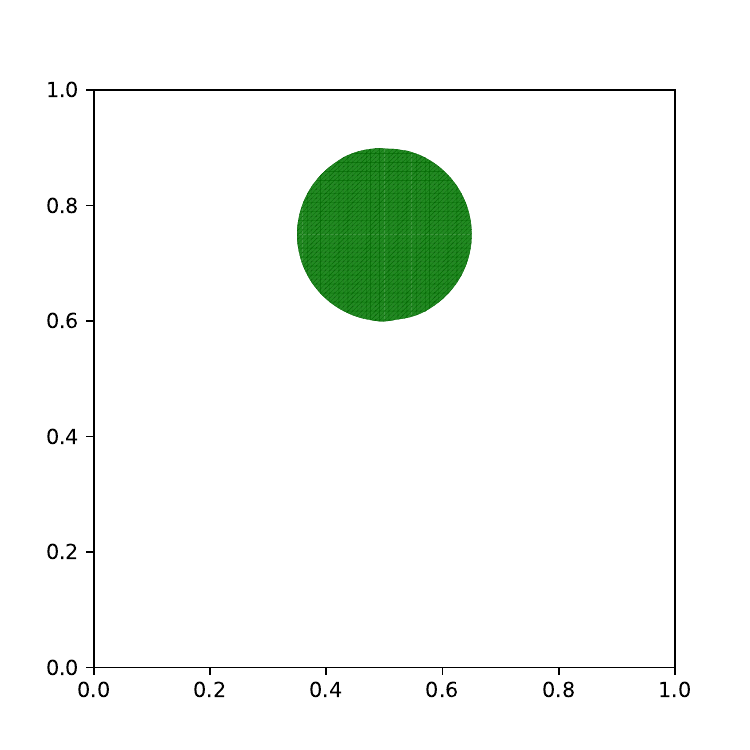}  & \includegraphics[width=4cm, align=c]{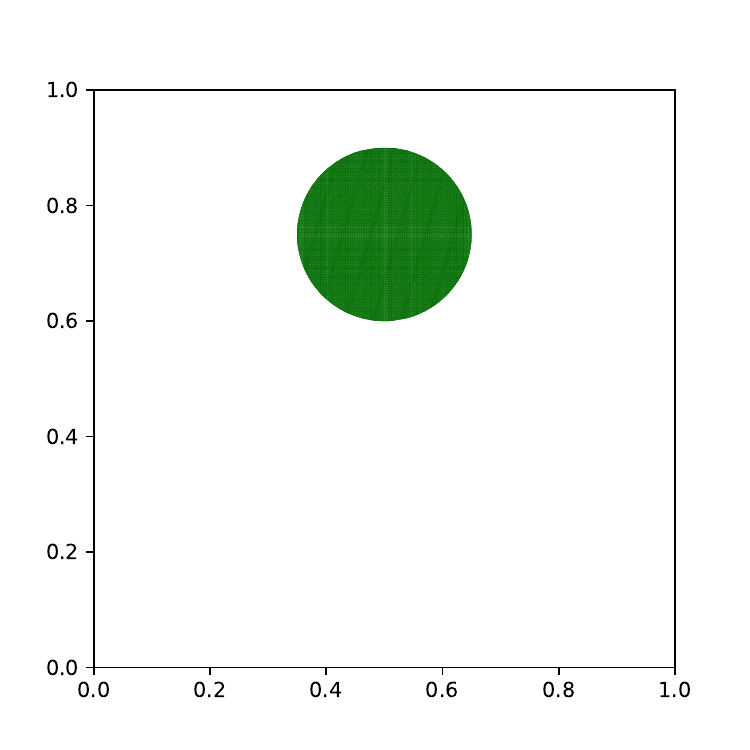}
    \end{tabular}
    \caption{Results of Rider–Kothe reversed single vortex, all $C_r=1$.}
    \label{tab:single-vortex}
\end{table}

We evaluate the efficacy of our method in handling shearing flows through the Rider-Kothe reversed single vortex test~\cite{rider1998reconstructing}. In this \revminor{test}, the velocity field is \revminor{derived from} the stream function
\begin{equation}
\psi(x,y) = \frac{1}{\pi}\sin^2(\pi x)\sin^2(\pi y)\cos\left(\frac{\pi t}{T}\right),
\end{equation}
where $T=8$.
The corresponding \revminor{velocity components} are computed as
\begin{equation}
\begin{aligned}
    u&=-2\cos(\pi t/T)\cos(\pi y)\sin^2 (\pi x)\sin(\pi y),\\
    v&= 2\cos(\pi t/T)\cos(\pi x)\sin(\pi x)\sin^2(\pi y).
\end{aligned}
\end{equation}
When subjected to this velocity field, a liquid circle initially centered at $(0.5,0.75)$ with a radius of $0.15$ undergoes significant stretching around the center, reaching its maximum extent at $T/2$, and subsequently reverses its trajectory back to its initial position. The absolute shape error is then measured using \eqref{eqn:abs-error}, with the initial shape serving as the ground truth. The advection results are summarized in Table~\ref{tab:single-vortex}. Remarkably, even at a relatively low resolution of $64\times 64$, our method exhibits \revminor{strong} performance, effectively restoring the initial state. As \revminor{the} resolution increases, \revminor{the} errors become imperceptible to the naked eye.

\begin{table}[!htbp]
    \centering
    \begin{tabular}{lcccc}
        \toprule
        Algorithm & Resolution & $E_m$ & $E_g$ & $O_n$ \\
        \midrule
        \multirow{3}{*}{Rider and Kothe~\cite{rider1998reconstructing}}  
        & $32\times 32$ & ---  & $4.78\times 10^{-2}$ & --- \\
        & $64\times 64$ & --- & $6.96\times 10^{-3}$ & $2.78$ \\
        & $128\times 128$ & ---& $1.44\times 10^{-3}$ & $2.27$\\
        \midrule
        \multirow{3}{*}{EMFPA-SIR~\cite{lopez2004volume}}  
        & $32\times 32$  & --- & $4.64\times 10^{-2}$ & --- \\
        & $64\times 64$ &  --- &$5.94\times 10^{-3}$ & $2.97$ \\
        & $128\times 128$ & --- & $5.39\times 10^{-4}$ & $3.46$\\
        \midrule
        \multirow{4}{*}{Double-PLIC~\cite{lopez2005improved}}  
        & $32\times 32$ & ---  & $5.78\times 10^{-3}$ & --- \\
        & $64\times 64$ & --- & $1.77\times 10^{-3}$ & $1.71$ \\
        & $128\times 128$ & --- & $3.30\times 10^{-4}$ & $2.42$\\
        & $256\times 256$ & --- & $8.69\times 10^{-5}$ & $1.93$\\
        \midrule
        \multirow{3}{*}{NIFPA-1~\cite{ivey2017conservative}}  
        & $64\times 64$ & $<10^{-12}$  & $1.14\times 10^{-2}$ & --- \\
        & $128\times 128$ & $<10^{-12}$ & $2.68\times 10^{-3}$ & $2.01$ \\
        & $256\times 256$ & $<10^{-12}$ & $5.37\times 10^{-4}$ & $2.32$\\
        \midrule
        \multirow{3}{*}{Owkes and Desjardins~\cite{owkes2017mass}}  
        & $64\times 64$ & $4.12\times 10^{-13}$ & $1.04\times 10^{-2}$ & --- \\
        & $128\times 128$ & $2.53\times 10^{-13}$ & $1.34\times 10^{-3}$ & $1.94$ \\
        & $256\times 256$ & $1.15\times 10^{-13}$ & $3.50\times 10^{-4}$ & $2.22$\\
        \midrule
        \multirow{4}{*}{CCU~\cite{comminal2015cellwise}}  
        & $32\times 32$ & ---  & $3.81\times 10^{-2}$ & --- \\
        & $64\times 64$ & --- & $4.58\times 10^{-3}$ & $3.06$ \\
        & $128\times 128$ & --- & $1.00\times 10^{-3}$ & $2.20$\\
        & $256\times 256$ & --- &$1.78\times 10^{-4}$ & $2.59$\\
        \midrule
        \multirow{3}{*}{iPAM~\cite{zhang2014fourth} ($h_L=0.1h$)}  
        & $32\times 32$ & ---  & $6.21\times 10^{-4}$ & --- \\
        & $64\times 64$ & --- &$1.05\times 10^{-5}$ & $2.99$ \\
        & $128\times 128$ & --- & $1.37\times 10^{-6}$ & $2.94$\\
        \midrule
        \multirow{3}{*}{AMR-MOF~\cite{ahn2009adaptive}}  
        & $32\times 32$ & ---  & $2.33\times 10^{-2}$ & --- \\
        & $64\times 64$ & --- &$3.15\times 10^{-3}$ & $2.88$ \\
        & $128\times 128$ & --- & $5.04\times 10^{-4}$ & $2.64$\\
        \midrule
        \multirow{4}{*}{Edge Cut (proposed)}  
        & $32\times 32$ & ---  & $8.75\times 10^{-3}$ & --- \\
        & $64\times 64$ & --- & $1.15\times 10^{-3}$ & $2.93$ \\
        & $128\times 128$ & --- & $1.76\times 10^{-4}$ & $2.71$\\
        & $256\times 256$ & --- & $4.61\times 10^{-5}$ & $1.93$\\
        \bottomrule
    \end{tabular}
    \caption{Shape errors of Rider–Kothe reversed single vortex tests at $t=T$ with different methods and parameters. The results of EMFPA-SIR are taken from~\cite{lopez2005improved}. The results of NIFPA-1 are taken from~\cite{maric2020unstructured}. The mass \revminor{errors} reported in~\cite{owkes2017mass} \revminor{are} the absolute \revminor{differences} in the liquid area, and we divide them by $\pi r^2$ where $r=0.15$ to obtain $E_m$ in the table.}
    \label{tab:vortex-errors-comparison}
\end{table}

The shape errors and comparisons \revminor{for} the single vortex test are detailed in Table~\ref{tab:vortex-errors-comparison} with \revminor{a} Courant number $C_r=1$.
To assess our method's ability to uphold mass conservation, we compute the relative area error:
\begin{equation}
E_m=\left|\frac{\sum_{i,j}A_{i,j}^0-\sum_{i,j}\tilde{A}_{i,j}}{\sum_{i,j}A_{i,j}^0}\right|,
\end{equation}
where $A_{i,j}^0$ denotes the liquid area in cell $(i,j)$ at $t=0$, and $\tilde{A}_{i,j}$ represents the advected liquid area in cell $(i,j)$.
Notably, our method outperforms prior methods, except for iPAM, which is an improved version of PAM. In comparison to PAM, our method is free of the high computational costs associated with maintaining and simplifying liquid polygons.

\subsection{Zalesak's Disk}

\begin{figure}[htbp]
    \centering
    \begin{subfigure}{0.45\textwidth}
        \includegraphics[width=\linewidth]{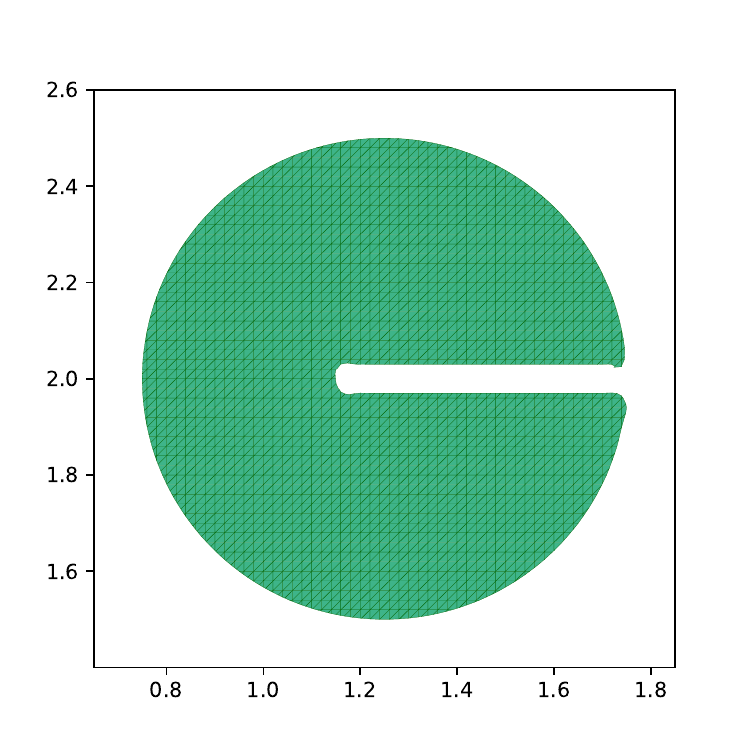}
        \caption{$t=\frac{1}{4}T$}
    \end{subfigure}
    \begin{subfigure}{0.45\textwidth}
        \includegraphics[width=\linewidth]{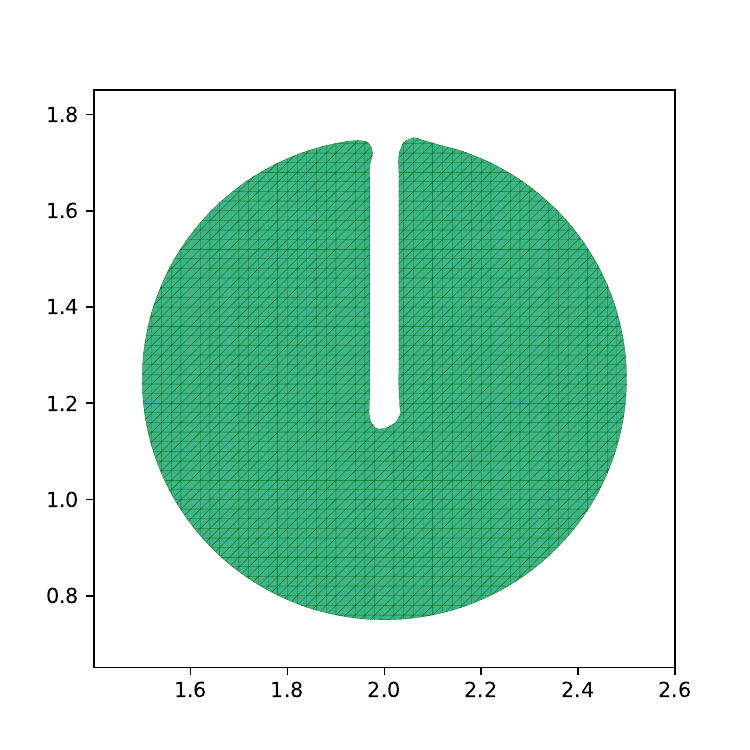}
        \caption{$t=\frac{1}{2}T$}
    \end{subfigure}

    \begin{subfigure}{0.45\textwidth}
        \includegraphics[width=\linewidth]{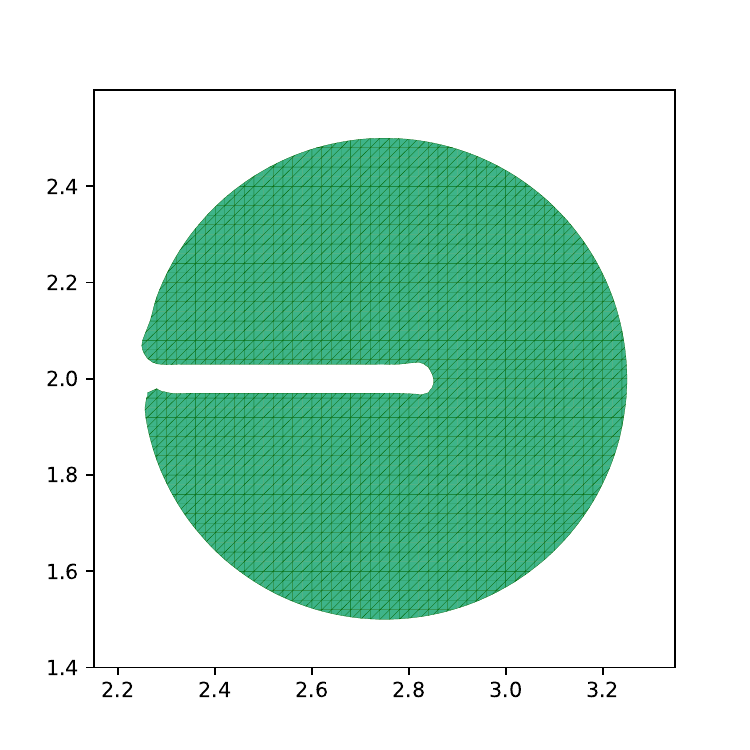}
        \caption{$t=\frac{3}{4}T$}
    \end{subfigure}
    \begin{subfigure}{0.45\textwidth}
        \includegraphics[width=\linewidth]{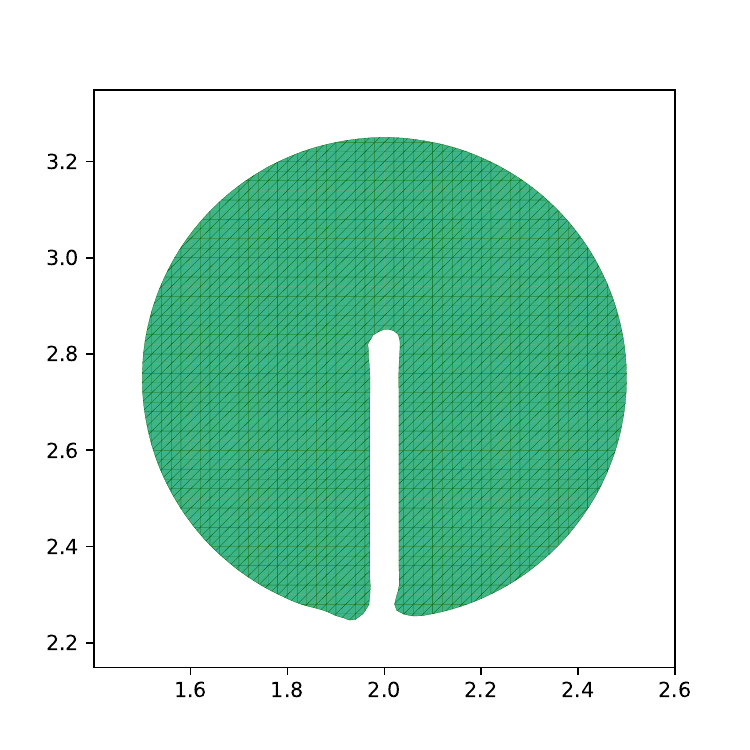}
        \caption{$t=T$}
    \end{subfigure}

    \caption{Zalesak's disk at $200\times 200$ resolution and $C_r=1$.}
    \label{fig:zalesak200}
\end{figure}

The Zalesak's disk advection test, originally proposed by Zalesak~\cite{zalesak1979fully} and then adopted by Rudman~\cite{rudman1997volume}, has become a widely recognized benchmark for evaluating the efficacy of various advection schemes in capturing sharp interfaces at different orientations. In the existing literature, this test has two different configurations, which we refer to as config A and config B. In config A, a notched disk centered at $(2.0,2.75)$ with radius $R=0.5$ is initialized inside a $[0,4]^2$ domain. The \revminor{notch width is} $s=0.06$ and the width of the bridge between the left and right halves is $r=0.4$~\cite{zhang2008new}. Therefore, we can compute its initial liquid area $A^0=0.7494$.
The stream function is expressed as~\cite{zhang2008new}
\begin{equation}
\psi(x,y)=-\frac{\omega}{2}[(x-x_o)^2+(y-y_o)^2],
\end{equation}
where $(x_o,y_o)=(2,2)$ denotes the center of the computational domain, and $\omega=0.5$ represents the angular velocity. Consequently, the velocity field is defined by:
\begin{equation}
\begin{aligned}
u&=-\omega(y-2),\\
v&=\omega(x-2).
\end{aligned}
\end{equation}
At $T=4\pi$, the notched disk completes a full circle and returns to its initial position. The advection results are depicted in Figure~\ref{fig:zalesak200}.

On the other hand, the computational domain in config B is $[-0.5,0.5]^2$, and the disk with $R=0.15$, $r=s=0.05$, centered at $(0,0.25)$~\cite{owkes2017mass}, has $A^0=0.05822$. To compensate for the difference between the two configurations, we define the relative shape error to the initial liquid area $A^0$ as \[E_r=\frac{\sum_{i,j} |A_{i,j} - \tilde{A}_{i,j}|}{|\sum_{i,j}A_{i,j}^0|}=\frac{E_g}{A^0}.\]
The relative area error $E_m$ and the relative shape error $E_g$ for the Zalesak's disk test are provided in Table~\ref{tab:zalesak-results}.
This demonstrates that our proposed algorithm achieves state-of-the-art performance in preserving geometric shapes among VOF-based methods. Especially, our method \revminor{effectively preserves} of the interior notch, \revminor{despite} slight distortions at sharp corners attributable to the simplified representation capacity of the triangle cuts. At the same time, our algorithm also accurately maintains mass conservation. \revminor{Moreover,} our method shows better accuracy at $C_r=1$ compared to $C_r=0.25$, because the former requires fewer time steps, \revminor{thus reducing} the accuracy loss caused by interface discretization. Specifically, our algorithm achieves near-perfect mass conservation in most cases, except for the uncommon case $4$, as discussed in Section~\ref{subsec:edgecut-correction}. Using fewer time steps helps reduce the occurrence of such cases. As a result, our algorithm attains high mass conservation accuracy at $C_r=1$.

\begin{table}[!htbp]
    \centering
    \begin{tabular}{lccccc}
        \toprule
        Algorithm & $C_r$ & Resolution & $E_m$ & $E_r$\\
        \midrule
        \multirow{1}{*}{EMFPA-SIR~\cite{lopez2004volume}}  & \multirow{1}{*}{0.25} & $200\times 200$ & --- & $8.74\times 10^{-3}$\\
        \hline
        \multirow{3}{*}{Owkes and Desjardins~\cite{owkes2017mass}}  & \multirow{3}{*}{} & $50\times 50$ & $4.41\times 10^{-14}$ & $6.83\times 10^{-2}$\\
        & & $100\times 100$ & $7.93\times 10^{-13}$ & $2.02\times 10^{-2}$\\
         & & $200\times 200$ & $7.78\times 10^{-12}$ & $9.09\times 10^{-3}$\\
        \hline
        \multirow{3}{*}{THINC/QQ~\cite{xie2017toward}($\beta=6$)}  & \multirow{3}{*}{0.25} & $50\times 50$ & --- & $8.96\times 10^{-2}$\\
        & & $100\times 100$ & --- & $3.22\times 10^{-2}$\\
         & & $200\times 200$ & --- & $1.67\times 10^{-2}$\\
        \hline
        \multirow{1}{*}{AMR-MOF~\cite{ahn2009adaptive}}  & \multirow{1}{*}{} & $200\times 200$ & --- & $2.32\times 10^{-4}$\\
        \hline
        PAM~\cite{zhang2008new} & \multirow{4}{*}{0.25} & $200\times 200$ & --- & $5.30\times 10^{-4}$ \\
        \multirow{3}{*}{iPAM~\cite{zhang2014fourth}($h_L=5h^2$)}  &  & $50\times 50$ & --- & $8.74\times 10^{-3}$\\
        & & $100\times 100$ & --- & $4.00\times 10^{-4}$\\
         & & $200\times 200$ & --- & $1.79\times 10^{-5}$\\
        \hline
        \multirow{6}{*}{Edge Cut (proposed)}  & \multirow{3}{*}{$1.0$} & $50\times 50$ & $4.68\times 10^{-3}$ & $2.05\times 10^{-2}$\\
         & & $100\times 100$ & $5.30\times 10^{-10}$ & $7.13\times 10^{-3}$\\
        &  & $200\times 200$ & $1.65\times 10^{-10}$ & $2.20\times 10^{-3}$\\
        & \multirow{3}{*}{$0.25$} & $50\times 50$ & $1.65\times 10^{-2}$ & $4.31\times 10^{-2}$\\
        & & $100\times 100$ & $2.46\times 10^{-4}$ & $1.43\times 10^{-2}$\\
       &  & $200\times 200$ & $5.97\times 10^{-7}$ & $4.88\times 10^{-3}$\\
        \bottomrule
    \end{tabular}
    \caption{Results of Zalesak's disk test. The result of AMR-MOF is taken from~\cite{zhang2014fourth}. Owkes \& Desjardins~\cite{owkes2017mass} and THINC/QQ~\cite{xie2017toward} use config B; other tests use config A.}
    \label{tab:zalesak-results}
\end{table}

\subsection{Deformation Field}

This deformation field test, originally proposed by Smolarkiewicz \cite{smolarkiewicz1982multi}, serves as a rigorous assessment of advection methods, evaluating their effectiveness in handling highly deformable flows.

The computational domain is consistent with the Rider-Kothe reversed vortex test, covering the unit square $[0,1]^2$. Initially, the liquid region \revminor{is} a circle centered at $(0.5,0.5)$. The governing stream function\revminor{, which describes the flow velocity field,} is expressed as \cite{zhang2008new}:
\begin{equation}
\psi(x,y)=\frac{1}{n\pi}\sin(n\pi(x+0.5))\cos(n\pi(y+0.5))\cos\left(\frac{\pi t}{T}\right),
\end{equation}
where $T=2$ represents the period, and the parameter $n=4$ denotes the number of \revminor{vortices}. As the evolution progresses to $t=T$, the liquid region undergoes deformation and ultimately returns to its initial state.
Consequently, the velocity field is characterized by:
\begin{equation}
\begin{aligned}
u&=-\cos\left(\frac{\pi t}{T}\right)\sin(n\pi (x + 0.5))\sin(n\pi(y + 0.5)),\\
v&=-\cos\left(\frac{\pi t}{T}\right)\cos(n\pi (x + 0.5))\cos(n\pi(y + 0.5)).
\end{aligned}
\end{equation}

\begin{figure}[ht]
    \centering

    \begin{subfigure}{0.29\textwidth}
        \includegraphics[width=\linewidth]{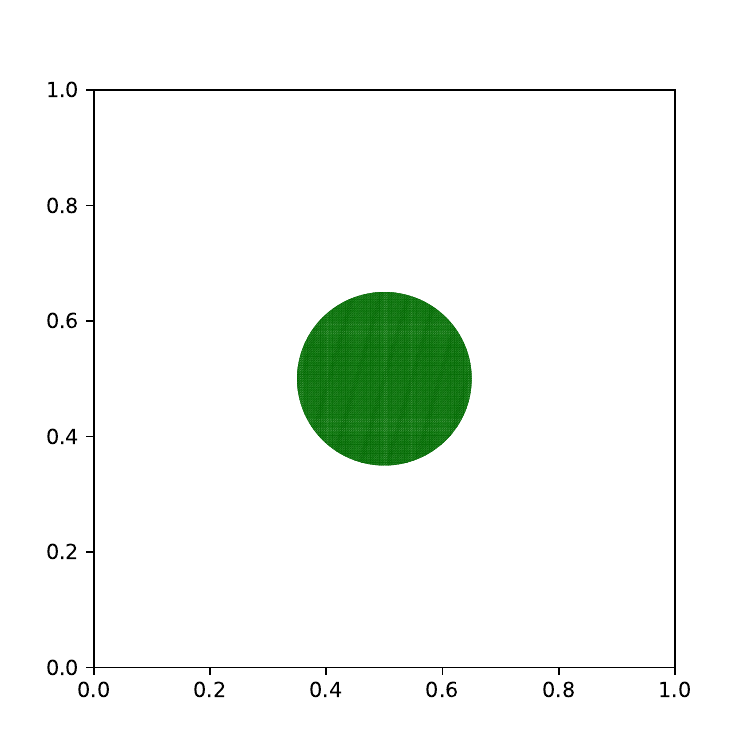}
        \caption{$t=0$}
    \end{subfigure}\hfill
    \begin{subfigure}{0.29\textwidth}
        \includegraphics[width=\linewidth]{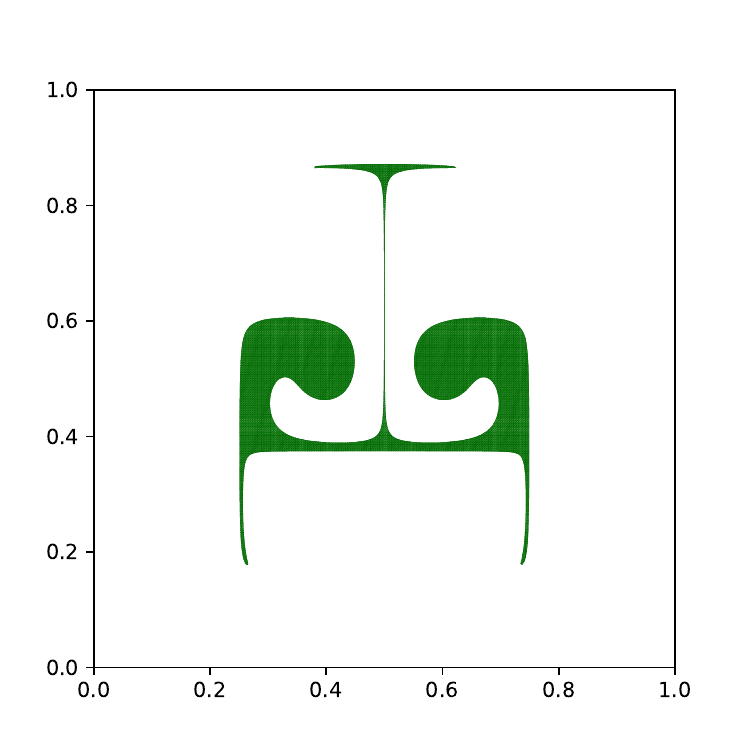}
        \caption{$t=\frac{1}{4}T$}
    \end{subfigure}\hfill
    \begin{subfigure}{0.29\textwidth}
        \includegraphics[width=\linewidth]{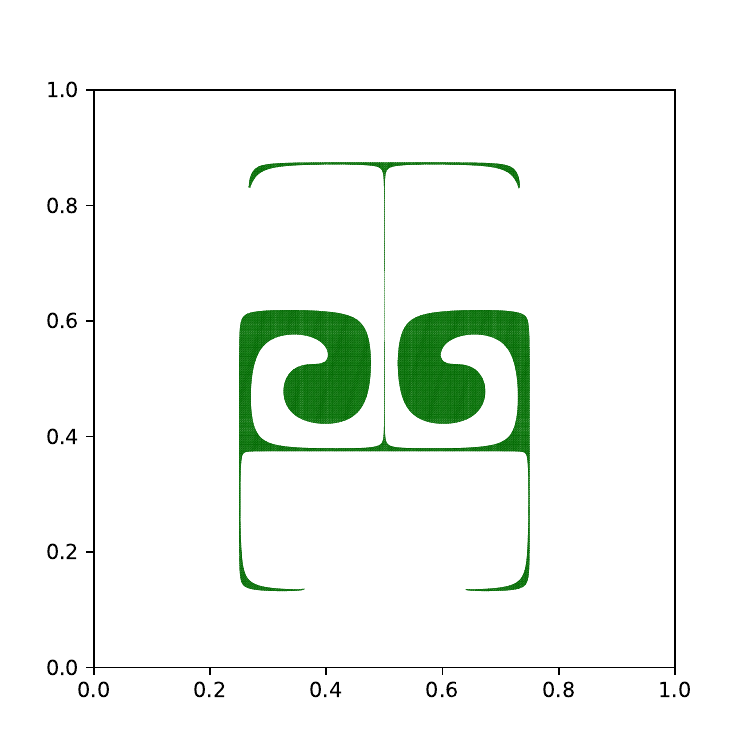}
        \caption{$t=\frac{1}{2}T$}
    \end{subfigure}

    \vspace{-0.5cm}

    \begin{subfigure}{0.29\textwidth}
        \includegraphics[width=\linewidth]{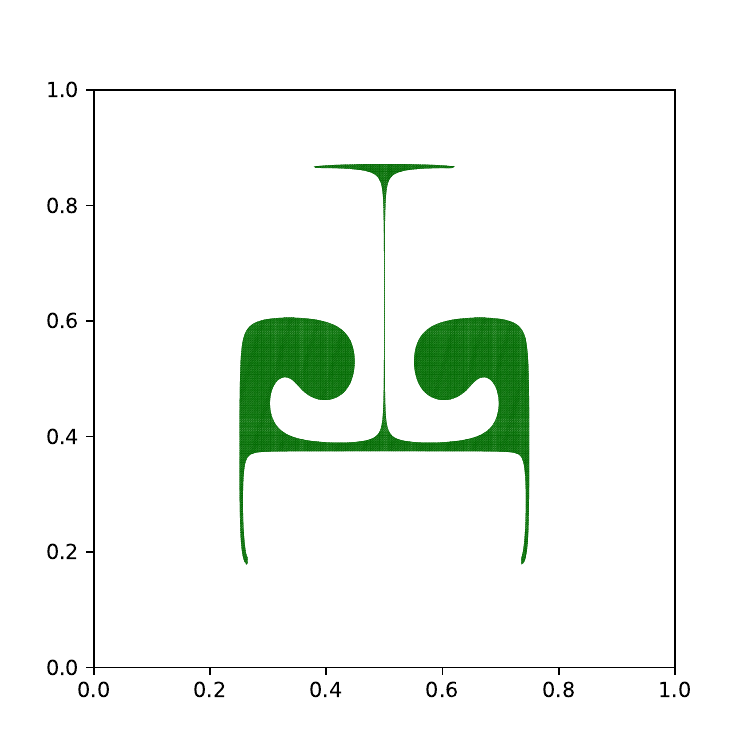}
        \caption{$t=\frac{3}{4}T$}
    \end{subfigure}\hfill
    \begin{subfigure}{0.29\textwidth}
        \includegraphics[width=\linewidth]{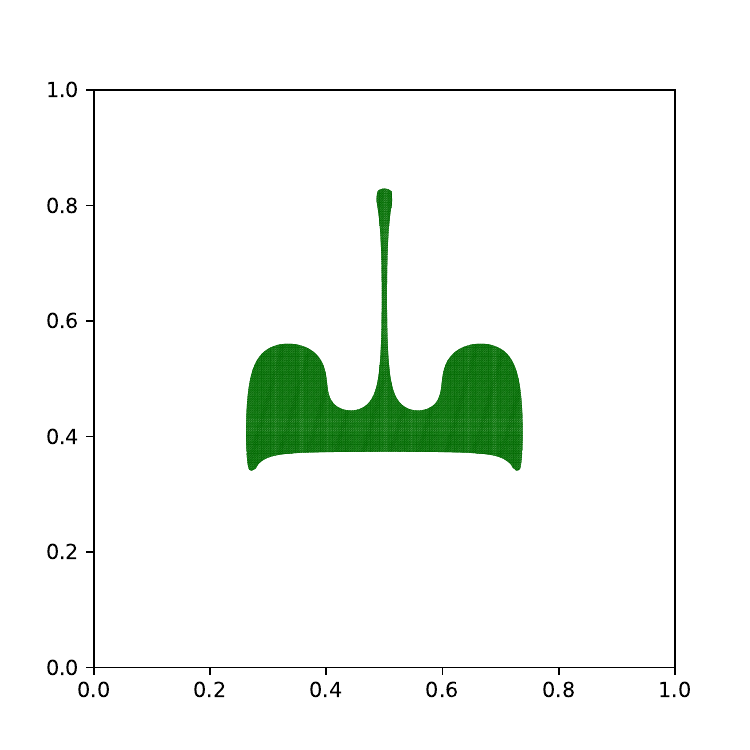}
        \caption{$t=\frac{7}{8}T$}
    \end{subfigure}\hfill
    \begin{subfigure}{0.29\textwidth}
        \includegraphics[width=\linewidth]{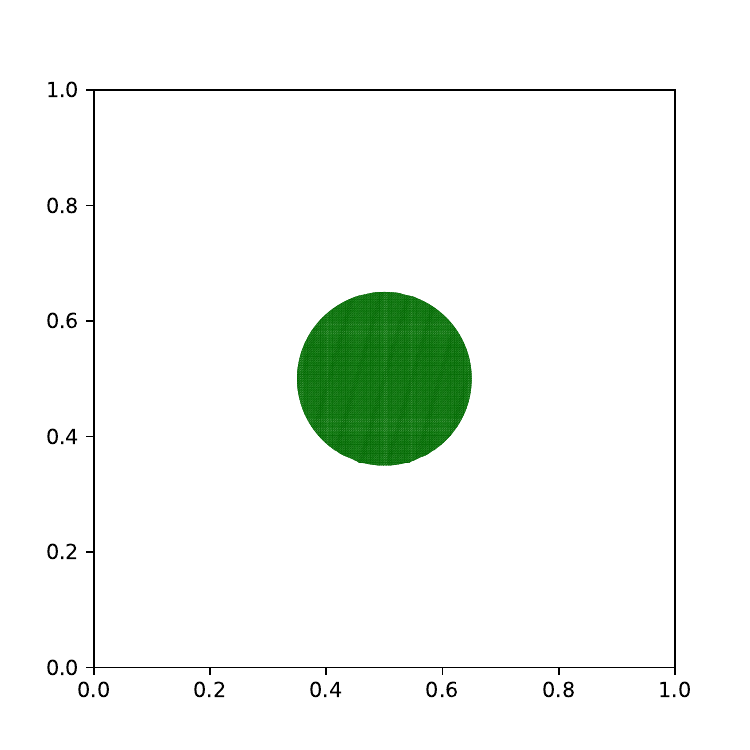}
        \caption{$t=T$}
    \end{subfigure}

    \caption{Deformation field test with $128\times 128$ grid and $C_r=1$.}
    \label{fig:deformation128}
\end{figure}

The shape errors are presented in Table~\ref{tab:deformation-results}, \revminor{along with} visual representations of the advected liquid regions in Figure~\ref{fig:deformation128}. Notably, our method exhibits an exceptional capability to accurately capture even the intricately thin sections of the liquid.

\begin{table}[htbp]
    \centering
    \begin{tabular}{lccccc}
        \toprule
        Algorithm & $C_r$ & Resolution & $E_m$ & $E_g$ & $O_h$ \\
        \midrule

        \multirow{2}{*}{Rider and Kothe~\cite{rider1998reconstructing}}  & \multirow{2}{*}{} & $64\times 64$ & --- & $1.12\times 10^{-2}$ & --- \\
        & & $128\times 128$ & --- & $5.95\times 10^{-3}$ & $0.91$ \\
        \hline
        \multirow{3}{*}{THINC~\cite{xiao2005simple}}  & \multirow{3}{*}{} & $32\times 32$ & --- & $2.60\times 10^{-2}$ & --- \\
        & & $64\times 64$ & --- & $1.38\times 10^{-2}$ & $0.91$ \\
        & & $128\times 128$ & --- & $9.26\times 10^{-3}$ & $0.58$ \\
        \hline
        \multirow{6}{*}{CCU~\cite{comminal2015cellwise}}  & \multirow{3}{*}{$1.0$} & $64\times 64$ & --- & $1.07\times 10^{-2}$ & --- \\
        & & $128\times 128$ & --- & $5.58\times 10^{-3}$ & $0.94$ \\
        & & $256\times 256$ & --- & $1.47\times 10^{-3}$ & $1.93$\\
        & \multirow{3}{*}{$0.1$} & $64\times 64$ & --- & $1.42\times 10^{-2}$ & --- \\
        & & $128\times 128$ & --- & $7.31\times 10^{-3}$ & $0.96$\\
        & & $256\times 256$ & --- & $1.60\times 10^{-3}$ & $2.19$ \\
        \hline

        \multirow{6}{*}{PAM~\cite{zhang2008new}}  & \multirow{3}{*}{$1.0$} & $64\times 64$ & $7.11\times 10^{-5}$ & $7.73\times 10^{-5}$ & --- \\
        & & $128\times 128$ & $7.44\times 10^{-6}$ & $7.19\times 10^{-6}$ & $3.43$ \\
        & & $256\times 256$ & $7.45\times 10^{-7}$ & $3.85\times 10^{-7}$ & $4.22$\\
        & \multirow{3}{*}{$0.1$} & $64\times 64$ & $7.16\times 10^{-6}$ & $8.89\times 10^{-5}$ & --- \\
        & & $128\times 128$ & $7.42\times 10^{-7}$ & $1.41\times 10^{-5}$ & $2.67$\\
        & & $256\times 256$ & $9.02\times 10^{-8}$ & $1.86\times 10^{-6}$ & $2.92$ \\
        \hline
        \multirow{3}{*}{iPAM~\cite{zhang2014fourth}($h_L=h^{3/2}$)}  & \multirow{3}{*}{} & $64\times 64$ & --- & $4.31\times 10^{-5}$ & --- \\
        & & $128\times 128$ & --- & $5.55\times 10^{-6}$ & $2.96$ \\
        & & $256\times 256$ & --- & $6.96\times 10^{-7}$ & $3.00$ \\
        \hline
        
        \multirow{6}{*}{Edge Cut (proposed)}  & \multirow{3}{*}{$1.0$} & $64\times 64$ & $6.25\times 10^{-5}$ & $1.79\times 10^{-3}$ & --- \\
        & & $128\times 128$ & $1.97\times 10^{-4}$ & $4.26\times 10^{-4}$ & $2.07$ \\
        & & $256\times 256$ & $1.31\times 10^{-5}$ & $7.63\times 10^{-5}$ & $2.48$\\
        & \multirow{3}{*}{$0.1$} & $64\times 64$ & $1.39\times 10^{-3}$ & $2.49\times 10^{-3}$ & --- \\
        & & $128\times 128$ & $1.36\times 10^{-3}$ & $8.20\times 10^{-4}$ & $1.60$\\
        & & $256\times 256$ & $1.12\times 10^{-4}$ & $1.46\times 10^{-4}$ & $2.49$ \\
        \bottomrule
    \end{tabular}
    \caption{Results of the deformation field test.}
    \label{tab:deformation-results}
\end{table}

\subsection{Time Cost}
\label{subsec:time-cost}

\begin{table}[ht]
\centering
\begin{tabular}{lcccc}
\toprule
\textbf{Algorithm} & \textbf{Running Time (s)} \\ 
\midrule
AMR-MOF~\cite{ahn2009adaptive}      & $107.36$                     \\
iPAM~\cite{zhang2014fourth} ($h_L=0.1h$)        & $16.32$                     \\
CCU~\cite{comminal2015cellwise}        & $5.98$                      \\
Owkes and Desjardins~\cite{owkes2017mass} & $0.17$ \\
Edge Cut (proposed) & $3.07$                      \\
\bottomrule
\end{tabular}
\caption{Comparison of \revminor{running time between different algorithms}. Timing data \revminor{for} iPAM~\cite{zhang2014fourth} and CCU~\cite{comminal2015cellwise} are scaled \revminor{for comparison}.}
\label{tab:runtime_comparison}
\end{table}

In this section, we will provide the timing data acquired on the $128\times 128$ single vortex test in Section~\ref{subsec:single-vortex} and discuss the algorithm's efficiency. Table~\ref{tab:runtime_comparison} shows the comparison of execution time between different methods. Our timing data \revminor{is} obtained from a workstation with a 3.6 GHz Intel i7-12700K CPU and 32 GB RAM, which we used to conduct all experiments. In terms of runtime efficiency, our method outperformed most previous methods with high interface-tracking accuracy, attributed to our simple interface representation data structure and advection algorithm, which is easy to implement and is free of costly operations like optimization processes in MOF, or geometric \revminor{editing} of polygons in PAM.

As a prototype program, our algorithm still has significant room for time optimization. In the interface advection (see Algorithm~\ref{alg:edgecut-advection-full}), the most time-consuming part is the polygon-polygon intersection in equation \eqref{eqn:preimage-intersection-polys}, which takes $2.95$s($96\%$) of the runtime, and all other parts only take $0.12$s. This is primarily due to our use of the exact geometry computation procedure of CGAL \cite{alliez2016cgal} to avoid computing errors when degeneracy cases are found, \textit{e.g.,} an edge in the pre-image is very close to a vertex of the liquid polygon, leading to mistakenly identifying two edge cuts on the connected line segments instead of one.
Since exact computations can reduce program speed by $1$ to $2$ orders of magnitude compared to floating-point geometric operations, especially in cases close to degeneracy \cite{bartels2021fast}, we anticipate that the runtime efficiency of our program can be significantly improved by adopting floating-point geometric algorithms that can handle degeneracy cases \cite{wang2014adaptive} to a similar level to Owkes and Desjardins\cite{owkes2017mass}.

We can further optimize our program by leveraging the fact that methods like MOF and PAM rely on actual polygon-polygon intersections to calculate the moments of fluid, while our algorithm only requires the intersection area for edge cut correction in Algorithm~\ref{alg:quadratic-area-correction}, and the triangle edge cuts advection (see Algorithm~\ref{alg:edgecut-simple-advection}) can be built only by querying segment-segment intersections and cross products. Although Algorithm~\ref{alg:case2-calc-vt} may require exact polygon-polygon intersections, it is seldom called. In the current implementation, we will calculate these actual intersections in all cases, which can be optimized in the future.

In the current implementation, thread-unsafe features \revminor{in} the CGAL library, such as lazy types \cite{de2019accelerating}, significantly hinder the parallelism of our algorithm. However, in our proposed Algorithm~\ref{alg:edgecut-advection-full}, each triangle is \revminor{processed} separately, only sharing the liquid polygons $\{\mathcal{P}_i^n\}$ as read-only data. \revminor{This allows for efficient parallelization using} thread-safe features \revminor{from} shared-memory programming platforms like OpenMP~\cite{chandra2001parallel} or TBB~\cite{voss2019pro}.

\section{Discussion on 3D Extension}
\label{sec:discussion-3d}

\subsection{Extending to 3D Space}

\begin{figure}[!htbp]
    \centering
    \begin{subfigure}[b]{0.30\textwidth}
        \centering
        \includegraphics[width=\textwidth]{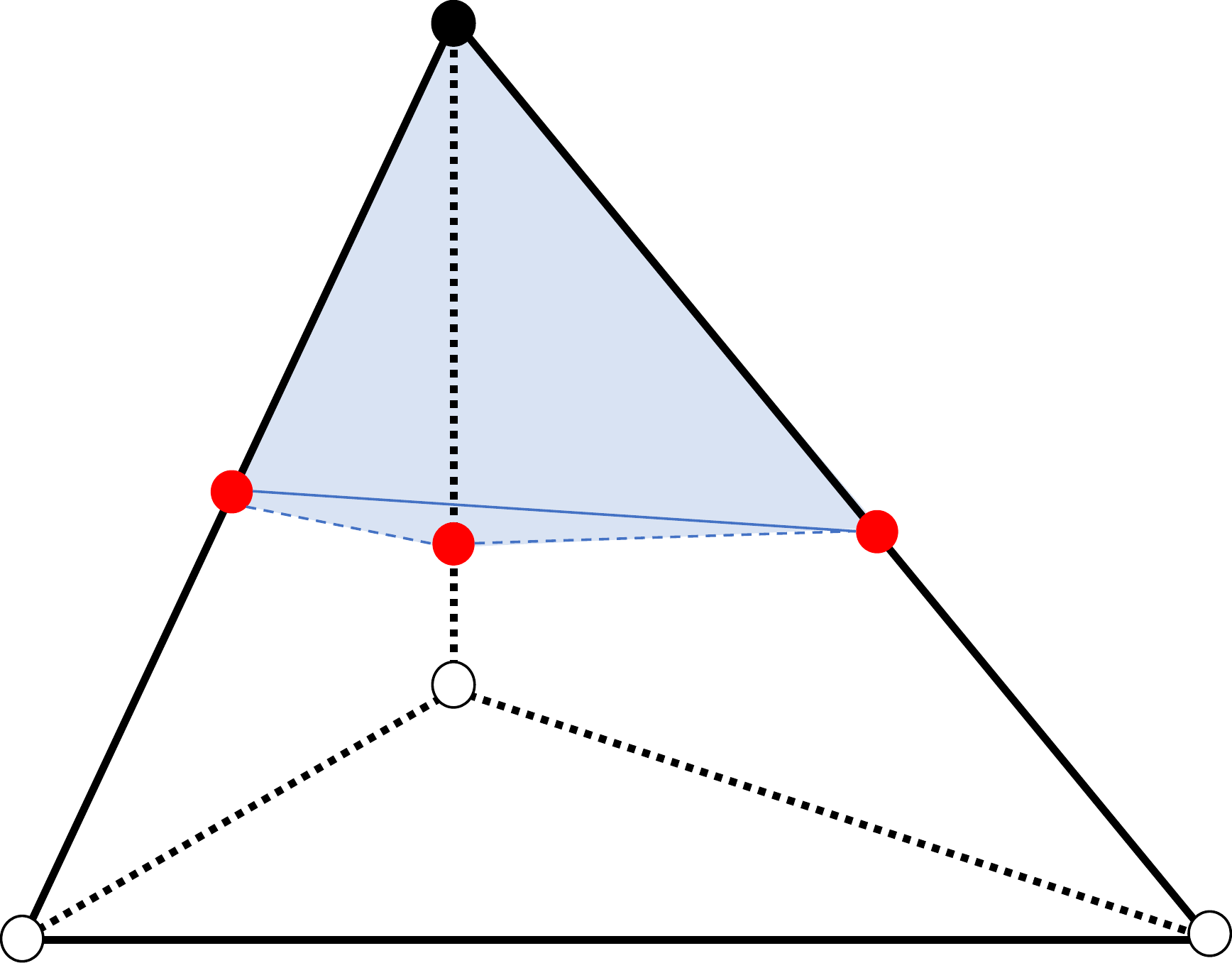}
        \caption{$0$ edges have two valid cuts.}
        \label{fig:1fluid_0double}
    \end{subfigure}
    \hspace{0.10\textwidth}
    \begin{subfigure}[b]{0.30\textwidth}
        \centering
        \includegraphics[width=\textwidth]{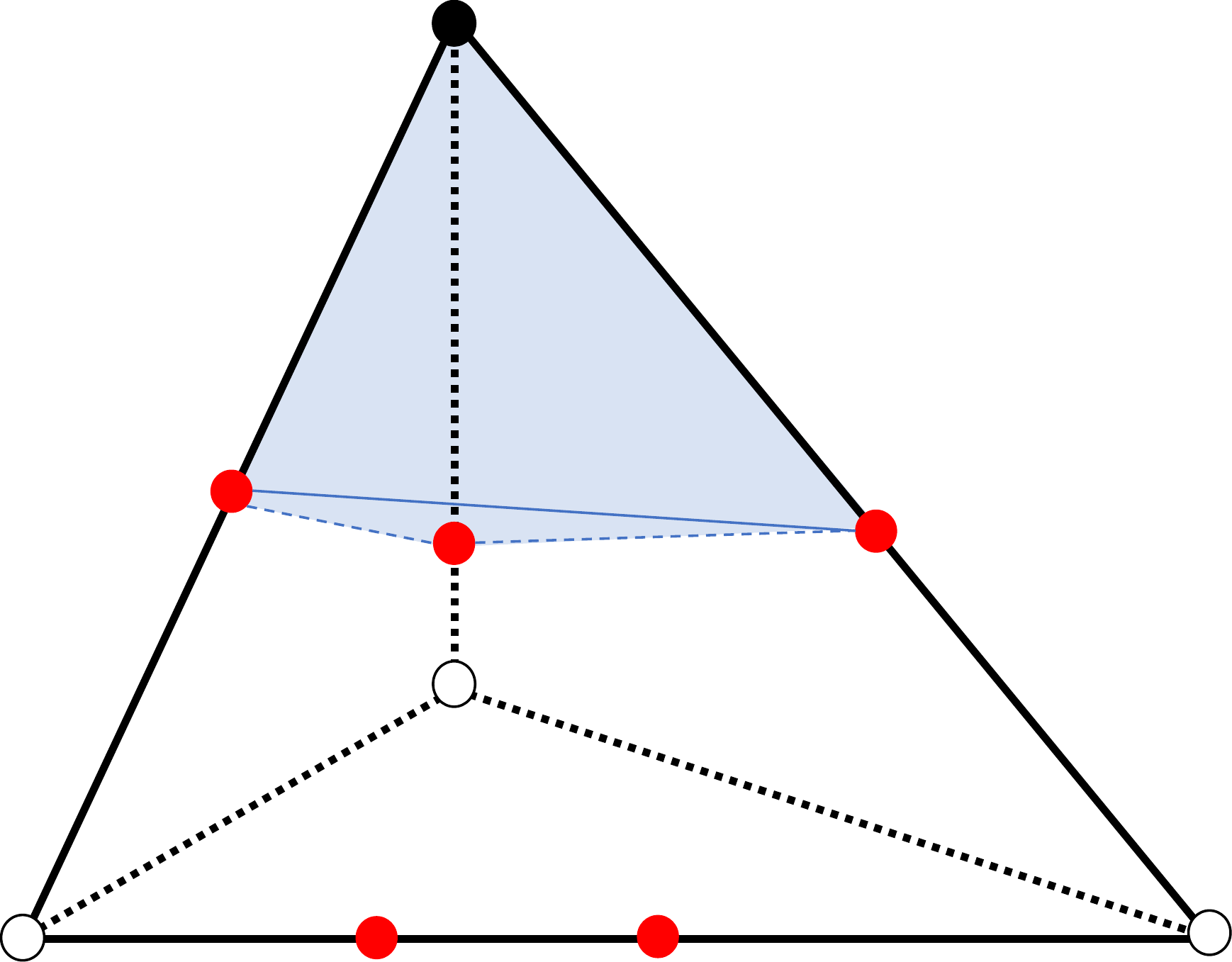}
        \caption{$1$ edge has two valid cuts.}
        \label{fig:1fluid_1double}
    \end{subfigure}

    \vspace{0.5cm}

    \begin{subfigure}[b]{0.30\textwidth}
        \centering
        \includegraphics[width=\textwidth]{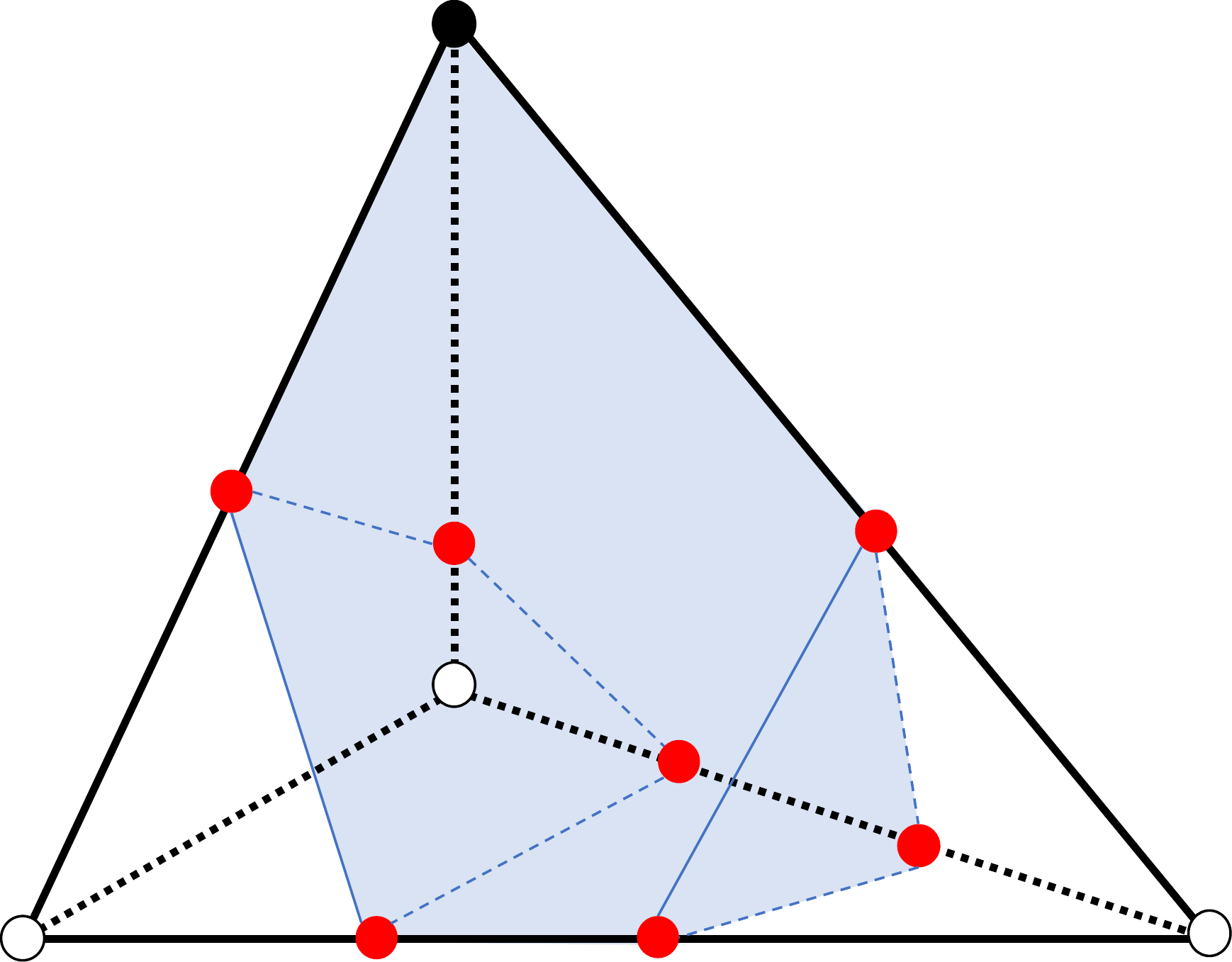}
        \caption{$2$ edges have two valid cuts.}
        \label{fig:1fluid_2double}
    \end{subfigure}
    \hspace{0.10\textwidth}
    \begin{subfigure}[b]{0.30\textwidth}
        \centering
        \includegraphics[width=\textwidth]{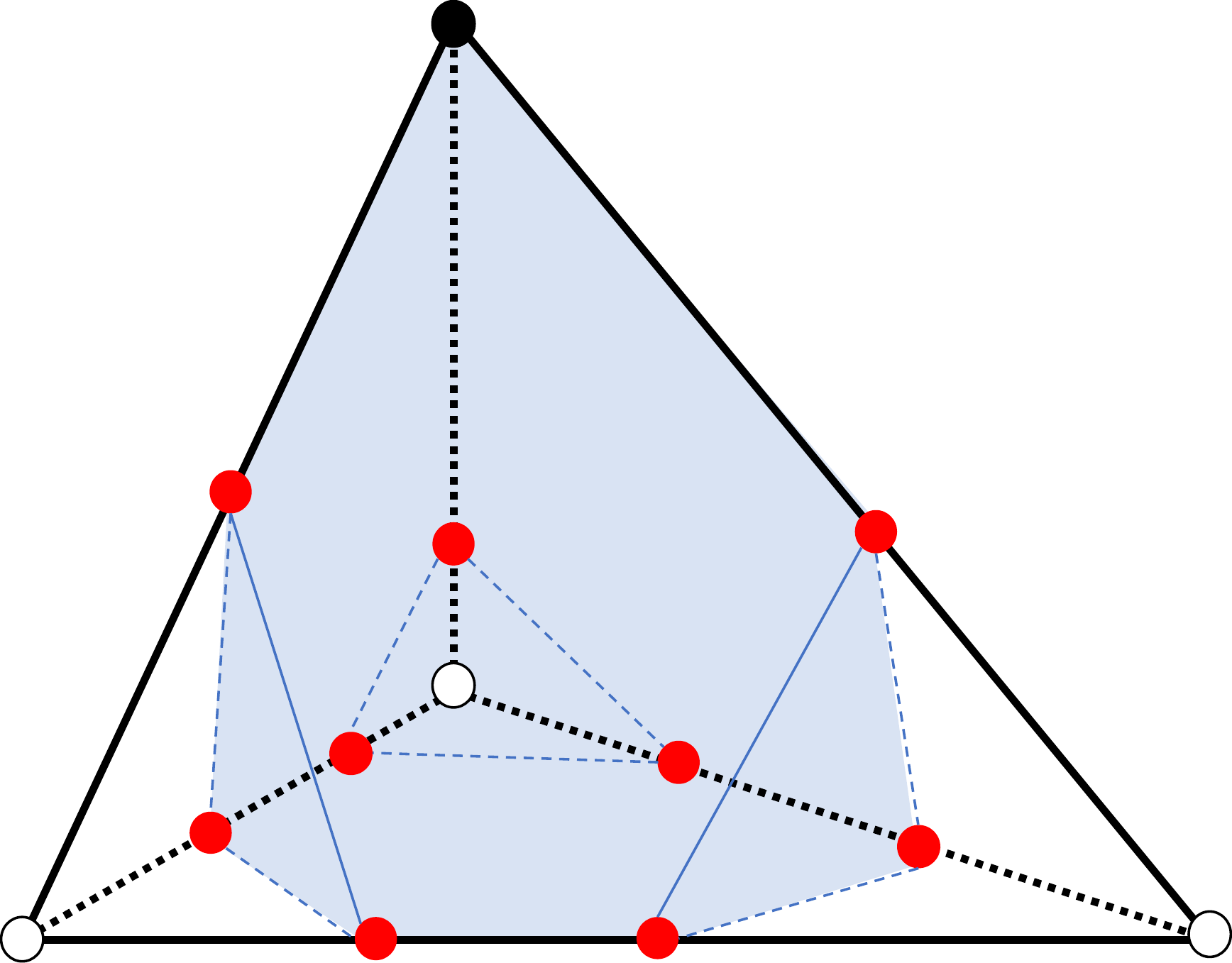}
        \caption{$3$ edges have two valid cuts.}
        \label{fig:1fluid_3double}
    \end{subfigure}
    \caption{Four possible cases with $1$ fluid vertex and $3$ air vertices. (a) No edges have two valid cuts. One interface is reconstructed in the tetrahedron. (b) One edge has two valid cuts. These two cuts are ignored and the interface reconstruction is the same as (a). (c) Two edges have two valid cuts. Two interfaces are reconstructed. (d) Three edges have two valid cuts. Three interfaces are reconstructed. }
    \label{fig:1fluid}
\end{figure}

In this section, we briefly explore the possibility of extending our algorithm to 3D space. Now consider performing \revminor{interface} reconstruction based on the edge cut information of a tetrahedron in 3D space, \revminor{assuming} that all interfaces are planes inside the tetrahedron. Similar to the discussion in Section~\ref{sec:triangle-edge-cut}, exchanging the roles of liquid and air, \revminor{or} rotating the tetrahedron, \revminor{does} not change the interface inside it. Therefore, among these equivalent cases, we can focus on discussing \revminor{only} one of them. Without loss of generality, we can assume that among the four vertices of the tetrahedron, there are $0$, $1$, or $2$ \revminor{fluid} vertices, \revminor{while} the remaining vertices being air. Next, we will discuss these three cases individually.

\begin{figure}[!htbp]
    \centering
    \begin{subfigure}[b]{0.30\textwidth}
        \centering
        \includegraphics[width=\textwidth]{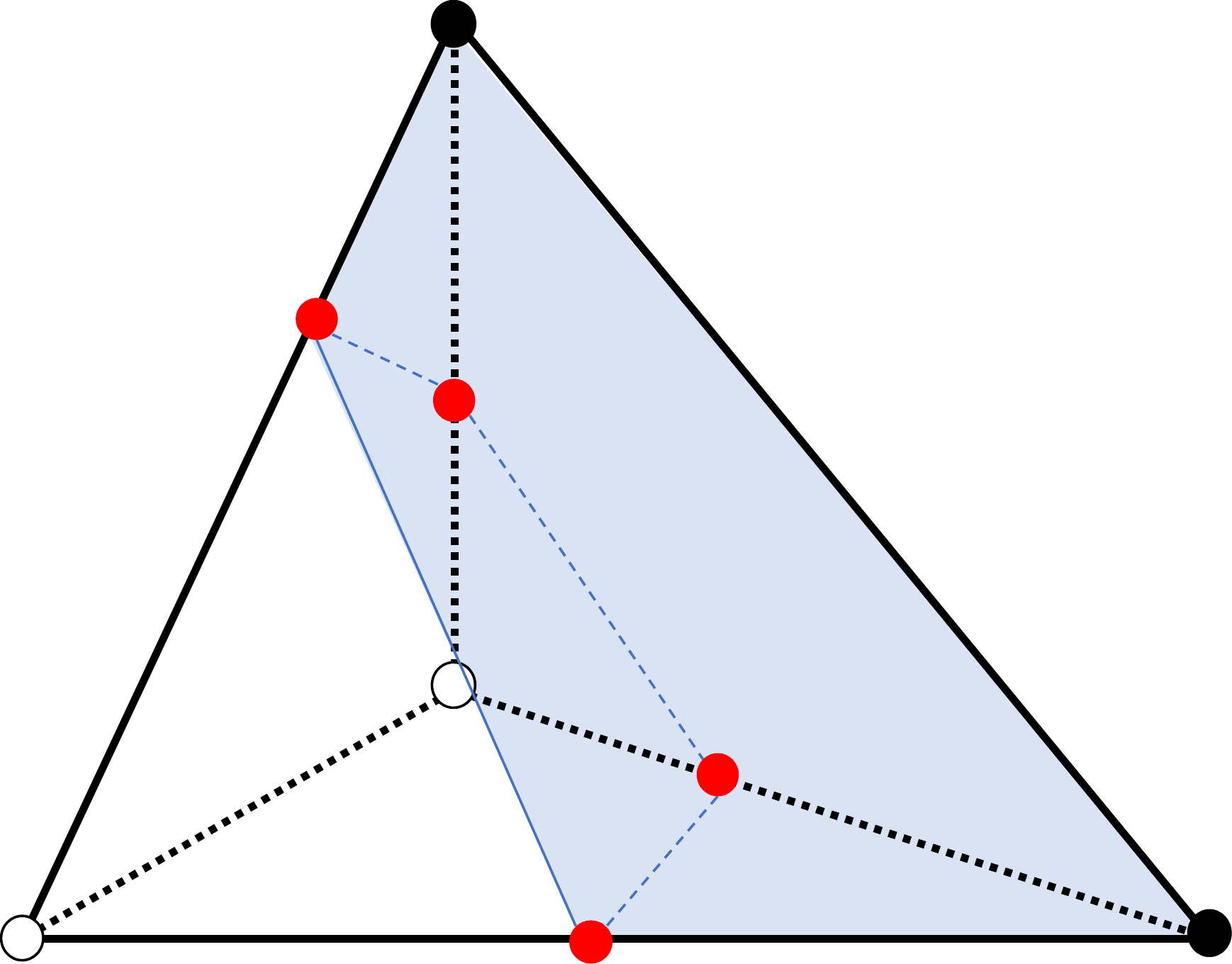}
        \caption{$0$ edges have two valid cuts.}
        \label{fig:2fluid_0double}
    \end{subfigure}
    \hspace{0.10\textwidth}
    \begin{subfigure}[b]{0.30\textwidth}
        \centering
        \includegraphics[width=\textwidth]{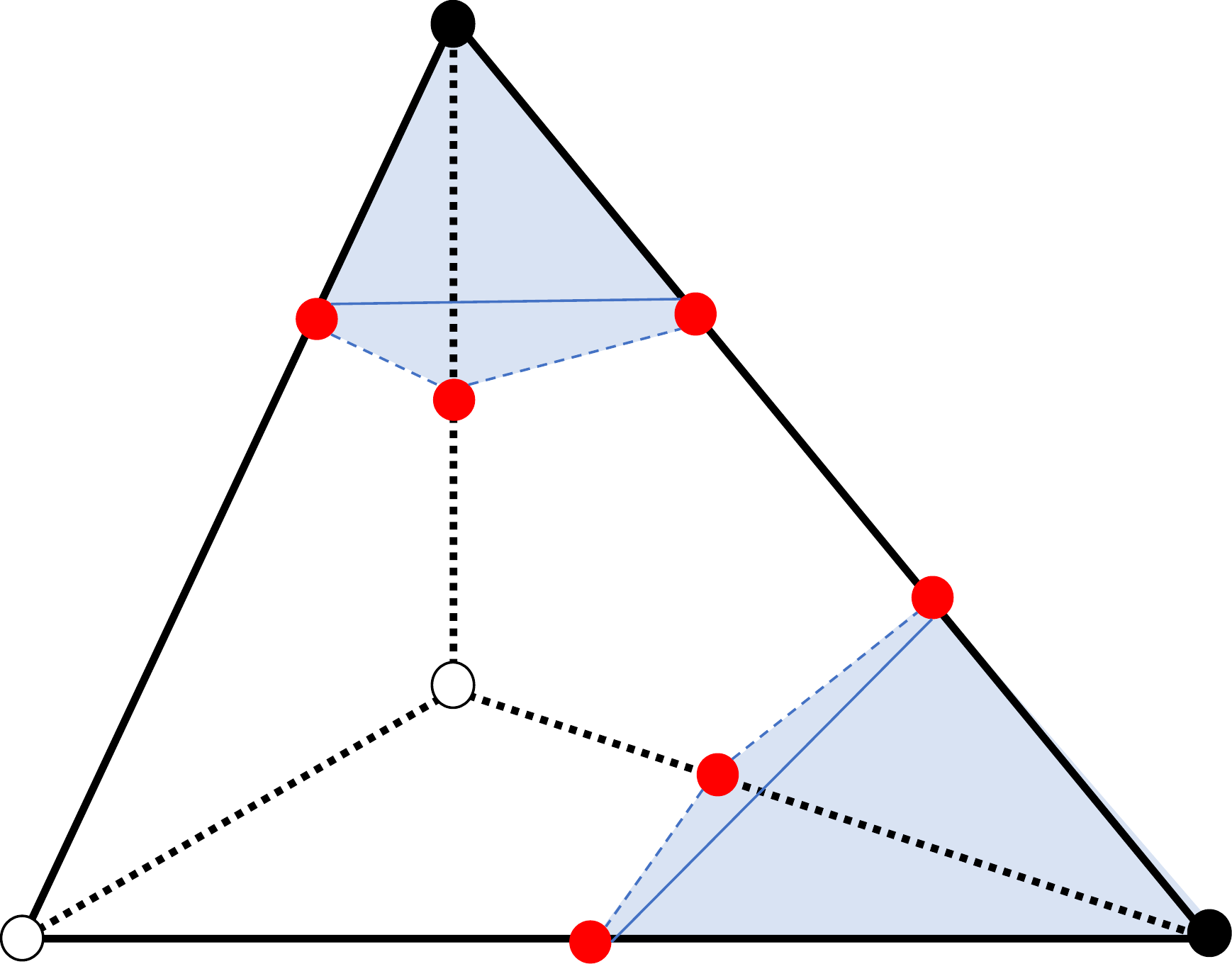}
        \caption{$1$ edge has two valid cuts.}
        \label{fig:2fluid_1double}
    \end{subfigure}

    \vspace{0.5cm}

    \begin{subfigure}[b]{0.30\textwidth}
        \centering
        \includegraphics[width=\textwidth]{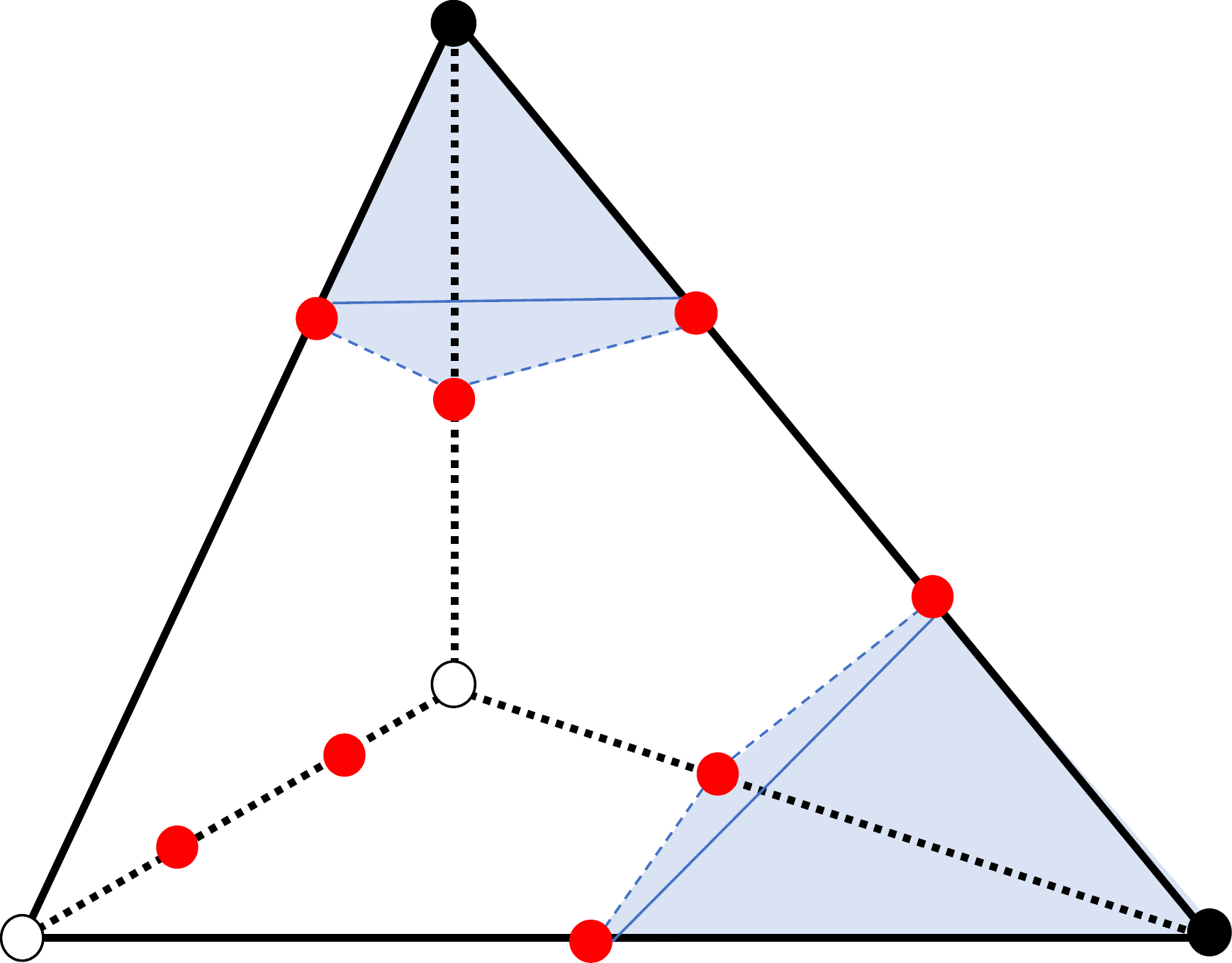}
        \caption{$2$ edges have two valid cuts.}
        \label{fig:2fluid_2double_a}
    \end{subfigure}
    \hspace{0.10\textwidth}
    \begin{subfigure}[b]{0.30\textwidth}
        \centering
        \includegraphics[width=\textwidth]{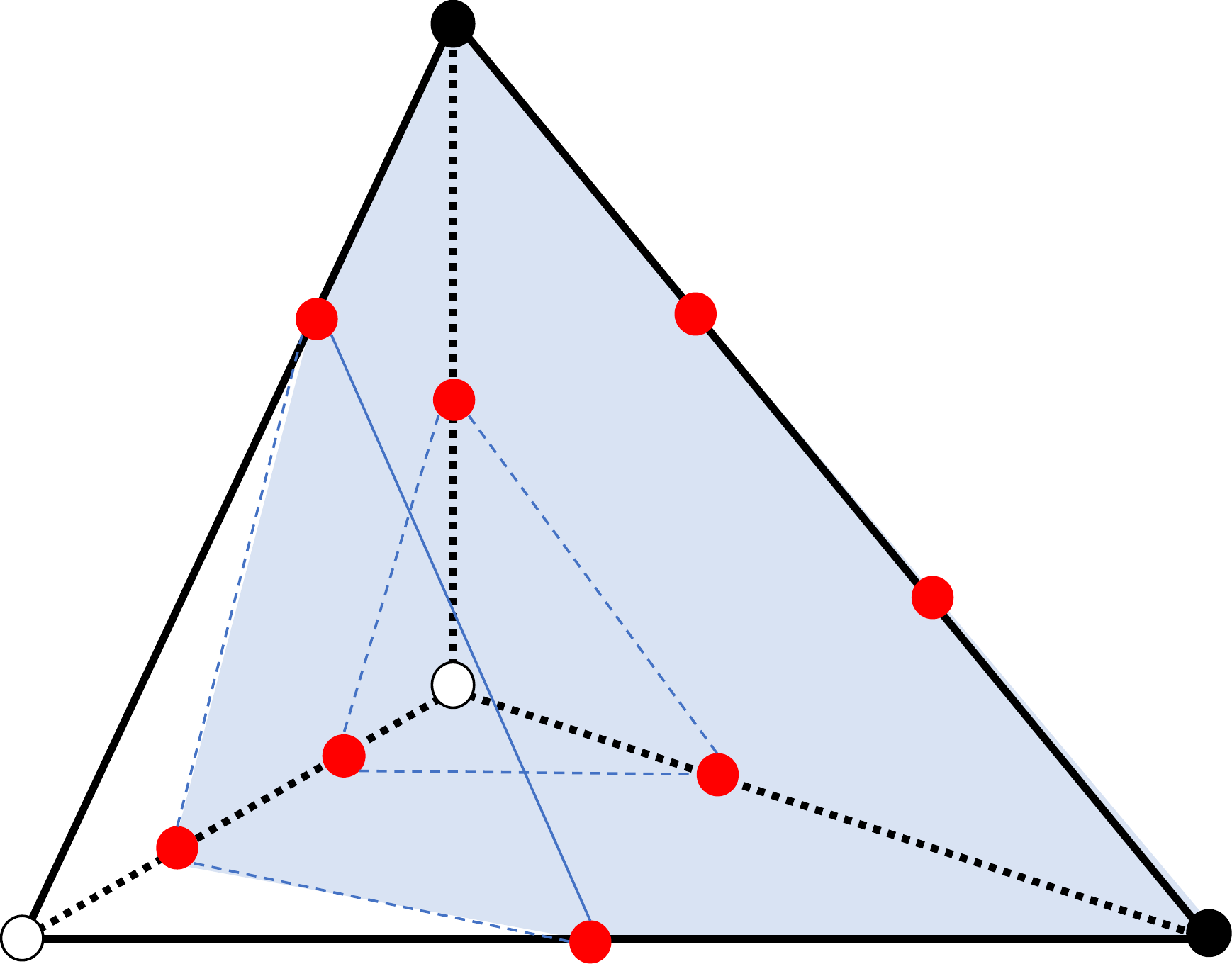}
        \caption{$2$ edges have two valid cuts.}
        \label{fig:2fluid_2double_b}
    \end{subfigure}
    \caption{Four possible cases with $2$ fluid vertices and $2$ air vertices. (a) No edges have two valid cuts. One interface is reconstructed from four cut points. (b) One edge has two valid cuts. Two interfaces are reconstructed. (c) Two edges have two valid cuts. Cuts on one of them are ignored and the reconstruction is the same as (b). (d) Another possible reconstruction of (c). }
    \label{fig:2fluid}
\end{figure}

\paragraph{Tetrahedron with $1$ fluid vertex.} Suppose that the four vertices of the tetrahedron are $\bm{v}_1,\bm{v}_2,\bm{v}_3$ and $\bm{v}_4$, and $\chi(\bm{v}_1)=1$ while $\chi(\bm{v}_2)=\chi(\bm{v}_3)=\chi(\bm{v}_4)=0$. We denote the edge connecting $\bm{v}_1$ and $\bm{v}_2$ as $e_{12}$, and so on. In this case, there must be one valid cut on edges $e_{12}$, $e_{13}$, $e_{14}$, and in the remaining three edges, there may be $\{0,1,2,3\}$ edges with two valid cuts. These four cases are summarized in Figure~\ref{fig:1fluid}. In Figure~\ref{fig:1fluid_1double}, we have to discard two cuts, introducing some geometric errors. In Figure~\ref{fig:1fluid_2double}, one interface is defined by four vertices, and we need to fit a plane based on their positions.

\begin{figure}[h]
    \centering
    \begin{subfigure}[b]{0.30\textwidth}
        \centering
        \includegraphics[width=\textwidth]{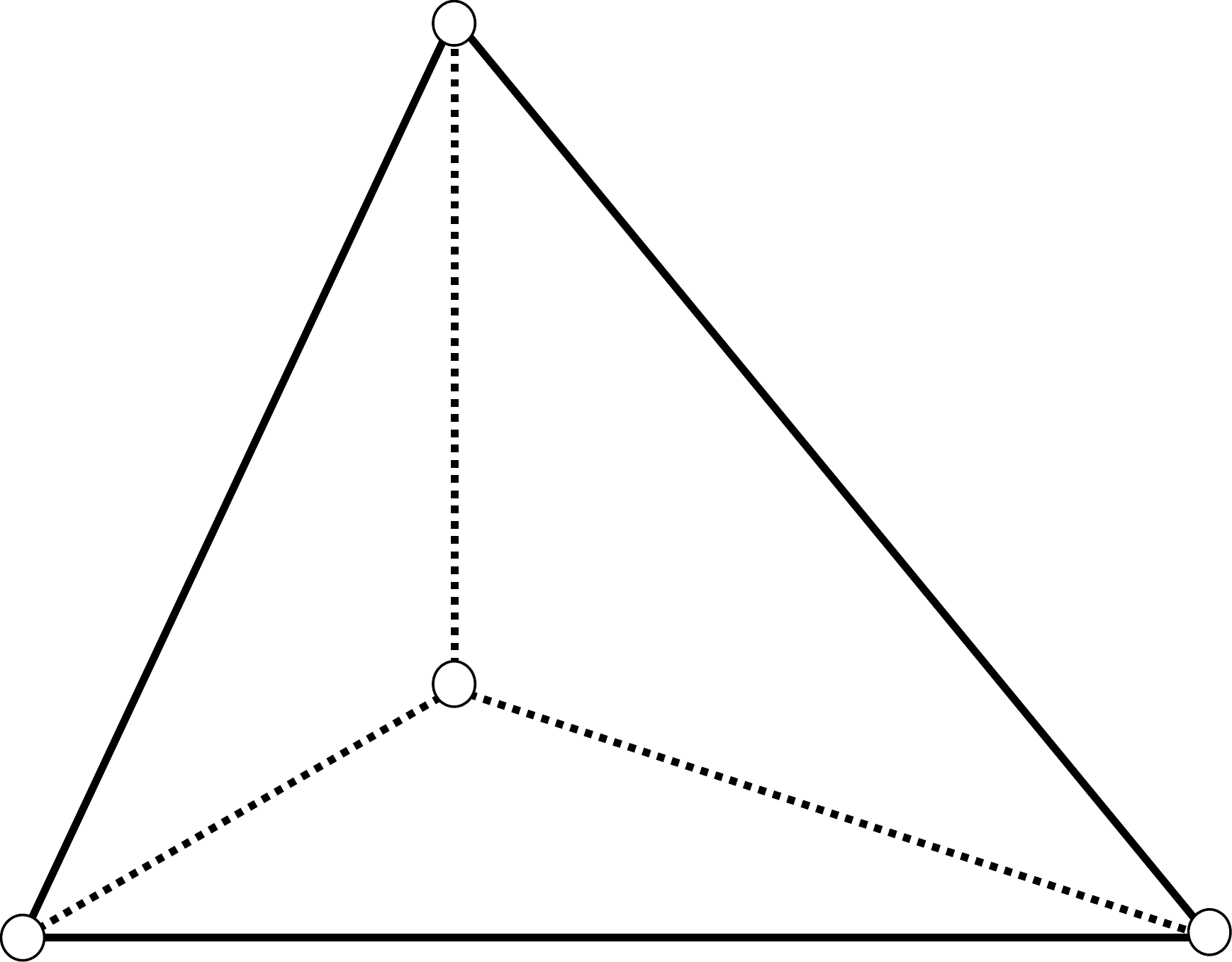}
        \caption{$0$ edges have two valid cuts.}
        \label{fig:0fluid_0double}
    \end{subfigure}
    \hspace{0.10\textwidth}
    \begin{subfigure}[b]{0.30\textwidth}
        \centering
        \includegraphics[width=\textwidth]{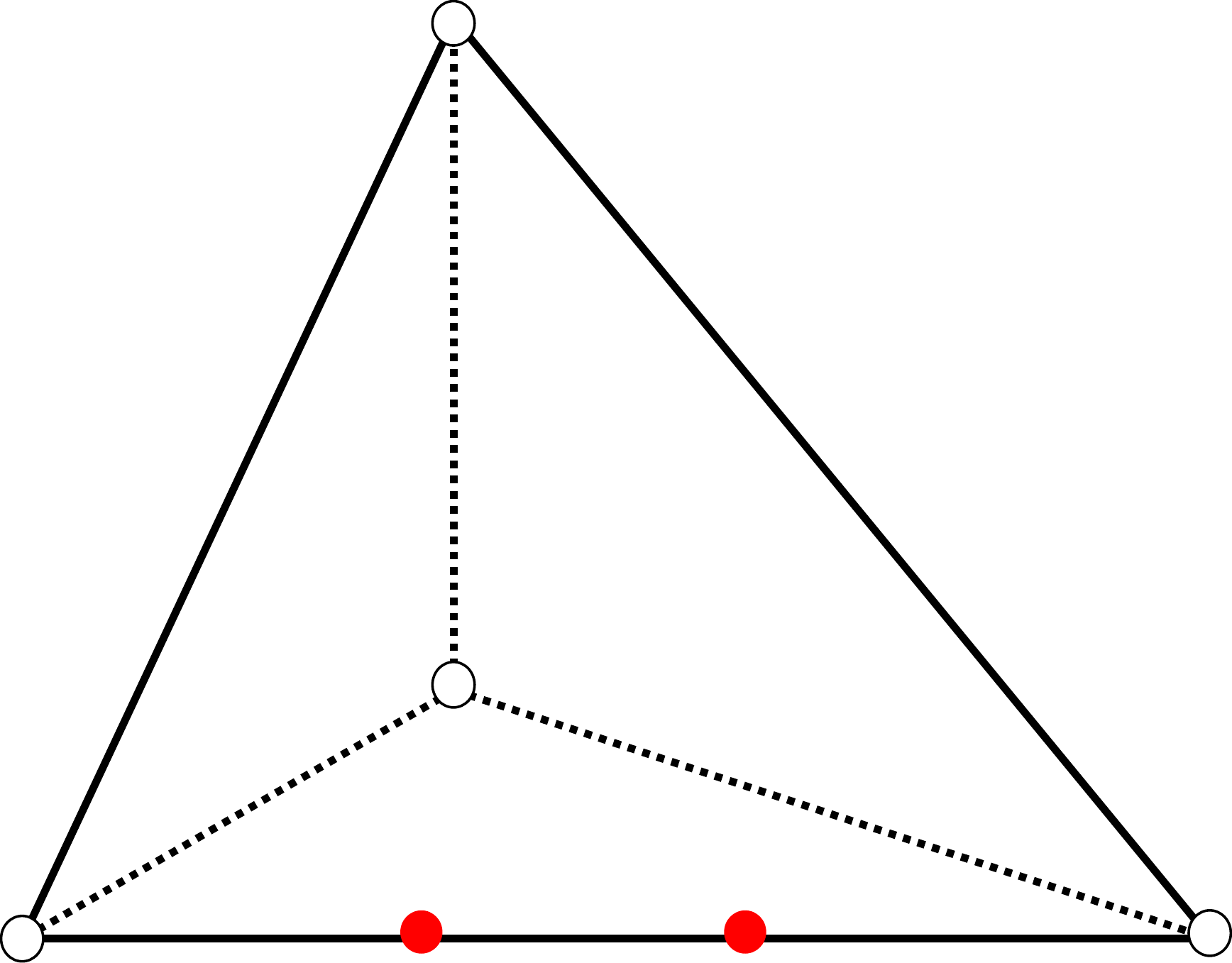}
        \caption{$1$ edge has two valid cuts.}
        \label{fig:0fluid_1double}
    \end{subfigure}

    \vspace{0.5cm}

    \begin{subfigure}[b]{0.30\textwidth}
        \centering
        \includegraphics[width=\textwidth]{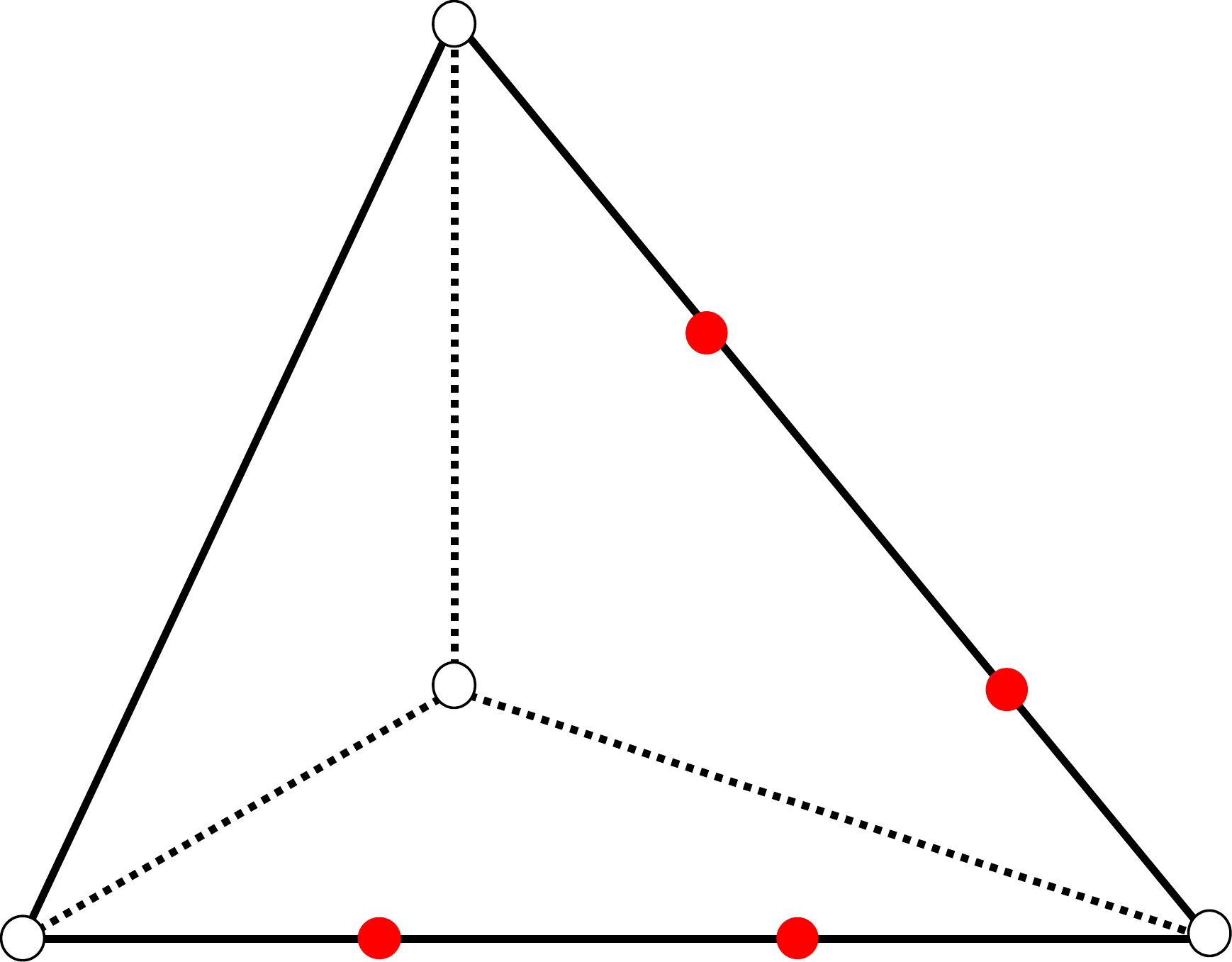}
        \caption{$2$ edges have two valid cuts.}
        \label{fig:0fluid_2double_a}
    \end{subfigure}
    \hspace{0.10\textwidth}
    \begin{subfigure}[b]{0.30\textwidth}
        \centering
        \includegraphics[width=\textwidth]{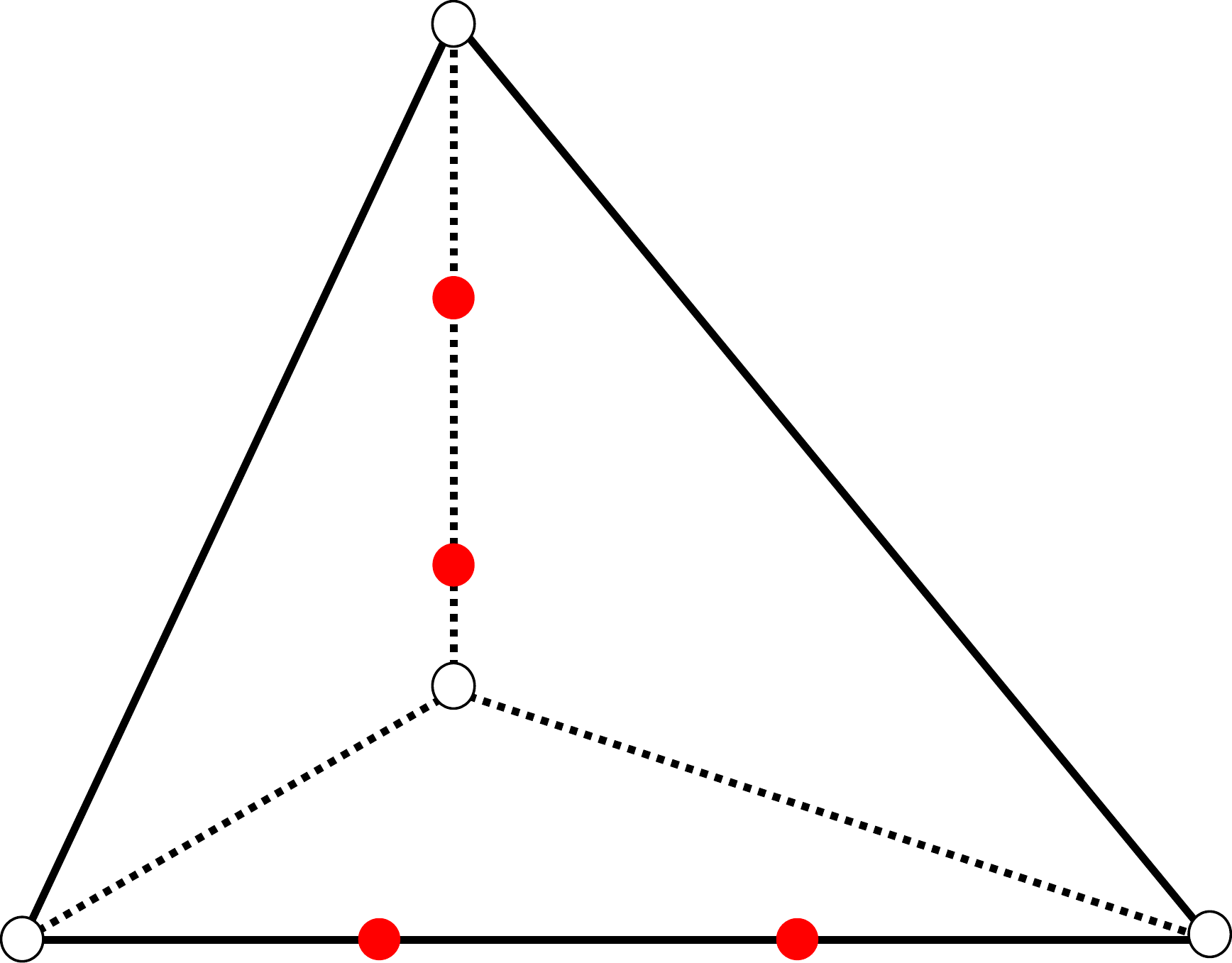}
        \caption{$2$ edges have two valid cuts.}
        \label{fig:0fluid_2double_b}
    \end{subfigure}
    \caption{Four possible cases with $0$ fluid vertices and $4$ air vertices, where there are $\{0,1,2\}$ edges with two valid cuts on them. (a) No edges have two valid cuts. (b) One edge has two valid cuts. (c) Two adjacent edges have two valid cuts. (d) Two non-adjacent edges have two valid cuts. \revminor{In} all cases, edge cuts are ignored and no interface is reconstructed.}
    \label{fig:0fluid_012double}
\end{figure}

\paragraph{Tetrahedron with $2$ fluid vertices.} Suppose that $\chi(\bm{v}_1)=\chi(\bm{v}_2)=1$ and $\chi(\bm{v}_3)=\chi(\bm{v}_4)=0$. There must be one valid cut on edges $e_{13}$, $e_{14}$, $e_{23}$ and $e_{24}$. In the remaining two edges $e_{12}$ and $e_{34}$, there might be $\{0,1,2\}$ edges with two valid cuts. Figure~\ref{fig:2fluid} summarizes these situations. Notably,  ambiguity arises in \revminor{Figures}~\ref{fig:2fluid_2double_a} and \ref{fig:2fluid_2double_b}  that they represent two different interface reconstructions for the same edge cut, which cannot be distinguished based on the edge cut information alone.

\begin{figure}[h]
    \centering
    \begin{subfigure}[b]{0.30\textwidth}
        \centering
        \includegraphics[width=\textwidth]{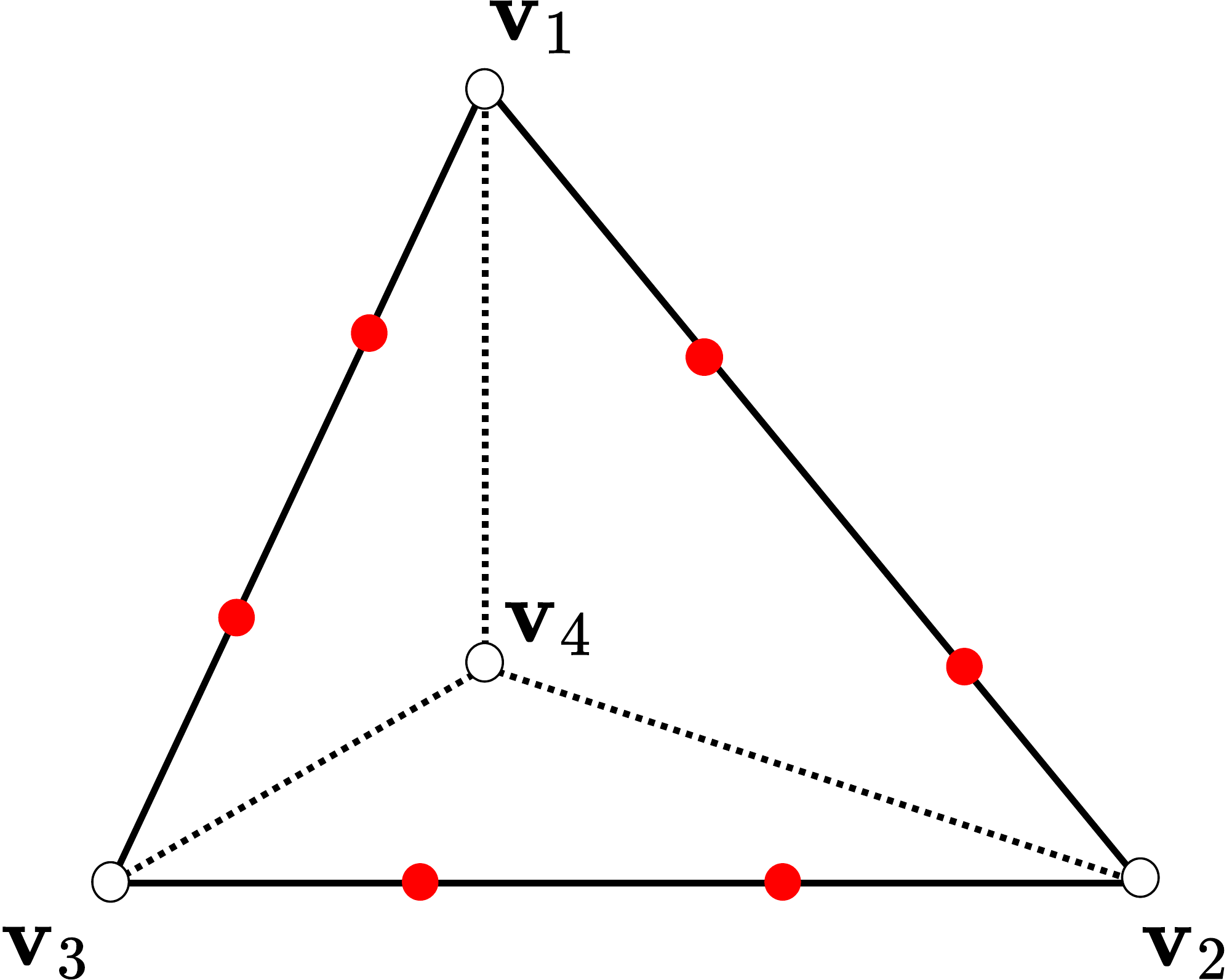}
        \caption{$3$ edges have two valid cuts.}
        \label{fig:0fluid_3double_a}
    \end{subfigure}
    %\hspace{0.05\textwidth}
    \begin{subfigure}[b]{0.30\textwidth}
        \centering
        \includegraphics[width=\textwidth]{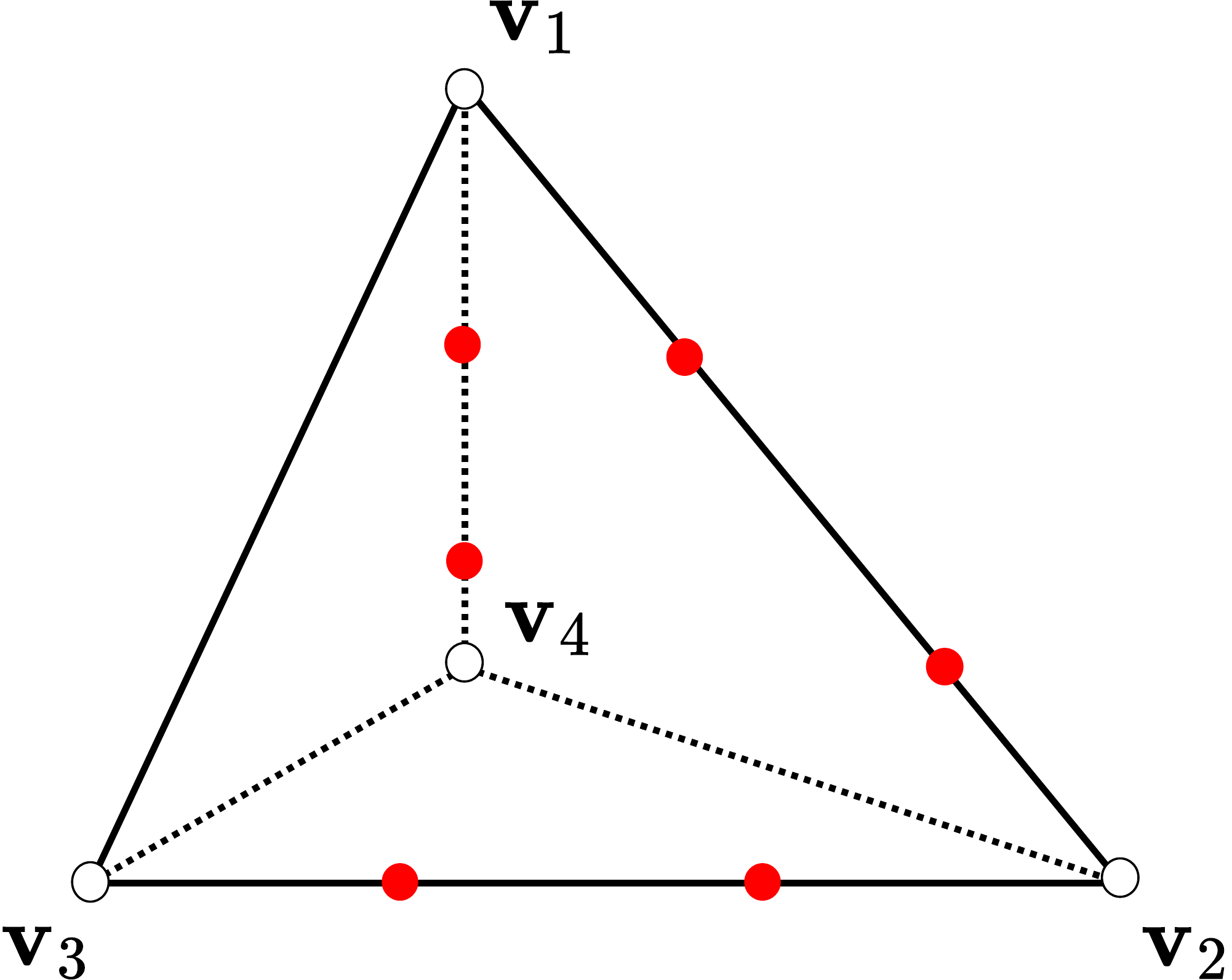}
        \caption{$3$ edge has two valid cuts.}
        \label{fig:0fluid_3double_b}
    \end{subfigure}
    %\hspace{0.05\textwidth}
    \begin{subfigure}[b]{0.30\textwidth}
        \centering
        \includegraphics[width=\textwidth]{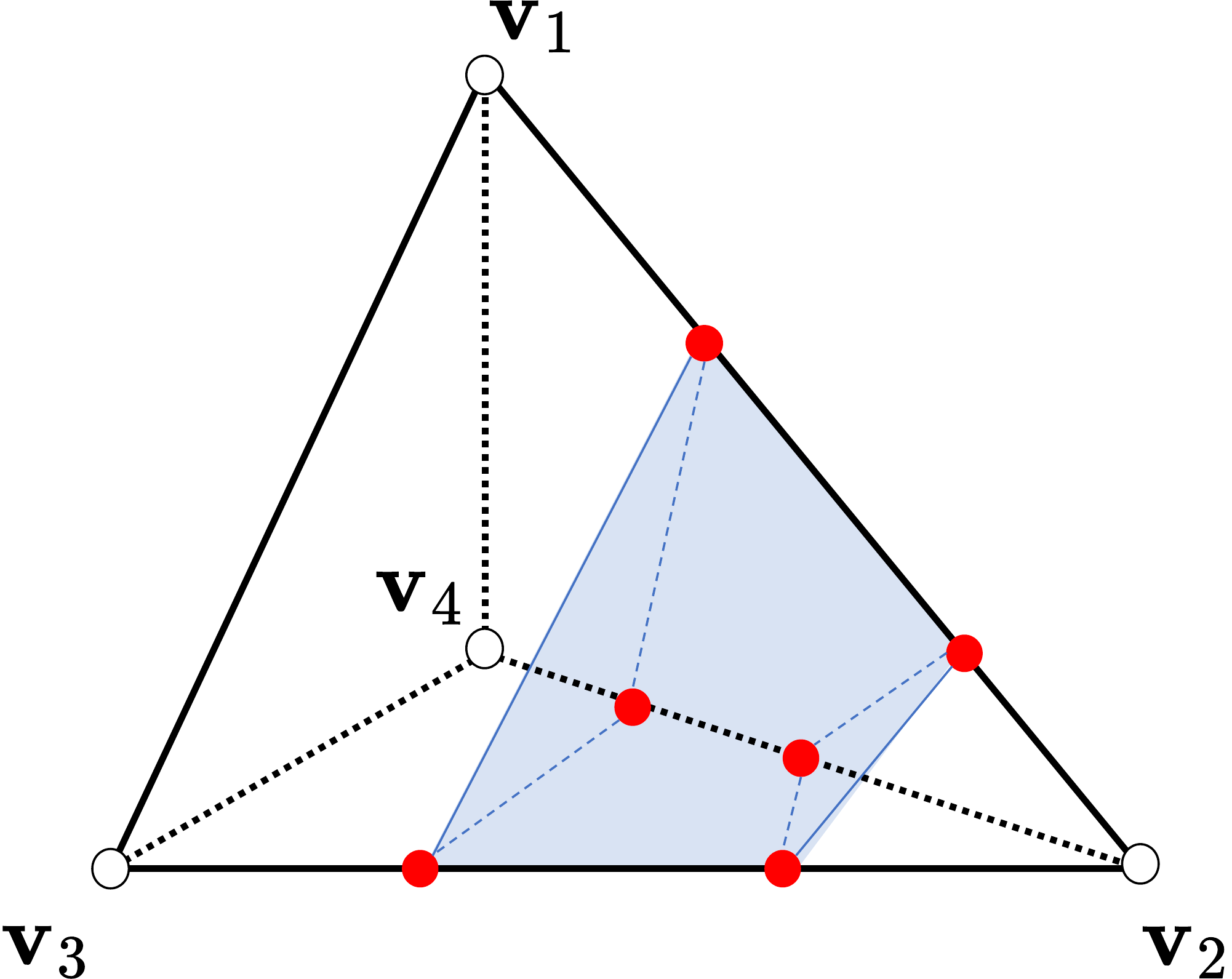}
        \caption{$3$ edges have two valid cuts.}
        \label{fig:0fluid_3double_c}
    \end{subfigure}
    
    \caption{Three possible cases with $0$ fluid vertices and there are $3$ edges with two valid cuts on them. Two of them are edges $e_{12}$ and $e_{23}$. (a) The third edge is $e_{13}$. (b) The third edge is $e_{14}$, which is equivalent to the case of $e_{34}$. (c) The third edge is $e_{24}$, where two interfaces are reconstructed.}
    \label{fig:0fluid_3double}
\end{figure}

\begin{figure}[h]
    \centering
    \begin{subfigure}[b]{0.30\textwidth}
        \centering
        \includegraphics[width=\textwidth]{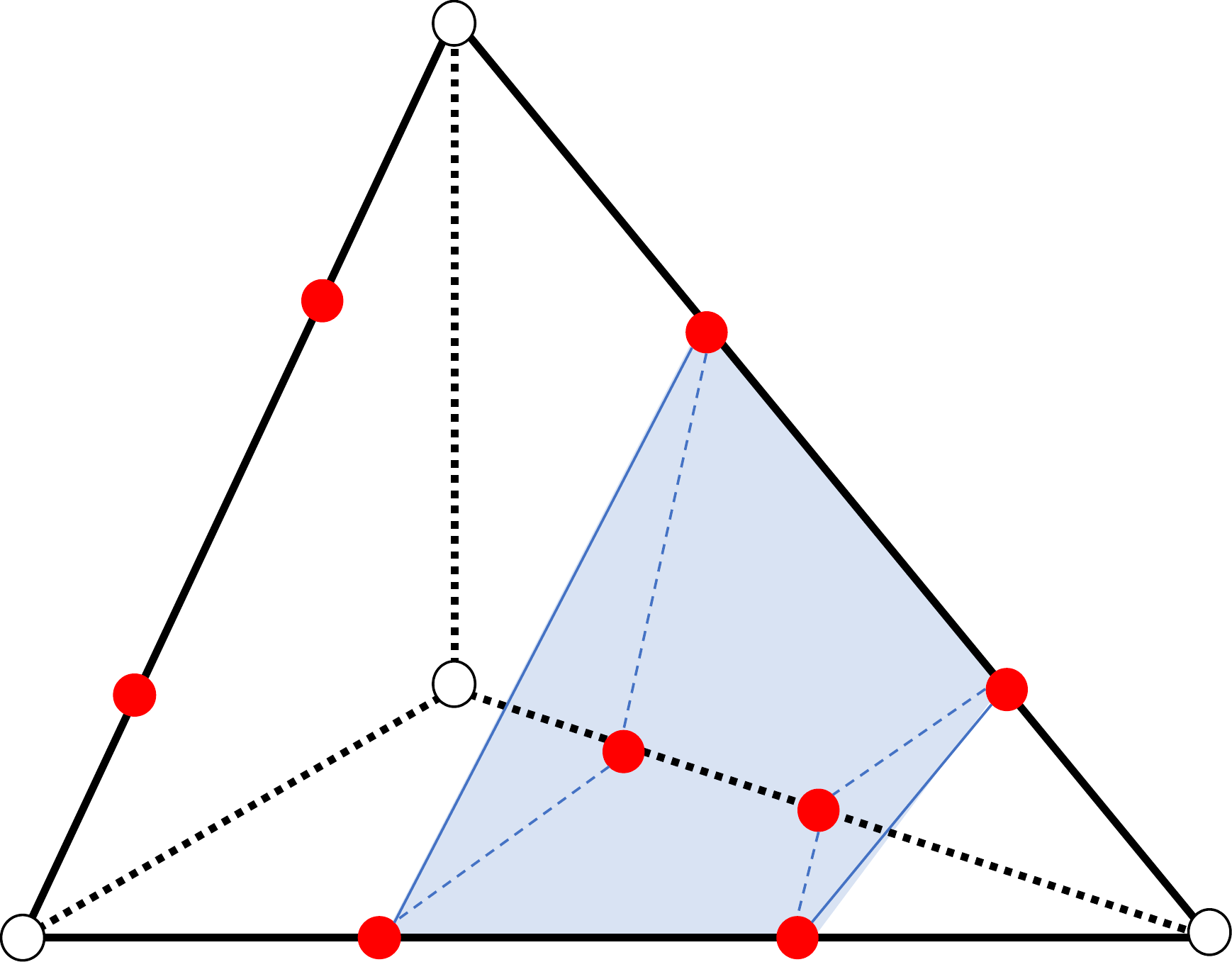}
        \caption{$4$ edges have two valid cuts.}
        \label{fig:0fluid_4double_a}
    \end{subfigure}
    \hspace{0.10\textwidth}
    \begin{subfigure}[b]{0.30\textwidth}
        \centering
        \includegraphics[width=\textwidth]{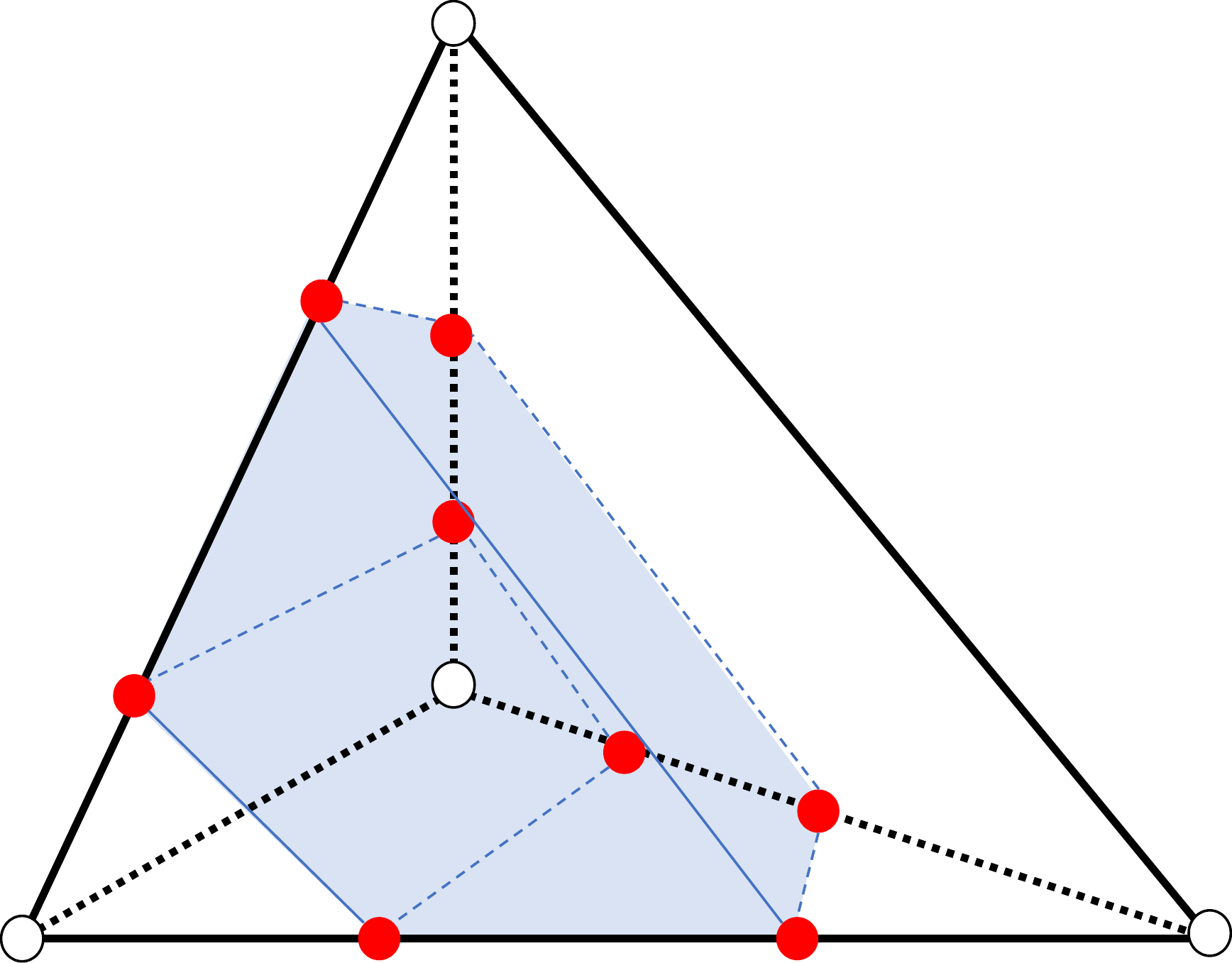}
        \caption{$4$ edges have two valid cuts.}
        \label{fig:0fluid_4double_b}
    \end{subfigure}

    \vspace{0.5cm}

    \begin{subfigure}[b]{0.30\textwidth}
        \centering
        \includegraphics[width=\textwidth]{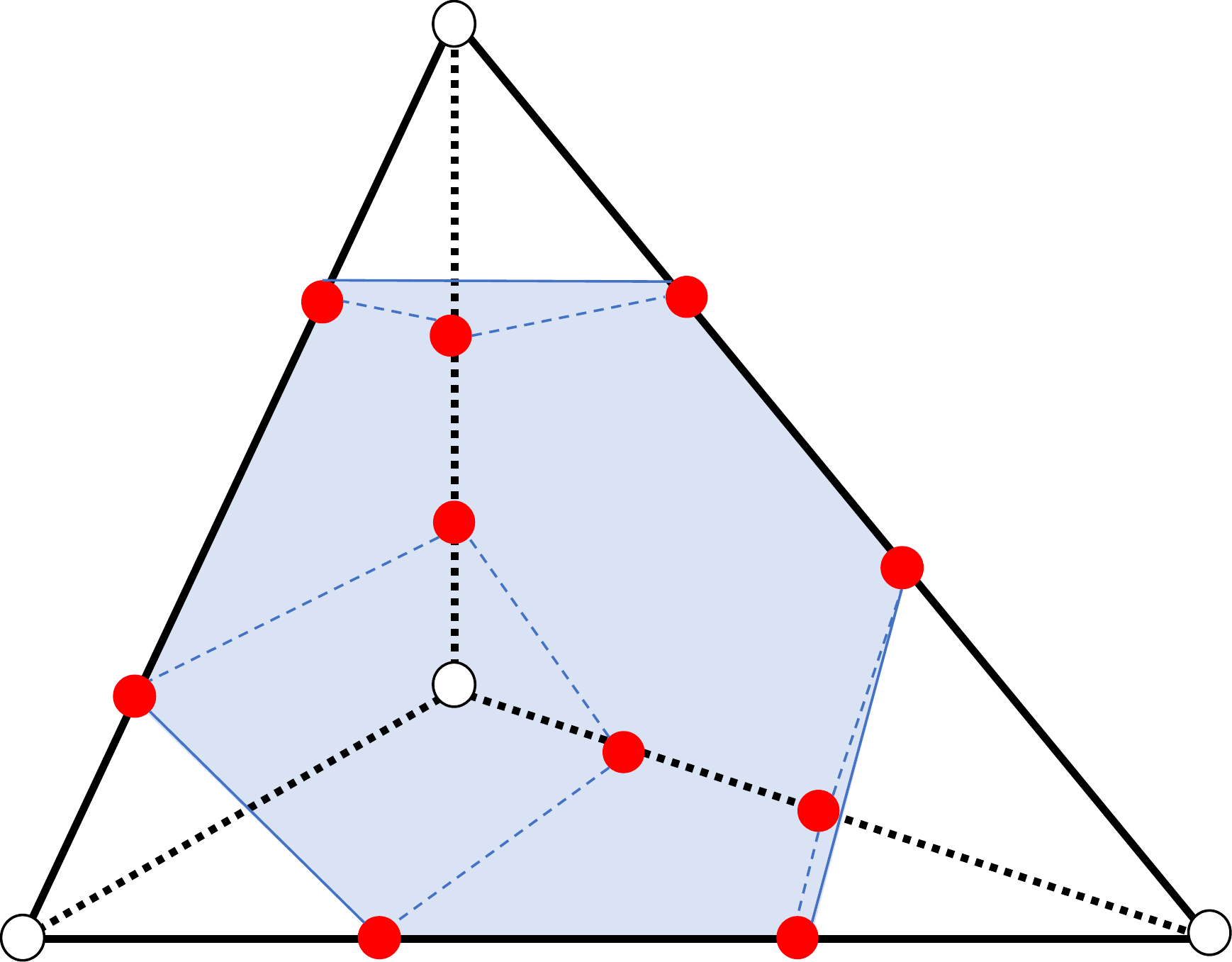}
        \caption{$5$ edges have two valid cuts.}
        \label{fig:0fluid_5double}
    \end{subfigure}
    \hspace{0.10\textwidth}
    \begin{subfigure}[b]{0.30\textwidth}
        \centering
        \includegraphics[width=\textwidth]{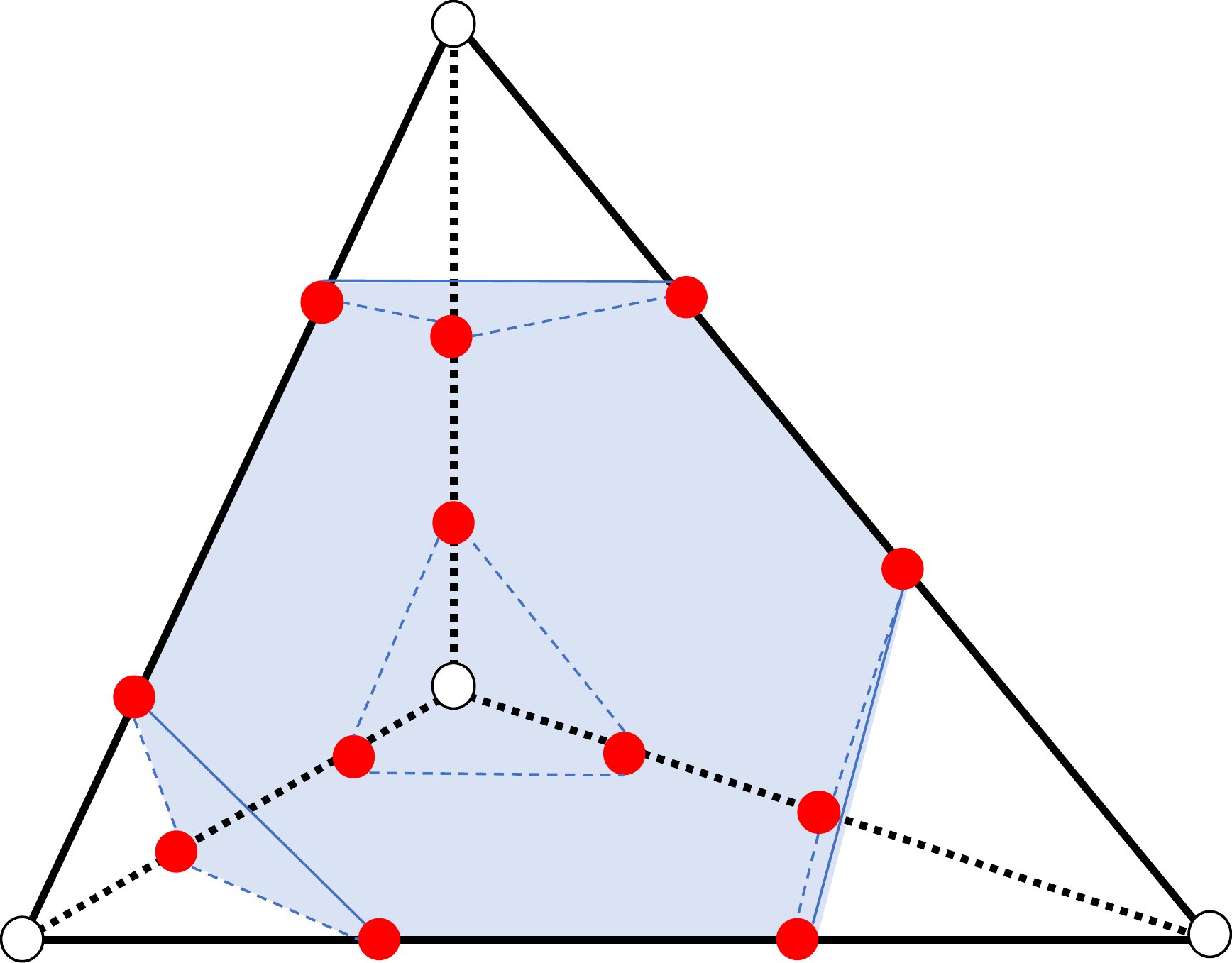}
        \caption{$6$ edges have two valid cuts.}
        \label{fig:0fluid_6double}
    \end{subfigure}
    \caption{Four possible cases with $0$ fluid vertices and $\{4,5,6\}$ edges with two valid cuts on them. (a) $4$ edges have two valid cuts and the other two adjacent edges have no valid cuts. (b) $4$ edges have two valid cuts and the other two non-adjacent edges have no valid cuts. (c) $5$ edges have two valid cuts. (d) $6$ edges have two valid cuts.}
    \label{fig:0fluid_456double}
\end{figure}

\paragraph{Tetrahedron with $0$ fluid vertices.} \revminor{In this case,} we have $\chi(\bm{v}_1)=\chi(\bm{v}_2)=\chi(\bm{v}_3)=\chi(\bm{v}_4)=0$, and there might be $\{0,1,2,3,4,5,6\}$ edges that have two valid cuts on them. No interface \revminor{is} reconstructed for $\{0,1\}$ edges, as shown in \revminor{Figures}~\ref{fig:0fluid_0double} and \ref{fig:0fluid_1double}. We will further simplify our discussion utilizing the fact that one tetrahedron edge is adjacent to $4$ other edges, and only $1$ edge is non-adjacent to it. If there are $2$ edges with two valid cuts, they can be either adjacent, like Figure~\ref{fig:0fluid_2double_a}, or non-adjacent, like Figure~\ref{fig:0fluid_2double_b}. In either case, there is no interface within the tetrahedron. Figure~\ref{fig:0fluid_3double} discusses three different cases in which $3$ edges have two valid cuts. Since an edge is non-adjacent to only one edge in the tetrahedron, there must be two adjacent edges in three, \revminor{which} we assume to be $e_{12}$ and $e_{23}$. The third edge may be $e_{13}$ (Figure~\ref{fig:0fluid_3double_a}), $e_{14}$ and $e_{34}$, which are equivalent (Figure~\ref{fig:0fluid_3double_b}), \revminor{or} $e_{24}$ (Figure~\ref{fig:0fluid_3double_c}). Only in the case of $e_{24}$, an interface is shown in the tetrahedron. Next, if there are $4$ edges with two cuts each, it implies that there are $2$ edges with no cuts. These two edges can either be adjacent (Figure~\ref{fig:0fluid_4double_a}) or non-adjacent (Figure~\ref{fig:0fluid_4double_b}). Finally, \revminor{Figures}~\ref{fig:0fluid_5double} and \ref{fig:0fluid_6double} show the cases that $5$ and $6$ edges have two valid cuts.

Based on the above discussion, there are a total of $18$ basic cases for reconstructing the interface within a tetrahedron using edge cut information. Among these, $8$ cases will ignore some cut points, leading to geometric errors in the interface reconstruction. Additionally, if there are $2$ fluid vertices and $2$ edges with two valid cuts, we encounter an ambiguity where it is not possible to determine whether it corresponds to Figure~\ref{fig:2fluid_2double_a} or Figure~\ref{fig:2fluid_2double_b}. While extending the algorithm to $\mathbb{R}^3$ is non-trivial due to these reasons, We anticipate addressing the aforementioned challenges in future work.

\section{Conclusion}
\label{sec:conclusion}

In this paper, we proposed an Eulerian interface tracking algorithm \revminor{for} unstructured triangle \revminor{meshes} based on the triangle edge cut interface representation. We also designed an area correction algorithm to further improve the accuracy of mass \revminor{conservation}. Our interface representation method features a low level of memory usage that can capture sub-triangle geometric features with $6$ DoFs in a triangle. The proposed interface advection algorithm, \revminor{combined} with area correction, can track the dynamic evolution of the interface on a low-cost, fully definitive basis, without any \revminor{expensive} optimization process. Our algorithm \revminor{handles} each triangle cell independently in a parallelization-friendly manner. On different numerical tasks, it's proven that our method outperforms traditional VOF methods, offering a promising \revminor{approach} for accurate and efficient interface tracking in complex numerical simulations.

\begin{figure}[htbp]
  \centering
  \includegraphics[width=0.6\textwidth]{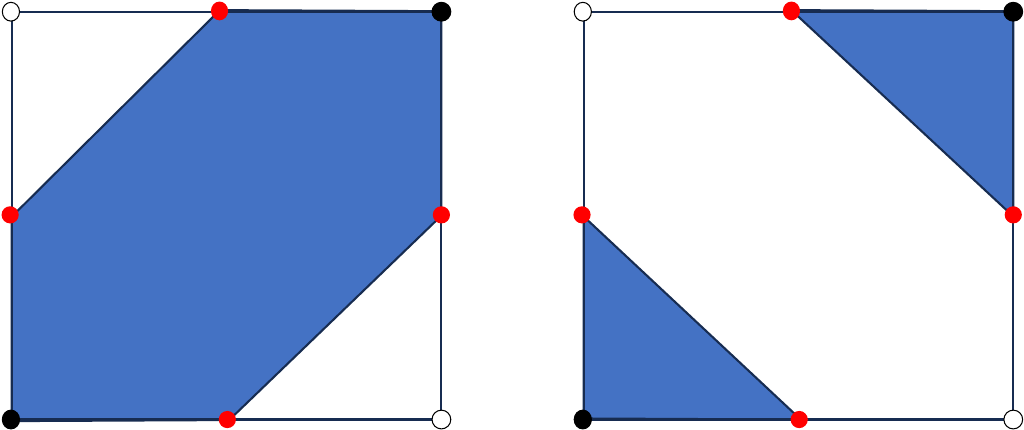}
  \caption{Ambiguities of \revminor{rectangular} cell edge cut reconstruction.}
  \label{fig:cell-ambiguity}
\end{figure}

The algorithm proposed in this paper requires an unstructured triangle mesh because \revminor{applying} this method \revminor{to} a lattice grid \revminor{can} introduce ambiguities \revminor{due to the rectangular cells}. For example, the two cases in Figure~\ref{fig:cell-ambiguity} have identical vertex materials and edge cuts, however exhibit different liquid polygons, \revminor{leading to} ambiguity. Another limitation is that the triangle edge cut representations will smooth out sharp corners inside triangles, producing some artifacts, which may be alleviated by adding more vertices within the triangles. Furthermore, the use of CGAL's exact geometry computation slows down runtime and hinders parallelization.

In the future, we plan to address the dimensionality challenges discussed in Section~\ref{sec:discussion-3d} and extend our proposed edge cut-based interface tracking method to 3D space. To enhance efficiency, we aim to adopt faster floating-point geometric algorithms to reduce the computational cost of polygon-polygon intersections and further accelerate the program through parallelization. We also intend to integrate the proposed algorithm with physical simulations to accurately model fluid phenomena with thin structures.

\section{Acknowledgment}

Georgia Tech authors acknowledge NSF IIS \#2433322, ECCS \#2318814, CAREER \#2433307, IIS \#2106733, OISE \#2433313, and CNS \#1919647 for funding support.

%\appendix

%\section{Sample Appendix Section}

 \bibliographystyle{elsarticle-num} 
 \bibliography{cas-refs}





% \end{thebibliography}
\end{document}